\documentclass[a4paper,11pt]{article}
\usepackage{jheppub}
\usepackage{mathrsfs}
\usepackage{verbatim}
\usepackage{slashed}

\newcommand{\R}{\mathbb{R}}

\newcommand{\C}{\mathbb{C}}
\newcommand{\CP}{\mathbb{CP}}

\newcommand{\nn}{\nonumber}
\newcommand{\mc}[1]{\mathcal{#1}}

\newcommand{\pd}[2]{\frac{\partial #1}{\partial #2}}
\newcommand{\tree}{\mathrm{tree}}
\newcommand{\spaa}[1]{\langle#1\rangle}
\newcommand{\spbb}[1]{[#1]}
\newcommand{\spab}[3]{\langle#1|#2|#3]}

\newcommand{\ket}[1]{| #1 \rangle}

\newcommand{\spvec}[3]{\langle #1|\hspace*{.2mm}#2\hspace*{.3mm}|#3 \rangle}
\newcommand{\Res}{\mathop{\rm Res}}
\newcommand{\X}{X_{2,1,1}^{**}}
\newcommand{\XX}{X_{2,1,1}}

\newcommand{\cut}{\operatorname{cut}}
\newcommand{\bxi}{\bar{\xi}}
\newcommand{\W}{\mathbf\Omega}
\newcommand{\tW}{\widetilde{\mathbf\Omega}}
\newcommand{\tS}{\widetilde{\mc S}}
\newcommand{\tG}{\widetilde{\mc G}}
\newcommand{\tomega}{\widetilde{\omega}}

\newcommand{\bc}{\begin{center}}
\newcommand{\ec}{\end{center}}

\newcommand{\LRa}{\Longrightarrow}

\newcommand{\LLra}{\Longleftrightarrow}

\newtheorem{thm}{Theorem}

\newtheorem{example}{Example} 

\allowdisplaybreaks[1]

\title{Massive Nonplanar Two-Loop Maximal Unitarity}
\author{Mads S{\o}gaard}
\author{and Yang Zhang}
\affiliation{
Niels Bohr International Academy and Discovery Center, Niels Bohr Institute, \\
University of Copenhagen, Blegdamsvej 17, DK-2100 Copenhagen, Denmark
}
\emailAdd{madss@nbi.dk}
\emailAdd{zhang@nbi.dk}

\abstract{We explore maximal unitarity for nonplanar two-loop integrals with up
to four massive external legs. In this framework, the amplitude is reduced to a
basis of master integrals whose coefficients are extracted from maximal cuts.
The hepta-cut of the nonplanar double box defines a nodal algebraic curve
associated with a multiply pinched genus-3 Riemann surface. All possible
configurations of external masses are covered by two distinct topological
pictures in which the curve decomposes into either six or eight Riemann spheres.
The procedure relies on consistency equations based on vanishing of integrals of
total derivatives and Levi-Civita contractions. Our analysis indicates that
these constraints are governed by the global structure of the maximal cut.
Lastly, we present an algorithm for computing generalized cuts of massive
integrals with higher powers of propagators based on the Bezoutian matrix
method.}

\begin{document}
\maketitle
\flushbottom
\clearpage

\section{Introduction}
Contemporary experimental high energy physics is concentrated on the Large
Hadron Collider (LHC) at CERN. Our ability to utilize the huge amount of data
delivered by the experiment towards further scientific progress relies on a
quantitative understanding of all relevant scattering processes in the Standard
Model. Otherwise, we are unable to extract signals of new physics from the
background. Precise theoretical predictions in Quantum Chromodynamics (QCD) at
the LHC require not only amplitudes at leading order (LO) and next-to-leading
order (NLO), but also next-to-next-to leading order (NNLO) corrections to comply
with the level of accuracy of the data. For some processes, two-loop amplitudes
are important already at NLO because the LO terms begin at one loop.

The text book approach to perturbative scattering amplitudes is through Feynman
rules and diagrams. Although it tracks interactions of particles very
intuitively and in principle always works, this method suffers from severe
computational problems with increasing loop level or number of external legs.
The main reason is that the gauge redundancy of the theory introduces virtual
intermediate states that are off-shell. Not even very powerful computers are
able to deal with many of the phenomenologically interesting processes without
new clever ways to attack the problem.

The last two decades have seen many attempts to surmount the computational
bottleneck. The lesson is to exploit analyticity and unitarity of the scattering
matrix. Analyticity allows for amplitudes to be reconstructed from their
singularity structure, whereas by unitarity, residues at the poles factorize
onto products of simple amplitudes. Two of the most successful advances are the
original unitarity method for loop amplitudes developed by Bern, Dixon, Dunbar
and Kosower \cite{Bern:1994cg,Bern:1994zx} and the Britto-Cachazo-Feng-Witten
(BCFW) recursion relations \cite{Britto:2004ap,Britto:2005fq} for trees. In
these works, striking and otherwise completely unexpected structure and
simplicity are revealed by virtue of retaining only physical on-shell
information in a Lagrangianless setting. In a nutshell, all trees may now be
constructed recursively and further fused into loops.

The basic idea of the unitarity method (see also later studies, e.g.
refs.~\cite{Bern:1995db,Bern:1997sc,Britto:2004nc,Britto:2004nj,
Bern:2005hh,Bidder:2005ri,Britto:2005ha,Britto:2006sj,Mastrolia:2006ki,
Brandhuber:2005jw,Ossola:2006us,Anastasiou:2006gt,Bern:2007dw,Forde:2007mi,
Badger:2008cm,Giele:2008ve,Britto:2006fc,Britto:2007tt,Bern:2010qa,
Anastasiou:2006jv}) is to reconstruct the amplitude from double cuts that place
internal lines in a given channel on their mass-shell and break it into a
product of trees. Many individual contributions share such minimal cuts and are
therefore hard to separate. Therefore intermediate algebraic steps are typically
needed. In that view, the generalized unitarity method
\cite{Britto:2004nc,Forde:2007mi} is more efficient because several propagators
are cut simultaneously and thus fewer integrals are isolated.  Thanks to the
unitarity method, otherwise unfeasible computations of $2\to 2$ massless
scattering processes in QCD have been carried out.

In the last couple of years, two-loop amplitudes have received substantial
attention in the literature. The integrand-level reduction method of Ossola,
Papadopoulos and Pittau (OPP) has been extended to multi loops using
computational algebraic geometry, and a general way of classifying high-loop
unitarity cut solutions is now available
\cite{Badger:2012dp,Mastrolia:2011pr,Zhang:2012ce,Feng:2012bm,
Mastrolia:2012an,Mastrolia:2012wf,Mastrolia:2012du,Kleiss:2012yv,
Huang:2013kh,Badger:2013gxa,Badger:2012dv,CaronHuot:2012ab,Mastrolia:2013kca}.
These techniques were used by Badger, Frellesvig and one of the present authors
to calculate the planar part of the all-plus two-loop five-gluon amplitude in
QCD \cite{Badger:2013gxa} and also demonstrated for the planar triple box
\cite{Badger:2012dv}. In ref.~\cite{Feng:2014nwa} the unitarity method was
applied to two-loop diagrams to determine their integral bases.

Working directly at the level of the integral basis, the maximal unitarity
formalism initiated by Kosower and Larsen in ref.~\cite{Kosower:2011ty} has
emerged as an extension of the quadruple and triple cut at one loop
\cite{Britto:2004nc,Forde:2007mi}. In maximal unitarity one expands the
amplitude in a basis of integrals and seeks to isolate the integral coefficients
by finding multidimensional complex integration contours that are uniquely
associated with each individual master integral. One of the major advantages of
maximal unitarity is that one may circumvent the integrand basis which is
typically considerably larger than the integral basis. After the reduction onto
master integrals is complete, each coefficient is extracted as a linear
combination of residues of the product of trees that arise when the diagram
falls apart on-shell. The tree-level data is easy to manage using superspace
techniques \cite{Sogaard:2011pr,Bern:2009xq}. The leading singularity method
\cite{Buchbinder:2005wp,Cachazo:2008vp} previously addressed hepta-cuts and
octa-cuts at two loops in $\mc N = 4$ super Yang-Mills theory.

Until now, maximal unitarity has remained relatively unexplored in the nonplanar
sector \cite{Sogaard:2013yga}. In this paper, we extend the framework to
two-loop crossed-box integrals with up to four external massive legs. Indeed,
inspection of the nonplanar part of the integral basis for, e.g., $gg \to
V_1V_2$ computed in ref.~\cite{Caola:2014lpa} shows that in practice the
two-loop crossed boxes constitute most of the nonplanar
basis.\footnote{Moreover, we have checked that the remaining crossed triangles
also have leading singularities, and they are therefore expected to be amenable
to the techniques used here.} Remarkably, we find that essentially all features
of maximal unitarity observed in the planar sector
\cite{Kosower:2011ty,Johansson:2012zv,Johansson:2013sda,CaronHuot:2012ab} carry
over directly to the nonplanar sector. In particular, in determining projectors
for the master integrals, we find that the global structure of the maximal cut
seems to govern consistency equations from integration-by-parts (IBP) identities
and the number of master integrals. Moreover, we show that the constraints are
inherited through chiral branchings between distinct classes of hepta-cut
solutions.

The paper is organized in the following way. In section 2 we review the maximal
unitarity method and the theory of multivariate residues. In sections 3--5 we
respectively parametrize the hepta-cut solutions using mutually projecting
kinematics, analyze the global structure of the maximal cut and impose
consistency relations in order to uniquely fix the projectors for the master
integral coefficients for all kinematically inequivalent configurations.
Finally, in section 6, we present an enhanced algorithm to compute degenerate
multivariate residues from generalized unitarity cuts and apply the technique to
massive integrals with doubled propagators.

\section{Maximal Unitarity}
The modern version of the unitarity method relies on the existence of a finite
basis of linearly independent master integrals $\{I_i\}$ onto which the
amplitude in consideration can be expanded, up to additional rational terms, 
\begin{align}
\label{MIEQ}
\mc A_n^{L\text{-loop}} = \sum_{i\,\in\,\mathrm{Basis}}c_iI_i
+\text{rational terms}\;.
\end{align}
Therefore, if the basis integrals are known explicitly in dimensional
regularization, calculating an amplitude boils down to determining the rational
coefficients $\{c_i\}$. The trick is to apply generalized unitarity cuts to
either side of eq.~\eqref{MIEQ},
\begin{align}
\Delta\mc A_n^{L\text{-loop}} = 
\sum_{i\,\in\,\mathrm{Basis}}c_i\Delta I_i\;,
\end{align}
and exploit that the cut amplitude factorizes onto simpler quantities.

At the one-loop level, the basis consists of boxes, triangles and bubbles with
scalar numerators only. Here the computation has already been fully automated,
see refs.~\cite{Britto:2004nc,Forde:2007mi,Badger:2008cm,Mastrolia:2009dr}.  For
instance, a box coefficient is isolated by a quadruple cut and thereby becomes
the product of the trees at the four corners, evaluated in on-shell kinematics.
The on-shell internal lines are complex valued for general external momenta.
This implies that the cut prescription in terms of Dirac Delta functions
necessarily must be reformulated by means of a multidimensional complex contour
integral encircling global poles \cite{Kosower:2011ty,CaronHuot:2012ab}.

At two loops and beyond, the situation is more intricate, one of the main
reasons being that a minimal integral basis is not yet known. Integrals with
numerator insertions are in general algebraically irreducible and the reduction
to master integrals inevitably involves IBP identities. As a consequence,
multiple contributions contaminate the unitarity cuts. Although maximal cuts for
four particles at two and three loops have superficial resemblance to the
quadruple cut at one loop, it is also challenging to extract the coefficients
because the cut does not localize integrals to a point, but rather an algebraic
curve \cite{Kosower:2011ty,CaronHuot:2012ab} or generally speaking, an algebraic
surface \cite{Huang:2013kh,Sogaard:2013fpa}.

In the last couple of years, maximal unitarity has been applied to the planar
double box with up to four external massive legs
\cite{Kosower:2011ty,Johansson:2012zv,Johansson:2013sda} in general theories and
to the massless two-loop crossed box \cite{Sogaard:2013yga}. Recently, the
formalism was also extended to amplitude contributions whose maximal cuts define
multidimensional algebraic varieties \cite{Sogaard:2013fpa}, exemplified for the
planar triple box at three loops. Along these lines, the unitarity cut
prescription has been extended to accommodate loop integrals with doubled or
higher powers of propagators \cite{Sogaard:2014ila}. 

\subsection{Multivariate Residues}
Inspired by the discussion in the introduction, we start by reviewing basic
theory of multivariate residues, with emphasis on computation of nondegenerate
residues. We also refer the reader to classical text books by Griffiths and
Harris \cite{MR507725} and Hartshorne \cite{MR0463157}.

Let $U = \{z\in\C^n:\|z-\xi\|<\epsilon\}$ for $\epsilon>0$ be a small ball
around $z = \xi$ and assume that $f$ and $h$ are holomorphic maps in a
neighborhood of the closure $\bar U$ of $U$. For our purposes, it is in fact
sufficient to think of each component of $f$ and $h$ as just being multivariate
polynomials of certain degrees. Furthermore, suppose that
$f^{-1}(0)\cap U = \{\xi\}$, i.e. the components of $f$ have exactly one
simultaneous zero $\xi\in U$. Then for the meromorphic $n$-form,
\begin{align}
\omega = \frac{h(z)dz_1\wedge\cdots\wedge
dz_n}{f_1(z)\cdots f_n(z)}\;,
\end{align}
the associated residue at $z = \xi$ is computed by an integral over a contour
that is topologically equivalent to a torus of real dimension $n$ embedded in
$\C^n$. In detail we have
\begin{align}
\Res{}_{\{f_1,\dots,f_n\},\xi}(\omega) = 
\frac{1}{(2\pi i)^n}\oint_{\Gamma_\epsilon}
\frac{h(z) dz_1\wedge\cdots\wedge dz_n}{f_1(z)\cdots f_n(z)}\;,
\end{align}
where $\Gamma_\epsilon = \{z\in\C^n:|f_i(z)|=\epsilon_i\}$.

We remark several elementary properties of the residue. The residue is linear in
$h$, but alternating in the $f_i$s. Moreover, the value of a residue is
invariant under nonsingular complex coordinate transformations. It is not hard
to prove by Stokes' theorem that if locally $h\in I_f = \spaa{f_1,\dots,f_n}$,
where $I_f$ is the ideal generated by the $f_i$s, that is,
\begin{equation}
h(z)= a_1(z)f_1(z)+\cdots+a_n(z) f_n(z)\;,
\end{equation}
for holomorphic functions $a_i$ in a neighborhood of $\xi$, then we have
\begin{equation}
\Res{}_{\{f_1,\dots, f_n\},\xi}(\omega)=0\;.
\end{equation}

For the calculation of a residue we distinguish between three classes of
increasing difficulty: factorizable, nondegenerate and degenerate residues. If
each $f_i$ defines a univariate polynomial, i.e. $f_i(z) = f_i(z_i)$, the
contour factorizes onto a product of univariate contours such that the residue
can be obtained in a manner that trivially resembles the one-dimensional case,
\begin{equation}
\Res{}_{\{f_1,\dots, f_n\},\xi}(\omega) = 
\frac{1}{(2\pi i)^n}
\oint_{|f_1(z_1)|=\epsilon_1}
\frac{dz_1}{f_1(z_1)}\cdots
\oint_{|f_n(z_n)|=\epsilon_n}\frac{dz_n}{f_n(z_n)}h(z)\;.
\end{equation}
If the $f_i$s are not univariate polynomials and the Jacobian determinant of
$f_1,\dots,f_n$ evaluated at $z = \xi$ is nonzero,
\begin{align}
J(\xi)\equiv\det_{i,j}\left(
\frac{\partial f_i}{\partial z_j}\right)\bigg|_{z=\xi}\neq 0\;,
\end{align}
the residue is said to be nondegenerate. In that case it is natural to define
the residue via appropriate coordinate transformation as
\begin{align}
\label{LOCALIZATIONPROPERTY}
\Res{}_{\{f_1,\cdots,f_n\},\xi}(\omega) = h(\xi)/J(\xi)\;.
\end{align}
We immediately recognize the localization property
\eqref{LOCALIZATIONPROPERTY} as the obvious generalization of the Dirac delta
function to several complex variables once we define
\begin{align}
\int\bigg[\prod_{k=1}^ndz_k\bigg]
h(z)\prod_{j=1}^n\delta(z_j-\xi_j)\equiv
\frac{1}{(2\pi i)^n}\oint_{\Gamma_\epsilon(\xi)}
dz_1\wedge\cdots\wedge dz_n\frac{h(z)}{\prod_{j=1}^n(z_j-\xi_j)}\;.
\end{align}
In particular, this observation allows us to define generalized unitarity cuts
of amplitude contributions that only factorize for complex kinematics. 

In general, a multivariate residue is neither nondegenerate nor factorizable and
we then proceed by means of computational algebraic geometry and use the
transformation law and Gr\"obner basis method as we will explain in
section~\ref{DOUBLEDSEC}. Examples of multiloop unitarity cuts that give rise to
degenerate multivariate residues include among others the three-loop planar
triple box \cite{Sogaard:2013fpa} and integrals with doubled or higher powers of
propagators \cite{Sogaard:2014ila}. In the two-loop crossed-box computation we
will mostly encounter nondegenerate residues.

\section{Parametrization of Hepta-Cut Solutions}
\begin{figure}[!h]
\bc
\includegraphics[scale=0.8]{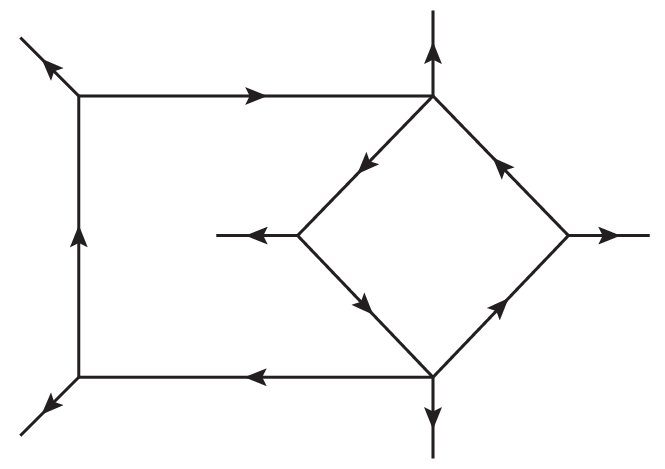}
\put(-263,7){$k_1$}
\put(-263,169){$k_2$}
\put(-188,87){$k_3$}
\put(0,87){$k_4$}
\put(-93,-9){$k_5$}
\put(-93,183){$k_6$}
\put(-161,18){$\ell_1$}
\put(-57,52){$\ell_2$}
\caption{The two-loop crossed-box integral. All external particles may be
massive.}
\label{XBOXDIAGRAM}
\ec
\end{figure}
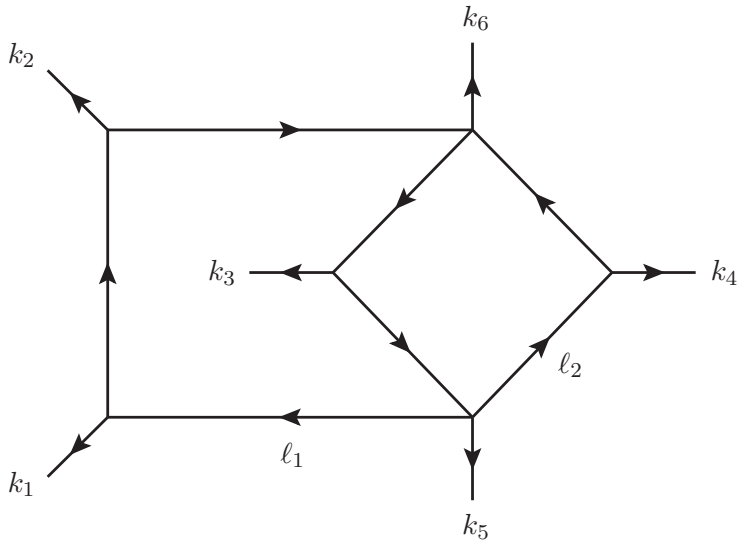

The dimensionally regularized Feynman scalar integral for the two-loop crossed
box amplitude contribution with possibly massive external legs $k_1,\dots,k_6$
distributed across all six vertices is
\begin{align}
\label{XBOXINT}
\XX[1] = {} &
\int_{\R^D}\!\frac{d^D\ell_1}{(2\pi)^D}
\int_{\R^D}\!\frac{d^D\ell_2}{(2\pi)^D}
\prod_{k=1}^7\frac{1}{f_k(\ell_1,\ell_2)} 
\end{align}
where the inverse propagators are
\begin{align}
f_1 = {} & \ell_1^2\;, &\;\;
f_2 = {} & (\ell_1-k_1)^2\;, &\;\;
f_3 = {} & (\ell_1-k_1-k_2)^2\;, \nn \\
f_4 = {} & \ell_2^2\;, &\;\;
f_5 = {} & (\ell_2-k_4)^2\;, &\;\; 
f_6 = {} & (\ell_1+\ell_2+k_5)^2\;, &\;\;
f_7 = {} & (\ell_1+\ell_2+k_3+k_5)^2\;. &
\label{INVPROP}
\end{align} 
Conventions and momentum flow are shown in fig.~\ref{XBOXDIAGRAM}. The Feynman
$i\epsilon$-prescription has been suppressed as it is irrelevant for our
purposes. Generally speaking, this integral will have a nontrivial polynomial
numerator function denoted $\Phi(\ell_1,\ell_2)$ and is in that situation
referred to as a tensor integral even though all Lorentz indices are properly
contracted. 

The seven inverse propagators $\{f_i\}$ generate a polynomial ideal 
$I = \spaa{f_1,\dots,f_7}$ and the hepta-cut equations define a complex
algebraic curve (or two-dimensional real surface) which is the zero locus of
$I$. In our notation,
\begin{align}
\mc S\equiv\mc Z(I) = 
\big\{\,(\ell_1,\ell_2)\in(\C^4)^{\otimes 2}\;|\; &
\ell_1^2 = 0\,,\;
(\ell_1-k_1)^2 = 0\,,\;
(\ell_1-k_1-k_2)^2 = 0\,,\; \nn \\[1mm] &
\ell_2^2 = 0\,,\;
(\ell_2-k_4)^2 = 0\,,\;
(\ell_1+\ell_2+k_5)^2 = 0\,,\; \nn \\[1mm] & \!
(\ell_1+\ell_2+k_3+k_5)^2 = 0\,\big\}\;.
\end{align}
The curve is generally reducible and the algebraic set $\mc S$ can always be
decomposed uniquely into a union of a finite number $n$ of irreducible
components which are in one-to-one correspondence with the inequivalent
hepta-cut solutions,
\begin{align}
\mc S = \bigcup_{i=1}^n\mc S_i\;, 
\quad \mc S_i\not{\!\!\subset}\,\mc S_j \;\; \mathrm{if}\;\; i\neq j\;.
\end{align}
The existence of such a decomposition can be proved by algebraic geometry and
primary decomposition of the polynomial ideal, see ref.~\cite{Zhang:2012ce}.
The number of solutions within an integral topology depends on the kinematic
configuration, in particular the distribution of massive and massless legs.

In the rest of this paper, we examine four-dimensional amplitude contributions
with two-loop crossed box topology with $k_5 = k_6 = 0$, allowing any other
configuration of massive and massless external legs. The Mandelstam invariants
are throughout this paper defined as
\begin{align}
s_{12}\equiv (k_1+k_2)^2\;, \quad
s_{13}\equiv (k_1+k_3)^2\;, \quad
s_{14}\equiv (k_1+k_4)^2\;,
\end{align}
so that momentum conservation can be stated as
\begin{align}
s_{12}+s_{13}+s_{14} = m_1^2+m_2^2+m_3^2+m_4^2\;.
\end{align}

The four-point massless two-loop crossed box was previously studied in terms of
residues and multidimensional contour integrals by one of the present authors in
ref.~\cite{Sogaard:2013yga} and by integrand-level reduction by Badger,
Frellesvig and the other author in ref.~\cite{Badger:2012dp}. 

\subsection{Mutually Projected Kinematics}
Scattering amplitudes of massless particles are naturally encoded in the spinor
helicity formalism by Lorentz invariant inner products of commuting spinors
$\lambda_i^\alpha$ and $\tilde\lambda_i^{\dot\alpha}$. For a momentum $k_i$ with
$k_i^2 = 0$ we have the representation 
$k_i^{\alpha\dot\alpha} = \lambda_i^\alpha\tilde\lambda_i^{\dot\alpha}$. 
It is then possible to define antisymmetric chiral and antichiral brackets,
\begin{align}
\spaa{ij} = -\spaa{ji} \equiv
\epsilon_{\alpha\beta}\lambda_i^\alpha\lambda_j^\beta\;, \quad
\spbb{ij} = -\spbb{ji} \equiv
\epsilon_{\dot\alpha\dot\beta}\lambda_i^{\dot\alpha}\lambda_j^{\dot\beta}\;,
\end{align}
and make contact to the Mandelstam invariants,
\begin{align}
s_{ij} = \spaa{ij}\spbb{ji} = 2k_i\cdot k_j\;.
\end{align}

We treat the massive hepta-cut equations for the two-loop crossed box using {\it
mutually projected} kinematics \cite{Ossola:2006us,Forde:2007mi}. Thereby the
spinor helicity formalism and massive momenta become compatible. Given a pair of
massive momenta $(k_i,k_j)$, the idea is to obtain massless momenta
$(k_i^\flat,k_j^\flat)$ each of which is the massless projection of one of the
massive legs in the direction of the other masslessly projected leg. Here we
will consider four-point kinematics with mutually projecting pairs $(k_1,k_2)$
and $(k_3,k_4)$. Other choices are of course also possible. Within each pair
we define
\begin{align}
\label{MUTUALLYPROJECTING}
k_{j,1}^{\flat,\mu} = k_{j,1}^\mu-
\frac{k_{j,1}^2}{2k_{j,1}\cdot k_{j,2}^\flat}k_{j,2}^{\flat,\mu}\;, \quad
k_{j,2}^{\flat,\mu} = k_{j,2}^\mu-
\frac{k_{j,2}^2}{2k_{j,2}\cdot k_{j,1}^\flat}k_{j,1}^{\flat,\mu}\;,
\end{align}
so that $(k_{j,1}^{\flat,\mu},k_{j,2}^{\flat,\mu})$ are massless momenta by
construction. It is easy to verify that
\begin{align}
k_{j,1}\cdot k_{j,2}^\flat = k_{j,1}^\flat\cdot k_{j,2} = 
k_{j,1}^\flat\cdot k_{j,2}^\flat\;.
\end{align}
We streamline notation and define the frequently used quantity
\begin{align}
\gamma_{j,12} = 2k_{j,1}^\flat\cdot k_{j,2}^\flat\;,
\end{align}
which upon identification in eq.~\eqref{MUTUALLYPROJECTING} leads to a quadratic
equation whose solutions are
\begin{align}
\gamma_{j,12}^\pm = k_{j,1}\cdot k_{j,2}\pm \big[(k_{j,1}\cdot
k_{j,2})^2-k_{j,1}^2k_{j,2}^2\big]^{1/2}\;.
\end{align}

It is perhaps more useful from a practical point of view to express the massless
projections in terms of the corresponding massive momenta. To that end we let
$(i,\bar{\imath})$ denote a mutually projecting pair and define
\begin{align}
\rho_{j,i} = \frac{k_{j,i}^2}{2k_{j,i}\cdot k_{j,\bar{\imath}}}\;.
\end{align}
Then it is straightforward to invert eq.~\eqref{MUTUALLYPROJECTING} with the
result
\begin{align}
\label{INVERTEDFLAT}
k_{j,i}^{\flat,\mu} = 
(1-\rho_{j,1}\rho_{j,2})^{-1}
(k_{j,i}^\mu-\rho_{j,i}k_{j,\bar{\imath}}^\mu)\;.
\end{align}
On the other hand, the decomposition of the massive legs in terms of a pair of
flattened momenta reads
\begin{align}
k_{j,i}^\mu = k_i^{\flat,\mu}+\rho_{j,i}k_{\bar{\imath}}^{\flat,\mu}\;.
\end{align}

It is convenient to introduce short hand notation $\rho_{12}\equiv\rho_{1,1}$,
$\rho_{21}\equiv\rho_{1,2}$ and $\gamma_{12}\equiv\gamma_{j,12}$ so that
$\rho_{12} = m_1^2/\gamma_{12}$ and $\rho_{21} = m_2^2/\gamma_{12}$, and
similarly for the other mutually projecting pair. If $m_1m_2 = 0$ we have
$\gamma_{12} = s_{12}$ and likewise for $\gamma_{34}$. Our final results can
therefore be expressed in terms of the nonzero masses among
$\{m_1,m_2,m_3,m_4\}$, $\gamma_{12}$ and $\gamma_{34}$ if respectively
$m_1m_2\neq 0$ and $m_3m_4\neq0$, along with two independent Mandelstam
invariants, say $s_{12}$ and $s_{14}$.

We adopt a loop momentum parametrization of the form,
\begin{align}
\ell_1^\mu =
\frac{1}{2}\spvec{\lambda_1^-}{\gamma^\mu}{\tilde\lambda_1^{'-}} = 
\frac{1}{2}\spab{\lambda_1}{\gamma^\mu}{\tilde\lambda_1'} \;, \quad
\ell_2^\mu =
\frac{1}{2}\spvec{\lambda_2^-}{\gamma^\mu}{\tilde\lambda_2^{'-}} = 
\frac{1}{2}\spab{\lambda_2}{\gamma^\mu}{\tilde\lambda_2'}\;,
\end{align}
so that $\ell_1^2 = \ell_2^2 = 0$. The various loop spinors are then constructed
from the spinors corresponding to the two mutually projecting pairs with general
complex coefficients,
\begin{align}
\ket{\lambda_1^+} = 
\xi_1\ket{1^{\flat,+}}+
\xi_2\frac{\spaa{4^\flat 1^\flat}}{\spaa{4^\flat 2^\flat}}\ket{2^{\flat,+}}\;, 
\quad
\ket{\tilde\lambda_1^{'-}} = 
\xi_1'\ket{1^{\flat,-}}+
\xi_2'\frac{\spbb{4^\flat 1^\flat}}{\spbb{4^\flat 2^\flat}}\ket{2^{\flat,-}}\;,
\nn \\
\ket{\lambda_2^+} = 
\xi_3\frac{\spaa{1^\flat 4^\flat}}{\spaa{1^\flat 3^\flat}}\ket{3^{\flat,+}}+
\xi_4\ket{4^{\flat,+}}\;,
\quad
\ket{\tilde\lambda_2^{'-}} = 
\xi_3'\frac{\spbb{1^\flat 4^\flat}}{\spbb{1^\flat 3^\flat}}\ket{3^{\flat,-}}+
\xi_4'\ket{4^{\flat,-}}\;.
\end{align}
In this way, massive external momenta are related to chiral and antichiral
spinors corresponding to their massless projections through
$2k_i^\mu = \spab{i^\flat}{\gamma^\mu}{i^\flat}$. Expanded explicitly in the
basis of four-vectors the two loop momenta read
\begin{align}
\label{PARAML1}
\ell_1^\mu(\xi_i,\xi_i') \,=\, {} & \xi_1\xi_1'k_1^{\flat,\mu}+\xi_2\xi_2'
\frac{k_1^\flat\cdot k_4^\flat}{k_2^\flat\cdot k_4^\flat}k_2^{\flat,\mu}
\nn \\ {} &
+\frac{\xi_1\xi_2'}{2}\frac{\spbb{1^\flat 4^\flat}}{\spbb{2^\flat 4^\flat}}
\spvec{1^{\flat,-}}{\gamma^\mu}{2^{\flat,-}}+
\frac{\xi_2\xi_1'}{2}\frac{\spaa{1^\flat 4^\flat}}{\spaa{2^\flat 4^\flat}}
\spvec{2^{\flat,-}}{\gamma^\mu}{1^{\flat,-}}\;, \\[2mm]
\label{PARAML2}
\ell_2^\mu(\xi_i,\xi_i') \,=\, {} & \xi_3\xi_3'
\frac{k_1^\flat\cdot k_4^\flat}{k_1^\flat\cdot k_3^\flat}k_3^{\flat,\mu}+
\xi_4\xi_4'k_4^{\flat,\mu}
\nn \\ {} &
+\frac{\xi_3\xi_4'}{2}\frac{\spaa{1^\flat 4^\flat}}{\spaa{1^\flat 3^\flat}}
\spvec{3^{\flat,-}}{\gamma^\mu}{4^{\flat,-}}+
\frac{\xi_4\xi_3'}{2}
\frac{\spbb{1^\flat 4^\flat}}{\spbb{1^\flat 3^\flat}}
\spvec{4^{\flat,-}}{\gamma^\mu}{3^{\flat,-}}\;,
\end{align}
and therefore we are able to eventually fix two of the complex parameters. This
freedom amounts sort of a gauge choice. We emphasize that this choice is not
necessarily the same for all on-shell branches. In fact, this is not possible in
the two-loop crossed box as opposed to the planar case \cite{Johansson:2013sda}.

\subsection{Four-Mass Hepta-Cut Equations}
To begin with we will derive the hepta-cut equations for nonzero external masses
in all four corners. Besides the on-shell constraints for $\ell_1$ and $\ell_2$
which are already satisfied automatically, it thus remains to examine the other
five massive hepta-cut equations. Three of them are very simple because they
only involve one of the loop momenta. Indeed, it takes little effort to realize
that
\begin{align}
(\ell_1-k_1)^2 = 0 \;\LRa\; {} &
m_1^2(1-\xi_1\xi_1')-
\frac{k_1^\flat\cdot k_4^\flat}{k_2^\flat\cdot k_4^\flat}
\xi_2\xi_2'\gamma_{12} = 0\;, \\[1mm]
(\ell_1-k_1-k_2)^2 = 0 \;\LRa\; {} &
s_{12}-m_1^2-\gamma_{12}\xi_1\xi_1'-
\frac{k_1^\flat\cdot k_4^\flat}{k_2^\flat\cdot k_4^\flat}
m_2^2\xi_2\xi_2' = 0\;, \\
\label{L2SIMPLE}
(\ell_2-k_4)^2 = 0 \;\LRa\; {} &
m_4^2(1-\xi_4\xi_4')-
\frac{k_1^\flat\cdot k_4^\flat}{k_1^\flat\cdot k_3^\flat}
\xi_3\xi_3'\gamma_{34} = 0\;.
\end{align} 
Lorentz products of flattened momenta are needed throughout this calculation.
Before we continue let us therefore for completeness derive the necessary
expressions. The trick is of course to apply eq.~\eqref{INVERTEDFLAT} and
thereby invoke the massive vectors whose contractions are well known. In fact,
\begin{align}
\label{KP1KP3}
k_1^\flat\cdot k_3^\flat = {} &
\gamma_{12}\gamma_{34} 
\left(2(\gamma_{12}^2-m_1^2m_2^2)(\gamma_{34}^2-m_3^2m_4^2)\right)^{-1} 
\nn \\[1mm] & \times\big\{
m_1^2\big[\gamma_{34}(m_2^2+m_3^2-s_{14})+
m_3^2(m_1^2+m_3^2-s_{12}-s_{14})\big] \nn \\[1mm] & \quad\;\;
+\gamma_{12}\big[m_3^2(m_1^2+m_4^2-s_{14})+
\gamma_{34}(m_2^2+m_4^2-s_{12}-s_{14})\big]
\big\}\;, \\[2mm]
\label{KP1KP4}
k_1^\flat\cdot k_4^\flat = {} & 
\gamma_{12}\gamma_{34} 
\left(2(\gamma_{12}^2-m_1^2m_2^2)(\gamma_{34}^2-m_3^2m_4^2)\right)^{-1} 
\nn \\[1mm] & \times\big\{
m_1^2\big[\gamma_{34}(m_1^2+m_3^2-s_{12}-s_{14})+
m_4^2(m_2^2+m_3^2-s_{14})\big] \nn \\[1mm] & \quad\;\;
+\gamma_{12}\big[m_4^2(m_2^2+m_4^2-s_{12}-s_{14})+
\gamma_{34}(m_1^2+m_4^2-s_{14})\big]
\big\}\;.
\end{align}
We also have
\begin{align}
\label{KP2KP3}
k_2^\flat\cdot k_3^\flat = {} &
\gamma_{12}\gamma_{34} 
\left(2(\gamma_{12}^2-m_1^2m_2^2)(\gamma_{34}^2-m_3^2m_4^2)\right)^{-1} 
\nn \\[1mm] & \times\big\{
m_2^2\big[m_3^2(m_1^2+m_4^2-s_{14})+
\gamma_{34}(m_2^2+m_4^2-s_{12}-s_{14})\big]
\nn \\[1mm] & \quad\;\;
+\gamma_{12}\big[m_3^2(m_1^2+m_3^2-s_{12}-s_{14})+
\gamma_{34}(m_2^2+m_3^2-s_{14})\big]
\big\}\;, \\[2mm]
\label{KP2KP4}
k_2^\flat\cdot k_4^\flat = {} & 
\gamma_{12}\gamma_{34} 
\left(2(\gamma_{12}^2-m_1^2m_2^2)(\gamma_{34}^2-m_3^2m_4^2)\right)^{-1} 
\nn \\[1mm] & \times\big\{
m_2^2\big[\gamma_{34}(m_1^2+m_4^2-s_{14})+
m_4^2(m_2^2+m_4^2-s_{12}-s_{14})\big]
\nn \\[1mm] & \quad\;\;
+\gamma_{12}\big[m_4^2(m_2^2+m_3^2-s_{14})+
\gamma_{34}(m_1^2+m_3^2-s_{12}-s_{14})\big]
\big\}\;.
\end{align}
In addition, we use the same technique to also provide explicit formulas for
various contractions of flattened momenta with the external massive legs needed
in one of the on-shell equations below,
\begin{align}
\label{KP1K3}
k_1^\flat\cdot k_3 = {} &
+\frac{\gamma_{12}[m_1^2(m_2^2+m_3^2-s_{14})+
\gamma_{12}(m_2^2+m_4^2-s_{12}-s_{14})]}{2(\gamma_{12}^2-m_1^2m_2^2)}\;,
\\[2mm]
\label{KP2K3}
k_2^\flat\cdot k_3 = {} &
-\frac{\gamma_{12}[\gamma_{12}(m_2^2+m_3^2-s_{14})+
m_2^2(m_2^2+m_4^2-s_{12}-s_{14})]}{2(\gamma_{12}^2-m_1^2m_2^2)}\;.
\end{align}
Another rather trivial, but useful, identity is $m_i^2 = 2k_i^\flat\cdot k_i$.
Finally, in what proceeds, we will also encounter the quantities
\begin{align}
\tau\equiv\frac{\spaa{1^\flat 4^\flat}\spaa{2^\flat 3^\flat}}
{\spaa{2^\flat 4^\flat}\spaa{1^\flat 3^\flat}}\;, \quad
\bar{\tau}\equiv\frac{\spbb{1^\flat 4^\flat}\spbb{2^\flat 3^\flat}}
{\spbb{2^\flat 4^\flat}\spbb{1^\flat 3^\flat}}\;,
\end{align}
which are complex conjugates of each other for real external momenta as
indicated. But actually $\tau = \bar{\tau}$. For completeness, $\tau$ can be
expanded and re-expressed in terms of the independent kinematic invariants
described above in the following way,
\begin{align}
\label{TAUEXP}
\tau = \bar{\tau} = {} &
-\frac{\gamma_{34}(\gamma_{12}+m_1^2)}{
(\gamma_{34}+m_3^2)[(\gamma_{12}\gamma_{34}-m_1^2m_3^2)
(\gamma_{12}\gamma_{34}-m_2^2m_4^2)
+\gamma_{12}\gamma_{34}s_{12}s_{14}]} \nn \\[2mm] {} &
\hspace*{.8cm}
\times\Big[(\gamma_{12}+m_2^2)(\gamma_{34}+m_3^2)
(m_2^2+m_3^2-s_{14})+2m_2^2m_3^2s_{12}+ \nn \\ {} &
\hspace*{1.45cm}
(\gamma_{12}+m_2^2)m_3^2(m_1^2-m_2^2-s_{12})
+(\gamma_{34}+m_3^2)m_2^2(m_4^2-m_3^2-s_{12})
\Big]\;.
\end{align}

Let us now return to the hepta-cut equations for loop momentum $\ell_1$. For
general masses $m_1 \neq 0 \neq m_2$ we obtain the solution
\begin{align}
\xi_1\xi_1' = 
\frac{\gamma_{12}s_{12}-(\gamma_{12}+m_2^2)m_1^2}
{\gamma_{12}^2-m_1^2m_2^2}
\equiv\bxi_1\;, \quad
\xi_2\xi_2' =
\frac{m_1^2(m_1^2+\gamma_{12}-s_{12})k_2^\flat\cdot k_4^\flat}
{(\gamma_{12}^2-m_1^2m_2^2)k_1^\flat\cdot k_4^\flat}
\equiv\bxi_2\;.
\label{MASSLESSLIMITREPLACE}
\end{align}

In contrast to the planar double box, there is only one additional nontrivial
on-shell constraint for loop momentum $\ell_2$. Rewriting eq.~\eqref{L2SIMPLE}
in the slightly more suggestive form,
\begin{align}
\label{XI3XI3P}
\xi_3\xi_3'+\mu(\xi_4\xi_4'-1) = 0\;, \quad 
\mu\equiv\frac{m_4^2k_1^\flat\cdot k_3^\flat}
{\gamma_{34}k_1^\flat\cdot k_4^\flat}\;,
\end{align}
we see that $\xi_3\xi_3' = 0$ if $m_4 = 0$ or $\xi_4\xi_4' = 1$.

The foresight in the choice of parametrization of $\ell_1$ and $\ell_2$ implies
that the coupled on-shell equations are also quite compact actually.  It happens
that one of them factorizes completely in a symmetric manner,
\begin{align}
\label{L1L2SUM}
(\ell_1+\ell_2)^2 \,=\, {} & 0 \;\LRa\; 
(\xi_1(\xi_3+\xi_4)+\xi_2(\tau\xi_3+\xi_4))\times
(\xi_1'(\xi_3'+\xi_4')+\xi_2'(\tau\xi_3'+\xi_4')) = 0\;,
\end{align}
whereas the other can be written
\begin{align}
\label{L1L2SUMK3}
(\ell_1+\ell_2+k_3)^2 \,=\, 0 \;\LRa\; \hspace*{10.5cm} \nn \\
m_3^2+2\bigg\{
k_1^\flat\cdot k_3 \xi_1\xi_1'+\bigg[
\tau k_1^\flat\cdot k_3^\flat+
\frac{m_3^2 k_1^\flat\cdot k_4^\flat}{\gamma_{34}}
\bigg](\xi_1\xi_2'+\xi_1'\xi_2) \hspace*{4cm} \nn \\ +
\frac{m_3^2k_1^\flat\cdot k_4^\flat}{2k_1^\flat\cdot k_3^\flat}\xi_3\xi_3'+
\frac{k_1^\flat\cdot k_4^\flat k_2^\flat\cdot k_3}{k_2^\flat\cdot k_4^\flat}
\xi_2\xi_2'+\frac{\gamma_{34}}{2}\xi_4\xi_4'
\bigg\} = 0\;.
\end{align}
Upon insertion of eqs.~\eqref{MASSLESSLIMITREPLACE} and \eqref{XI3XI3P} into
eq.~\eqref{L1L2SUMK3} we immediately find 
\begin{align}
\xi_4\xi_4' = 
-\bigg(\gamma_{34}-\frac{m_3^2m_4^2}{\gamma_{34}}\bigg)^{-1}\bigg\{
m_3^2\Big(1+\frac{m_4^2}{\gamma_{34}}\Big)+
2\bigg[k_1^\flat\cdot k_3\bxi_1+
\frac{k_1^\flat\cdot k_4^\flat}{k_2^\flat\cdot k_4^\flat}
k_2^\flat\cdot k_3\bxi_2 \hspace*{2.9cm} \nn \\
+\bigg(\tau k_1^\flat\cdot k_3^\flat+
\frac{m_3^2 k_1^\flat\cdot k_4^\flat}{\gamma_{34}}\bigg)(
\xi_1\xi_2'+\xi_1'\xi_2)\bigg]\bigg\}\;,
\end{align}
or in the slightly more appealing form,
\begin{align}
\label{XI4PSHORT}
\xi_4\xi_4' =
1+\frac{\tau(\bxi_1+\bxi_2+\xi_1\xi_2'+\xi_1'\xi_2)}
{(1-\tau)(\bxi_1-\tau\bxi_2)}\;.
\end{align}
In the last step we recast the equation by means of the two identities
\begin{align}
\frac{\tau}{(\tau-1)(\bxi_1-\tau\bxi_2)} =
\frac{2k_1^\flat\cdot k_4^\flat}{\gamma_{34}^2-m_3^2m_4^2}
\bigg[m_3^2+\gamma_{34}\tau
\frac{k_1^\flat\cdot k_3^\flat}{k_1^\flat\cdot k_4^\flat}\bigg]\;, \\[2mm]
\frac{\bxi_1+\tau^2\bxi_2}{(\tau-1)(\bxi_1-\tau\bxi_2)} = 
\frac{\gamma_{34}}{\gamma_{34}^2-m_3^2m_4^2}\bigg[
m_3^2\Big(1+\frac{m_4^2}{\gamma_{34}}\Big)+
2\bigg(k_1^\flat\cdot k_3\bxi_1+\frac{k_1^\flat\cdot k_4^\flat}{
k_2^\flat\cdot k_4^\flat}k_2^\flat\cdot k_3\bxi_2\bigg)
\bigg]\;,
\end{align}
which can be verified through eqs.~\eqref{KP1KP3}-\eqref{KP2K3} and
\eqref{TAUEXP} along with expressions for $\bxi_1$ and $\bxi_2$ given in
eq.~\eqref{MASSLESSLIMITREPLACE}, although that task is rather tedious.

In order to solve the hepta-cut equations, we decompose the reducible ideal
generated by the list of rewritten inverse propagators into an intersection of
six prime ideals. Then we compute generating sets that form Gr\"obner bases over
the field of rational functions in each irreducible ideal and obtain the
associated zero loci by hand. The six distinct hepta-cut solutions are really
three pairs of parity conjugates $(\mc S_1,\mc S_2)$, $(\mc S_3,\mc S_4)$,
$(\mc S_5,\mc S_6)$ and each on-shell branch is topologically equivalent to a
Riemann sphere, parametrized by a free variable $z\in\C$. 

We choose to make the behavior under parity conjugation manifest and present the
solutions in a symmetric manner such that $\mc S_{2k-1}$ and $S_{2k}$ for
$k = 1,2,3$ are related to each other by simply interchanging
$(\xi_1,\dots,\xi_4)\longleftrightarrow(\xi_1',\dots,\xi_4')$. In terms of the
parameters $(\xi_1,\xi_2,\xi_3,\xi_4)$ and
$(\xi_1',\xi_2',\xi_3',\xi_4')$, the solutions are
\begin{align}
\mc S_1\,:\;\; &
(\xi_1,\xi_2,\xi_3,\xi_4) = \left(
1,z,\frac{\mu\tau(1+z)}{1+\tau z},1\right) \nn \\ &
(\xi_1',\xi_2',\xi_3',\xi_4') = \left(
\bxi_1,\frac{\bxi_2}{z},-\frac{(\bxi_2+\bxi_1 z)(1+\tau z)}{
z(1-\tau)(\bxi_1-\tau\bxi_2)},
\frac{(\bxi_2+\bxi_1 z)(1+\tau z)}{
z(1-\tau)(\bxi_1-\tau\bxi_2)}\right)\;, \\[2mm]
\mc S_2\,:\;\; &
(\xi_1,\xi_2,\xi_3,\xi_4) = \left(
\bxi_1,\frac{\bxi_2}{z},-\frac{(\bxi_2+\bxi_1 z)(1+\tau z)}{
z(1-\tau)(\bxi_1-\tau\bxi_2)},
\frac{(\bxi_2+\bxi_1 z)(1+\tau z)}{
z(1-\tau)(\bxi_1-\tau\bxi_2)}\right) \nn \\ &
(\xi_1',\xi_2',\xi_3',\xi_4') = \left(
1,z,\frac{\mu\tau(1+z)}{1+\tau z},1\right)\;, \\[2mm]
\mc S_3\,:\;\; &
(\xi_1,\xi_2,\xi_3,\xi_4) = \left(
1,-\frac{\bxi_2}{\bxi_1},z,1\right)\,, \;\;\;
(\xi_1',\xi_2',\xi_3',\xi_4') = \left(
\bxi_1,-\bxi_1,0,1\right)\;, \\[2mm]
\mc S_4\,:\;\; &
(\xi_1,\xi_2,\xi_3,\xi_4) = \left(
\bxi_1,-\bxi_1,0,1\right)\,, \;\;\;
(\xi_1',\xi_2',\xi_3',\xi_4') = \left(
1,-\frac{\bxi_2}{\bxi_1},z,1\right)\;, \\[2mm]
\mc S_5\,:\;\; &
(\xi_1,\xi_2,\xi_3,\xi_4) = \left(
\bxi_1,-\tau\bxi_2,\mu,z\right)\,, \;\;\;
(\xi_1',\xi_2',\xi_3',\xi_4') = \left(
1,-1/\tau,1,0\right)\;, \\[2mm]
\mc S_6\,:\;\; &
(\xi_1,\xi_2,\xi_3,\xi_4) = \left(
1,-1/\tau,1,0\right)\,, \;\;\;
(\xi_1',\xi_2',\xi_3',\xi_4') = \left(
\bxi_1,-\tau\bxi_2,\mu,z\right)\;.
\end{align}
The branches are written in terms of independent kinematic constants $\bxi_1$,
$\bxi_2$, $\tau$ and $\mu$ \eqref{TAUEXP}-\eqref{XI3XI3P}. For completeness we
include the explicit forms of the variables in the four-vector expansions of the
two loop momenta for all hepta-cut solutions in
appendix~\ref{EXPLICITSOLUTIONSAPP}. 

\subsection{Massless Limits}
\label{MASSLESSLIMITS}
We will also analyze the crossed double box with one, two and three massless
legs. Integrals for this kinematics are also relevant for higher-multiplicity
scattering processes, starting already at five external massless particles.

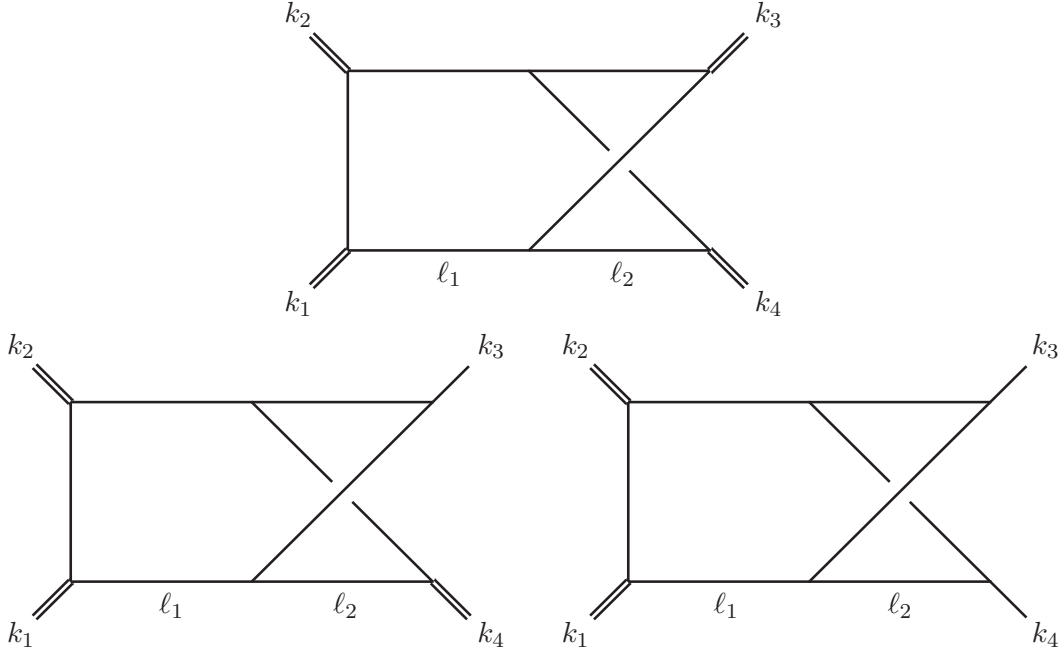
\begin{figure}[!h]
\bc
\includegraphics[scale=0.75]{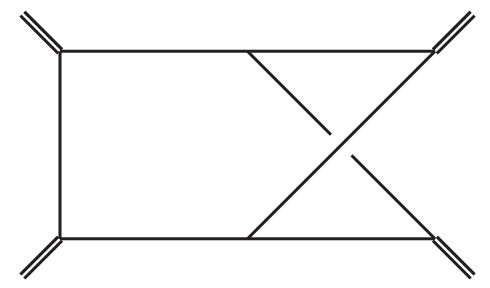}
\put(-179,-5){$k_1$}
\put(-179,104){$k_2$} 
\put(-3,104){$k_3$} 
\put(-3,-5){$k_4$}
\put(-122,6){$\ell_1$}
\put(-57,6){$\ell_2$}
\\[.2cm]
\includegraphics[scale=0.75]{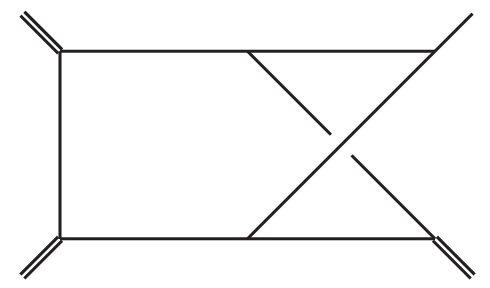}
\put(-179,-5){$k_1$}
\put(-179,104){$k_2$} 
\put(-3,104){$k_3$} 
\put(-3,-5){$k_4$}
\put(-122,6){$\ell_1$}
\put(-57,6){$\ell_2$}
\hspace*{1cm}
\includegraphics[scale=0.75]{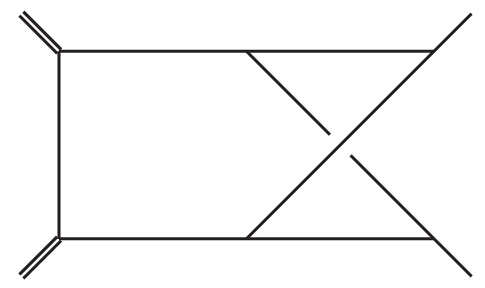}
\put(-179,-5){$k_1$}
\put(-179,104){$k_2$} 
\put(-3,104){$k_3$} 
\put(-3,-5){$k_4$}
\put(-122,6){$\ell_1$}
\put(-57,6){$\ell_2$}
\ec
\caption{The first class of two-loop crossed-box integrals includes the
four-mass case and related massless limits, i.e. three-mass and short-side
two-mass with massless legs in the nonplanar end of the diagram.  Massless and
massive external legs are denoted by single and doubled lines respectively.}
\label{XBOXINTSCLASSA}
\end{figure}

Let us look more closely at the hepta-cut equations and their solutions. We
focus on the two momenta in the crossed end of the diagram and assume that
$m_1m_2\neq 0$. The only dependence on $m_3$ is implicitly through other
parameters, e.g. $\tau$ and $\mu$. In particular, all on-shell equations and
their solutions have the correct limits for $m_3\to0$. Moreover, it is clear
that $\mu\to0$ for $m_4\to0$ so that eq.~\eqref{XI3XI3P} should be replaced by
$\xi_3\xi_3' = 0$. It can be shown that the number of branches remains six and
the explicit solutions follow from the four-mass case once we let $\mu\to0$.
The configurations corresponding to this class of kinematics are illustrated in
fig.~\ref{XBOXINTSCLASSA}.

The situation is slightly more complicated when momenta $k_1$ and $k_2$ in the
planar end become massless. Here, the massless limits are not smooth and hence
they should be treated carefully. We see that $\bxi_1\to 1$ as $m_2\to 0$ and
$m_1$ arbitrary. Also, $\bxi_1\to 1+m_2^2/\gamma_{12}$ when $m_1\to 0$.
Therefore the equation $\xi_1\xi_1'-\bxi_1 = 0$ will not give rise to
branchings. But if at least one mass among $\{m_1,m_2\}$ is zero, $\bxi_2 = 0$
and we instead get the equation $\xi_2\xi_2' = 0$ and
eq.~\eqref{MASSLESSLIMITREPLACE} must be replaced by a pair of solutions, i.e.
$\xi_2' = 0$ and $\xi_2$ free or vice versa. 

We solved the hepta-cut equations for this class of kinematics and found eight
distinct solutions that can be parametrized as follows,
\begin{align}
\tS_1\,:\;\; &
(\xi_1,\xi_2,\xi_3,\xi_4) = \left(
1,z,\frac{\mu\tau(1+z)}{1+\tau z},1\right)\,, \;\;\;
(\xi_1',\xi_2',\xi_3',\xi_4') = \left(
\bxi_1,0,-\frac{1+\tau z}{1-\tau},
\frac{1+\tau z}{1-\tau}\right)\,, \nn \\[1mm]
\tS_2\,:\;\; &
(\xi_1,\xi_2,\xi_3,\xi_4) = \left(
\bxi_1,0,-\frac{1+\tau z}{1-\tau},
\frac{1+\tau z}{1-\tau}\right)\,, \;\;\;
(\xi_1',\xi_2',\xi_3',\xi_4') = \left(
1,z,\frac{\mu\tau(1+z)}{1+\tau z},1\right)\,, \nn \\[1mm]
\tS_3\,:\;\; &
(\xi_1,\xi_2,\xi_3,\xi_4) = \left(
1,z,-\frac{1+z}{1+\tau z},1\right)\,, \;\;\;
(\xi_1',\xi_2',\xi_3',\xi_4') = \left(
\bxi_1,0,\frac{\mu\tau(1+\tau z)}{\tau-1},\frac{1+\tau z}{1-\tau}
\right)\,, \nn \\[1mm]
\tS_4\,:\;\; &
(\xi_1,\xi_2,\xi_3,\xi_4) = \left(
\bxi_1,0,\frac{\mu\tau(1+\tau z)}{\tau-1},\frac{1+\tau z}{1-\tau}
\right)\,, \;\;\;
(\xi_1',\xi_2',\xi_3',\xi_4') = \left(
1,z,-\frac{1+z}{1+\tau z},1\right)\,, \nn \\[1mm]
\tS_5\,:\;\; &
(\xi_1,\xi_2,\xi_3,\xi_4) = \left(
1,0,z,1\right)\,, \;\;\;
(\xi_1',\xi_2',\xi_3',\xi_4') = \left(
\bxi_1,-\bxi_1,0,1\right)\,, \nn \\[2mm]
\tS_6\,:\;\; &
(\xi_1,\xi_2,\xi_3,\xi_4) = \left(
\bxi_1,-\bxi_1,0,1\right)\,, \;\;\;
(\xi_1',\xi_2',\xi_3',\xi_4') = \left(
1,0,z,1\right)\,, \nn \\[2mm]
\tS_7\,:\;\; &
(\xi_1,\xi_2,\xi_3,\xi_4) = \left(
\bxi_1,0,\mu,z\right)\,, \;\;\;
(\xi_1',\xi_2',\xi_3',\xi_4') = \left(
1,-1/\tau,1,0\right)\,, \nn \\[2mm]
\tS_8\,:\;\; &
(\xi_1,\xi_2,\xi_3,\xi_4) = \left(
1,-1/\tau,1,0\right)\,, \;\;\;
(\xi_1',\xi_2',\xi_3',\xi_4') = \left(
\bxi_1,0,\mu,z\right)\,.
\end{align}

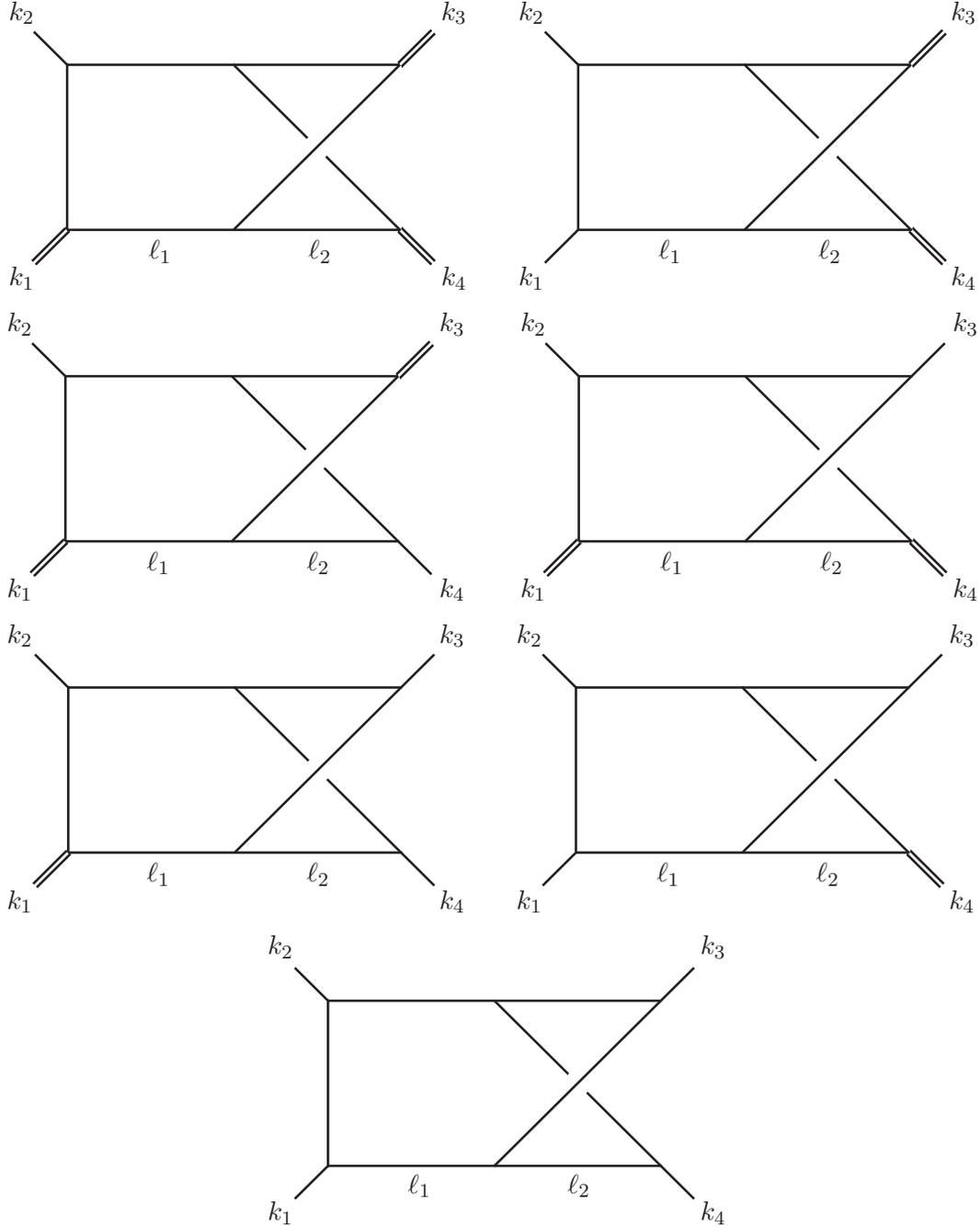
\begin{figure}[!h]
\bc
\hspace*{-.3cm}
\includegraphics[scale=0.75]{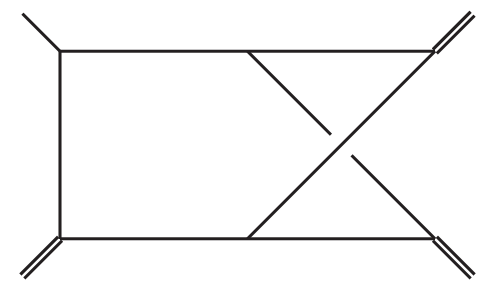}
\put(-179,-5){$k_1$}
\put(-179,104){$k_2$} 
\put(-3,104){$k_3$} 
\put(-3,-5){$k_4$}
\put(-122,6){$\ell_1$}
\put(-57,6){$\ell_2$}
\hspace{1cm}
\includegraphics[scale=0.75]{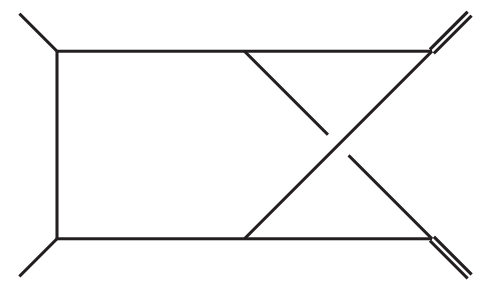} 
\put(-179,-5){$k_1$}
\put(-179,104){$k_2$} 
\put(-3,104){$k_3$} 
\put(-3,-5){$k_4$}
\put(-122,6){$\ell_1$}
\put(-57,6){$\ell_2$}
\\[.3cm]
\hspace*{-.3cm}
\includegraphics[scale=0.75]{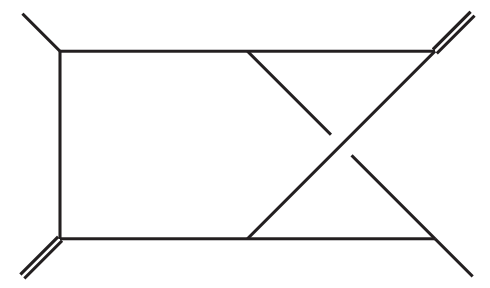}
\put(-179,-5){$k_1$}
\put(-179,104){$k_2$} 
\put(-3,104){$k_3$} 
\put(-3,-5){$k_4$}
\put(-122,6){$\ell_1$}
\put(-57,6){$\ell_2$}
\hspace{1cm}
\includegraphics[scale=0.75]{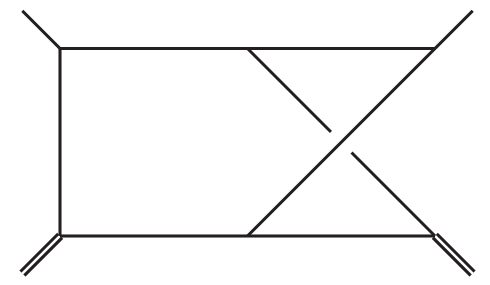}
\put(-179,-5){$k_1$}
\put(-179,104){$k_2$} 
\put(-3,104){$k_3$} 
\put(-3,-5){$k_4$}
\put(-122,6){$\ell_1$}
\put(-57,6){$\ell_2$}
\\[.3cm]
\hspace*{-.3cm}
\includegraphics[scale=0.75]{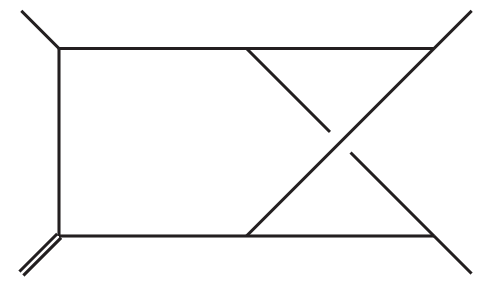}
\put(-179,-5){$k_1$}
\put(-179,104){$k_2$} 
\put(-3,104){$k_3$} 
\put(-3,-5){$k_4$}
\put(-122,6){$\ell_1$}
\put(-57,6){$\ell_2$}
\hspace{1cm}
\includegraphics[scale=0.75]{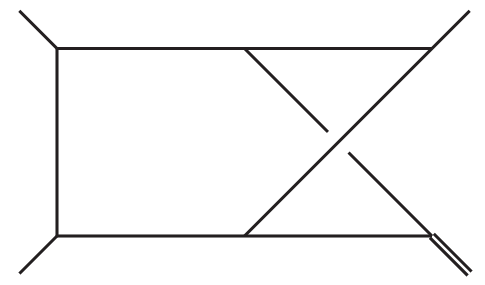}
\put(-179,-5){$k_1$}
\put(-179,104){$k_2$} 
\put(-3,104){$k_3$} 
\put(-3,-5){$k_4$}
\put(-122,6){$\ell_1$}
\put(-57,6){$\ell_2$}
\\[.3cm]
\includegraphics[scale=0.75]{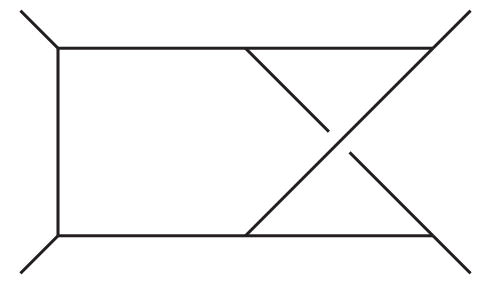}
\put(-179,-5){$k_1$}
\put(-179,104){$k_2$} 
\put(-3,104){$k_3$} 
\put(-3,-5){$k_4$}
\put(-122,6){$\ell_1$}
\put(-57,6){$\ell_2$}
\ec
\caption{The three-mass and the two-mass short-side integrals with massless legs
in the planar end of the diagram together with diagonal and long-side two-mass
integrals, one-mass integrals and finally the zero-mass integral correspond to
degenerate massless limits.}
\label{XBOXINTSCLASSB}
\end{figure}

Notice that six of the tilded solutions are inherited from the four-mass case in
the limit $\bxi_2\to 0$. The new branches are denoted $\tS_3$ and $\tS_4$. We
invite the reader to refer to appendix~\ref{EXPLICITSOLUTIONSAPP} for the full
expressions of the loop momenta on each branch. The various kinematic
configurations are depicted in figs.~\ref{XBOXINTSCLASSA} and
\ref{XBOXINTSCLASSB}.

\subsection{Classification of Kinematic Solutions}
The degeneracy of maximal-cut solutions is customarily understood
diagrammatically from the possible distributions of chiralities at the on-shell
vertices. Momentum conservation in a three-point vertex forces either square or
angle spinor products to align. Phrased slightly differently, the positive or
negative chirality spinors of the momenta are collinear,
$\lambda_a\propto\lambda_b\propto\lambda_c$ or
$\tilde\lambda_a\propto\tilde\lambda_b\propto\tilde\lambda_c$. We choose to
depict such vertices as $\circ$ and $\bullet$ and refer to them as antichiral
($\overline{\text{MHV}}$) and chiral (MHV) respectively. Vertices involving more
than three particles or massive momenta do not have a well-defined chirality.

In recent studies, a one-to-one correspondence between diagrams and kinematic
solutions was found for the planar double box with up to four massive legs
\cite{Larsen:2012sx,Johansson:2012zv,Johansson:2013sda,CaronHuot:2012ab}, the
two-loop crossed box \cite{Sogaard:2013yga} and the planar three-loop triple
box \cite{Sogaard:2013fpa}.

\begin{figure}[!h]
\bc
\includegraphics[scale=0.7]{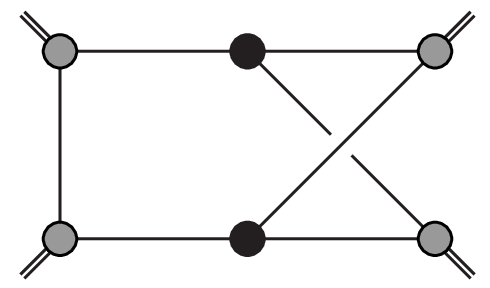}
\put(-108,-5){Solution $\mc S_1$}
\hspace*{.3cm}
\includegraphics[scale=0.7]{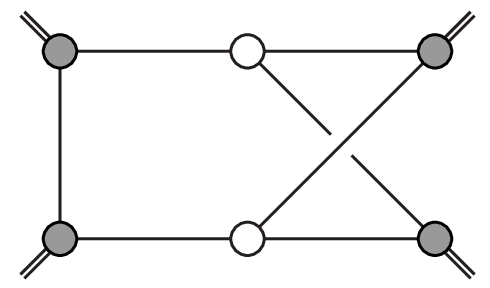}
\put(-108,-5){Solution $\mc S_2$} \\[5mm]
\includegraphics[scale=0.7]{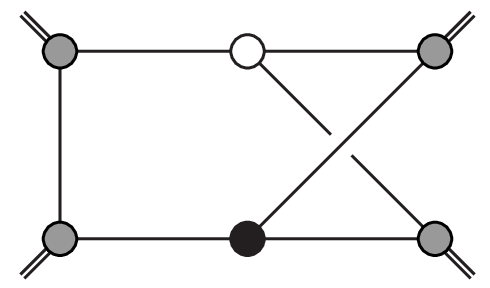}
\put(-112,-5){Solutions $\mc S_{3,5}$}
\hspace*{.3cm}
\includegraphics[scale=0.7]{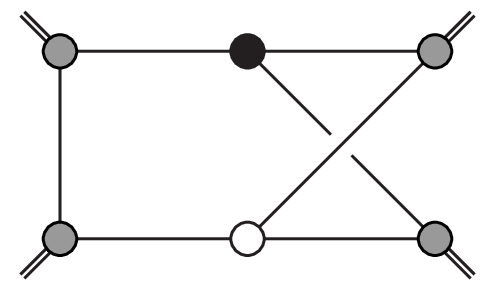}
\put(-112,-5){Solutions $\mc S_{4,6}$}
\ec
\caption{The four possible vertex configurations of the four-mass two-loop
crossed box. Black, white and gray blobs denote chiral, antichiral and nonchiral
vertices respectively. The opposite-chirality diagrams are not in one-to-one
correspondence with the hepta-cut solutions.}
\label{VERTICES2MEASY}
\end{figure}

The situation is different for the two-loop crossed box. Consider the four-mass
case which we know gives rise to six classes of hepta-cut solutions. The maximal
cut leaves two massless on-shell three-vertices in a 2-mass-easy sub-box and
hence we would erroneously predict only four solutions based on the diagrams in
fig.~\ref{VERTICES2MEASY}. The cause of this mismatch is that
solutions $\mc S_3$ and $\mc S_5$ actually correspond to the same
opposite-chirality diagram and similarly for $\mc S_4$ and $\mc S_6$. 

\section{Residues of the Loop Integrand and Topological Structure}
The internal momenta have eight degrees of freedom in strictly four dimensions
and therefore we are left with one free parameter, after imposing the hepta-cut
constraints. This implies that after promoting the original real slice contours
of integration to seven-tori encircling the simultaneous zeros of the
denominators in the crossed box integral, the integrand is localized onto the
Riemann sphere associated with one of the hepta-cut equations found in the
previous section.

The maximal cut involves an inverse Jacobian that has multiple poles in the
remaining variable $z$. These poles are known as composite leading
singularities. For integrals with tensor numerators, additional poles in the
integrand must be taken into account. Once the pole structure is properly
understood, we can define the hepta-cut integral by a weighted residue expansion
over a minimal set of all poles. We return to this part of the computation
shortly.

In order to actually perform the hepta-cut contour integrals, it is necessary to
break the automatic satisfaction of the on-shell constraints 
$\ell_1^2 = \ell_2^2 = 0$ which instead should be imposed by localization. To
that end it is advantageous to introduce new parameters $\zeta_1,\zeta_2\in\C$
and two null-vectors $\eta_1,\eta_2$ with the obvious properties,
\begin{align}
\slashed{\eta}_1^\mu\ket{\lambda_1^+} \neq 0 \neq 
\slashed{\eta}_1^\mu\ket{\tilde\lambda_1^{'-}}\;, \quad
\slashed{\eta}_2^\mu\ket{\lambda_2^+} \neq 0 \neq 
\slashed{\eta}_1^\mu\ket{\tilde\lambda_2^{'-}}\;,
\end{align}
and then shift the loop momentum parametrization such that
\begin{align}
\ell_1^\mu = 
\frac{1}{2}\spab{\lambda_1}{\gamma^\mu}{\tilde\lambda_1'}
+\zeta_1\eta_1^\mu\;, \quad
\ell_2^\mu = 
\frac{1}{2}\spab{\lambda_2}{\gamma^\mu}{\tilde\lambda_2'}
+\zeta_2\eta_2^\mu\;.
\end{align}
In general, this choice is tied to the specific on-shell solution in question. 
An appropriate choice in any case is to take the masslessly projected legs
because it simplifies the calculation. For instance, let $\eta_1 = k_2^\flat$
and $\eta_2 = k_3^\flat$. Then for all four-mass solutions $\mc S_1,\dots,\mc
S_4$ it is clear that
\begin{align}
\ell_1^2 = \gamma_{12}\xi_1\xi_1'\zeta_1 = 0\;, \quad
\ell_2^2 = \gamma_{34}\xi_4\xi_4'\zeta_2 = 0\;,
\end{align}
and therefore we obtain the desired solution $\zeta_1 = \zeta_2 = 0$ since
$\xi_1\xi_1' \neq 0 \neq \xi_4\xi_4'$ for general momenta. The same arguments
show that a valid choice for solutions $\mc S_5$ and $\mc S_6$ is 
$\eta_1 = k_2^\flat$ and $\eta_2 = k_4^\flat$ because in that situation 
$\xi_1\xi_1'\neq 0 \neq \xi_3\xi_3'$.

We also need to include the Jacobian $J^L\times J^R$ due to coordinate
transformation from loop four-momenta to parameter space. This can easily be
done by Wick rotation to Euclidean spacetime where the volume of the
4-parallelotope spanned by a set of four vectors $q_i^\mu$ can be computed from
the corresponding Gram determinant up to an overall sign, which we eventually
determine numerically along with potential factors of $i$ introduced by the
analytic continuation,
\begin{align}
\det_{\mu,i} q^\mu_i = \pm\big(\det_{i,j} q_i\cdot q_j\big)^{1/2}\;.
\end{align}
The set of variables parametrizing the two loop momenta $\ell_1$ and $\ell_2$
after removal of redundant degrees of freedom also depend on the branch in
question. In $\mc S_1$ and $\mc S_3$ we keep the variables 
$\alpha = (\zeta_1,\xi_1',\xi_2,\xi_2')$ and 
$\beta = (\zeta_2,\xi_3,\xi_3',\xi_4)$ and thus fix $\xi_1 = \xi_4' = 1$.  The
corresponding Jacobians to appear in the numerator are then
\begin{align}
J_{\mc S_{1,3}}^L = \det_{\mu,i}\pd{\ell_1^\mu}{\alpha_i} = 
i\frac{\gamma_{12}^2 k_1^\flat\cdot k_4^\flat\xi_1'}
{4k_2^\flat\cdot k_4^\flat}\;, \quad
J_{\mc S_{1,3}}^R = \det_{\mu,i}\pd{\ell_2^\mu}{\beta_i} = 
i\frac{\gamma_{34}^2 k_1^\flat\cdot k_4^\flat\xi_4'}
{4k_1^\flat\cdot k_3^\flat}\;.
\end{align}
Notice that these forms are not constant because the parametrization is not
linear in the parameters as opposed to previous work \cite{Sogaard:2013yga} in
the purely massless case. Similarly for $\mc S_2$ and $\mc S_4$ where
$\alpha = (\zeta_1,\xi_1,\xi_2,\xi_2')$ and 
$\beta = (\zeta_2,\xi_3,\xi_3',\xi_4')$ so that
\begin{align}
J_{\mc S_{2,4}}^L = \det_{\mu,i}\pd{\ell_1^\mu}{\alpha_i} = 
i\frac{\gamma_{12}^2 k_1^\flat\cdot k_4^\flat\xi_1}
{4k_2^\flat\cdot k_4^\flat}\;, \quad
J_{\mc S_{2,4}}^R = \det_{\mu,i}\pd{\ell_2^\mu}{\beta_i} = 
i\frac{\gamma_{34}^2 k_1^\flat\cdot k_4^\flat\xi_4}
{4k_1^\flat\cdot k_3^\flat}\;.
\end{align}
We finally compute the Jacobians appropriate to solutions $\mc S_5$ and 
$\mc S_6$,
\begin{align}
J_{\mc S_5}^L = \det_{\mu,i}\pd{\ell_1^\mu}{\alpha_i} = 
i\frac{\gamma_{12}^2 k_1^\flat\cdot k_4^\flat\xi_1}
{4k_2^\flat\cdot k_4^\flat}\;, \quad
J_{\mc S_5}^R = \det_{\mu,i}\pd{\ell_2^\mu}{\beta_i} = 
i\gamma_{34}^2\bigg(\frac{k_1^\flat\cdot k_4^\flat}{k_1^\flat\cdot
k_3^\flat}\bigg)^2\xi_3\;, \\
J_{\mc S_6}^L = \det_{\mu,i}\pd{\ell_1^\mu}{\alpha_i} = 
i\frac{\gamma_{12}^2 k_1^\flat\cdot k_4^\flat\xi_1'}
{4k_2^\flat\cdot k_4^\flat}\;, \quad
J_{\mc S_6}^R = \det_{\mu,i}\pd{\ell_2^\mu}{\beta_i} = 
i\gamma_{34}^2\bigg(\frac{k_1^\flat\cdot k_4^\flat}{k_1^\flat\cdot
k_3^\flat}\bigg)^2\xi_3'\;.
\end{align}

\subsection{Hextuply Pinched Genus-3 Curve}
The preceding discussions now lead us to the definition of localization of the
two-loop crossed-box integral onto the Riemann sphere associated with the $i$th
branch. Changing integration variables from loop momenta
$(\ell_1^\mu,\ell_2^\mu)$ to parameters $(\xi_i,\xi_i')$, replacing real slice
integration contours by a multidimensional tori encircling the joint solution of
the hepta-cut constraints and subsequently performing seven contour integrals
using eq.~\eqref{LOCALIZATIONPROPERTY} give rise to a total Jacobian $J_i$,
\begin{align}
\X[1]_{\mc S_i}\equiv\oint_{\Gamma_i} dzJ_i(z)\;.
\end{align}
The generic form of this Jacobian is a product of $n_i$ simple-pole factors
associated with the pinching points or intersections with neighboring Riemann
spheres, 
\begin{align}
J_i\equiv
\frac{h(z)}{\prod_{k=1}^{n_i}(z-z_{i;k})}\;,
\end{align}
where $h(z)$ is a regular function of $z$. For the two-loop crossed box
integrals with up to four massive legs and no doubled propagators, the Jacobian
will at most define a quartic polynomial because $n_i\leq 4$ for all $i$ as we
shall see below.

It is straightforward to obtain the Jacobians explicitly after the inverse
propagators have been expanded in parameter space. For brevity we merely state
the results here. We refer the reader to e.g.
refs.~\cite{Sogaard:2013yga,Sogaard:2013fpa,Kosower:2011ty} for related
examples. In advance of calculations below, we identify a frequently occurring
kinematic constant along the lines of ref.~\cite{Johansson:2013sda}, 
\begin{align}
\gamma_\star\equiv
\frac{\gamma_{12}\gamma_{34}}{
32(\gamma_{12}^2-m_1^2m_2^2)(\gamma_{34}^2-m_3^2m_4^2)
k_1^\flat\cdot k_4^\flat
}\;. 
\end{align}
The multivariate residues evaluated at the simultaneous zeros of the
denominators are,
\begin{gather}
\X[1]_{\mc S_{1,2}} = 
+\frac{\gamma_\star(1-\tau)(\bxi_1-\tau\bxi_2)}{
\tau\bxi_1^2(1+\mu\tau)}
\oint dz\frac{z}{
(z+1)\left(z+\frac{1}{\tau}\right)\left(
z+\frac{\bxi_2}{\bxi_1}\right)\left(
z+\frac{\tau\bxi_2}{\bxi_1}\right)
} \;, \\[1mm]
\X[1]_{\mc S_{3,4}} = 
-\frac{\gamma_\star}{\bxi_1-\tau\bxi_2}
\oint\frac{dz}{
\left(z+\frac{\bxi_1-\bxi_2}{\bxi_1-\tau\bxi_2}\right)
\left(z-\frac{\mu\tau(\bxi_1-\bxi_2)}{\bxi_1-\tau\bxi_2}\right)
} \;, \\[1mm]
\!\!\X[1]_{\mc S_{5,6}} = 
+\frac{\gamma_\star}{\bxi_1-\tau\bxi_2}
\oint\frac{dz}{
\left(z-\frac{\bxi_1-\tau^2\bxi_2}{\tau(\bxi_1-\tau\bxi_2)}\right)
\left(z+\frac{\mu(\bxi_1-\tau^2\bxi_2)}{\bxi_1-\tau\bxi_2)}\right)
} \;, 
\label{4MJACOBIANS}
\end{gather}
where the pole locations in the Jacobians are directly exposed,
\begin{align}
\label{ZPOLES}
\{z_{1;1},\dots,z_{1;4}\} = \left\{
-1,-\frac{1}{\tau},-\frac{\bxi_2}{\bxi_1},-\frac{\tau\bxi_2}{\bxi_1}
\right\} = \{z_{2;1},\dots,z_{2;4}\}\;, \nn \\[1mm]
\{z_{3;1},z_{3;2}\} = \left\{
-\frac{\bxi_1-\bxi_2}{\bxi_1-\tau\bxi_2},
\frac{\mu\tau(\bxi_1-\bxi_2)}{\bxi_1-\tau\bxi_2}
\right\} = \{z_{4;1},z_{4;2}\}\;, \nn \\[1mm]
\{z_{5;1},z_{5;2}\} = \left\{
\frac{\bxi_1-\tau^2\bxi_2}{\tau(\bxi_1-\tau\bxi_2)},
-\frac{\mu(\bxi_1-\tau^2\bxi_2)}{\bxi_1-\tau\bxi_2}
\right\} = \{z_{6;1},z_{6;2}\}\;.
\end{align}
Solutions $\mc S_1$ and $\mc S_2$ give rise to additional singularities where
either of the loop momenta become infinite for a finite value of the post
hepta-cut degree of freedom $z$. Finally, in each solution there is a pole at
$z = \infty$ which may be encircled. The union of the singular point loci for
all six branches therefore contains 24 points.

This leads to the definition of the octa-cut of a general tensor integral with
numerator insertion $\Phi(\ell_1(z),\ell_2(z))$,
\begin{align}
\X[\Phi(\ell_1(z),\ell_2(z)]\big|_{8-\cut}\equiv
\sum_{i=1}^{6} \oint_{\Gamma_i}
dzJ_i(z)\Phi(\ell_1(z),\ell_2(z))\;,
\end{align}
Here, $\Gamma_i$ is a weighted linear combination of small circles around the
poles in the remaining variable $z$ chosen so that the integrals extract the
residues of the loop integrand. Denoting the weight of the residue evaluated at
$z = \xi$ for the $i$th branch by $\omega(i,\xi)$, we have
\begin{align}
\X[\Phi(\ell_1(z),\ell_2(z))]\big|_{8-\cut} = 
\sum_{i=1}^{|\mc S|}\sum_{\xi\in\text{poles}}
\omega(i,\xi)\Res_{z=\xi} J_i(z)\Phi(z)\big|_{\mc S_i}\;.
\end{align}

\begin{figure}[!h]
\bc
\includegraphics[scale=0.5]{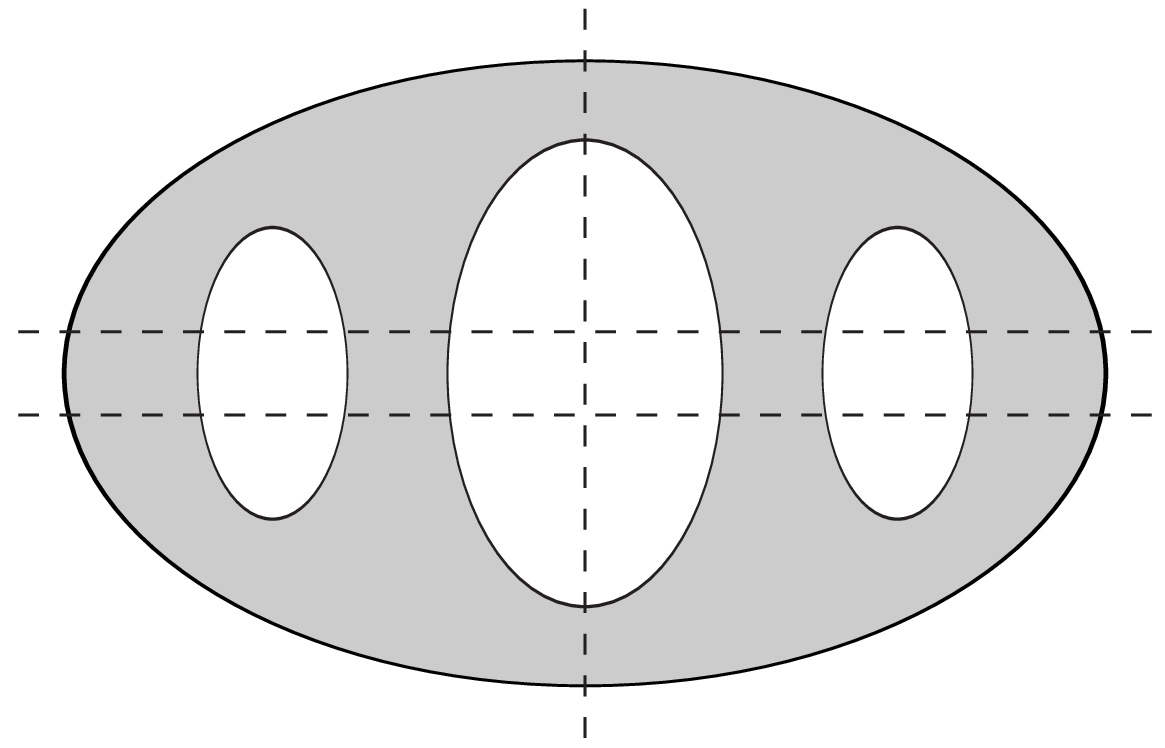}
\ec
\vspace*{-.5cm}
\caption{Topological depiction of the genus-3 algebraic curve defined by the
hepta-cut of the prime configuration of the two-loop crossed box primitive
amplitude. The one-dimensional complex curve should be understood as the filled
two-dimensional real surface. Degeneracies appropriate to specific kinematics
arise upon contraction of tubes along straight horizontal and vertical lines in
the paper plane through the handles of the surface.}
\label{GENUS3}
\end{figure}

\begin{figure}[!h]
\bc
\hspace*{.5cm}
\includegraphics[scale=0.65]{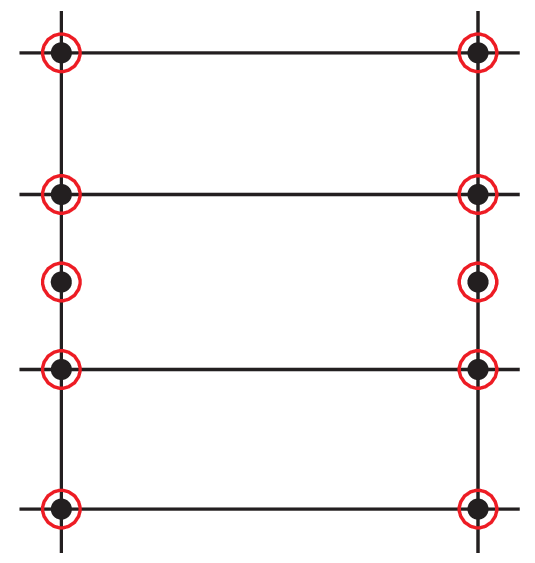}
\put(-189,14){$1\cap 4$}
\put(-189,58){$1\cap 6$}
\put(-189,113){$1\cap 5$}
\put(-189,158){$1\cap 3$}
\put(-1,14){$2\cap 4$}
\put(-1,58){$2\cap 6$}
\put(-1,113){$2\cap 5$}
\put(-1,158){$2\cap 3$}
\put(-182,87){$\infty_{LR}$}
\put(-8,87){$\infty_{LR}$}
\hspace*{1cm}
\includegraphics[scale=0.65,angle=90]{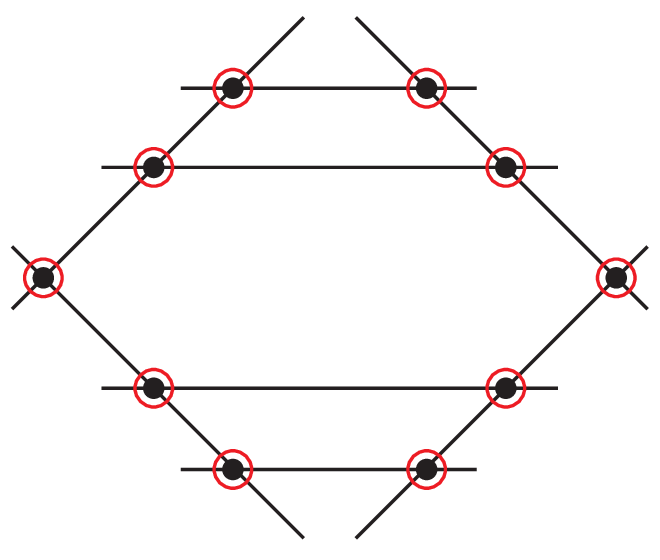}
\put(-54,145){$4\cap 6$}
\put(-29,120){$4\cap 8$}
\put(-179,145){$1\cap 6$}
\put(-205,120){$1\cap 8$}
\put(-205,50){$3\cap 7$}
\put(-179,25){$3\cap 5$}
\put(-54,25){$2\cap 5$}
\put(-29,50){$2\cap 7$}
\put(-184,85){$1\cap 3$}
\put(-50,85){$2\cap 4$}
\ec
\caption{Global structure of the hepta-cut of the two-loop crossed box; the
figures show intersections of the irreducible components of the algebraic curve
defined by the zero locus of the polynomial ideal generated by the inverse
propagators in the case of four external massive legs (left) and at least one
massless particle (right). All degenerate configurations considered in this
paper fall within these two topological pictures. The straight lines are drawn
for simplicity and should be interpreted as Riemann spheres.}
\label{GLOBALCUT}
\end{figure}
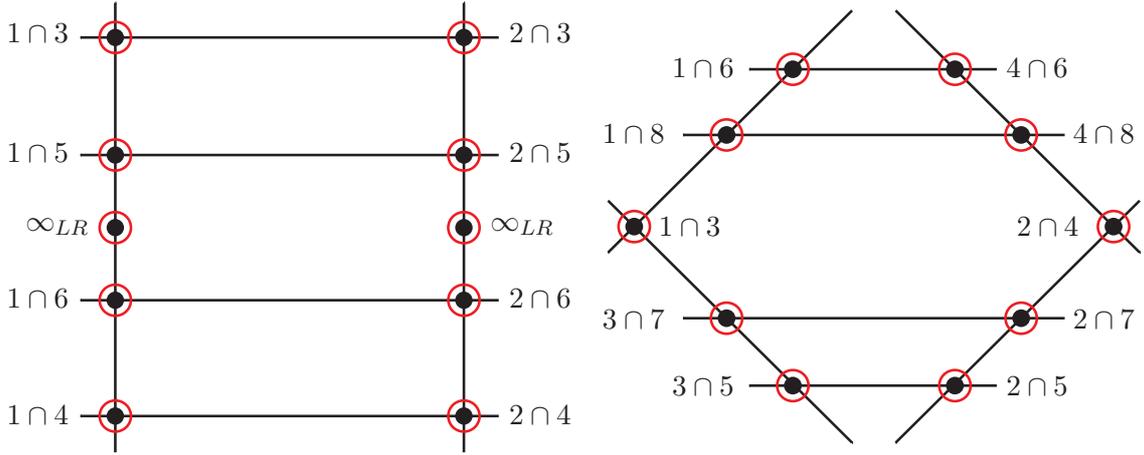

Not all of these residues are independent though, as can be explained from the
global structure of the unitarity cut \cite{CaronHuot:2012ab,Huang:2013kh}.
Indeed, consider an arbitrary integrand test function of the two loop momenta,
$\Phi(\ell_1(z),\ell_2(z))$, and assume regularity on the Jacobian poles.
It is then very easy to prove the residue relation,
\begin{align}
\Res_{\mc S_i\cap S_j}J(z)\Phi(\ell_1(z),\ell_2(z))\big|_{\mc S_i} = 
-\Res_{\mc S_i\cap S_j}J(z)\Phi(\ell_1(z),\ell_2(z))\big|_{\mc S_j}\;,
\end{align}
where the left and right hand sides of the equation are understood to be
evaluated in local coordinates on solutions $\mc S_i$ and $\mc S_j$
respectively. Other choices are equally valid, e.g symmetric in $i$ and $j$.
For the purpose of completeness, let us state all such identities:

\noindent
\begin{minipage}{0.50\textwidth}
\begin{align}
\Res_{z=z_{1,1}}J_1(z)\Phi(z)|_{\mc S_1} = 
-\Res_{z=z_{4,1}}J_4(z)\Phi(z)|_{\mc S_4}\;, \nn \\ 
\Res_{z=z_{1,2}}J_1(z)\Phi(z)|_{\mc S_1} = 
-\Res_{z=z_{6,2}}J_6(z)\Phi(z)|_{\mc S_6}\;, \nn \\ 
\Res_{z=z_{1,3}}J_1(z)\Phi(z)|_{\mc S_1} = 
-\Res_{z=z_{3,2}}J_3(z)\Phi(z)|_{\mc S_3}\;, \nn \\ 
\Res_{z=z_{1,4}}J_1(z)\Phi(z)|_{\mc S_1} = 
-\Res_{z=z_{5,1}}J_5(z)\Phi(z)|_{\mc S_5}\;, \nn \\ \nn
\end{align}
\vspace*{.1mm}
\end{minipage}
\begin{minipage}{0.50\textwidth}
\begin{align}
\Res_{z=z_{2,1}}J_2(z)\Phi(z)|_{\mc S_2} = 
-\Res_{z=z_{3,1}}J_3(z)\Phi(z)|_{\mc S_3}\;, \nn \\ 
\Res_{z=z_{2,2}}J_2(z)\Phi(z)|_{\mc S_2} = 
-\Res_{z=z_{5,2}}J_5(z)\Phi(z)|_{\mc S_5}\;, \nn \\ 
\Res_{z=z_{2,3}}J_2(z)\Phi(z)|_{\mc S_2} = 
-\Res_{z=z_{4,2}}J_4(z)\Phi(z)|_{\mc S_4}\;, \nn \\ 
\Res_{z=z_{2,4}}J_2(z)\Phi(z)|_{\mc S_2} = 
-\Res_{z=z_{6,1}}J_6(z)\Phi(z)|_{\mc S_6}\;. \nn \\
\end{align}
\vspace*{.1mm}
\end{minipage}
This pattern of intersections confirms the global topological structure of the
hepta-cut in fig.~\ref{GLOBALCUT}. This picture follows by pinching the tubes of
the genus-3 surface six times along a horizontal and a vertical line passing
through the center of the object, see fig.~\ref{GENUS3}. The number of
independent residues is reduced to 16. The residues at infinity and in numerator
insertions are not shared. However, within each Riemann sphere the
one-dimensional global residue theorem ensures that the sum of all residues
vanish. 

The upshot is that we only need to encircle 10 global poles to produce a
complete basis of homology for $\mc S_1\cup\cdots\cup\mc S_6$. In particular,
the contributions from branches $\mc S_3,\dots,\mc S_6$ are redundant, because
each of them only has a single pole (at $z = \infty$) besides those located at
intersections with $\mc S_1$ and $\mc S_2$. To minimize an overcomplete basis we
may simply set extraneous residue weights to zero. It is natural to choose an
ordered set of winding numbers, call it $\W$, so that we encircle all Jacobian
poles along with poles in numerator insertions where both loop momenta become
infinite simultaneously. Following the notation of ref.~\cite{CaronHuot:2012ab},
the weights of the ten global poles can be written
\begin{align}
\W = 
(\omega_{1\cap4},
\omega_{1\cap6},
\omega_{1\cap3},
\omega_{1\cap5},
\omega_{1,\infty_{LR}},
\omega_{2\cap3},
\omega_{2\cap5},
\omega_{2\cap4},
\omega_{2\cap6},
\omega_{1,\infty_{LR}}
) \equiv
(\omega_1,\dots,\omega_{10})\;.
\end{align}
By convention, a residue with weight $\omega_{i\cap j}$ is evaluated on the
$i$th branch. Later it may be convenient to instead encircle infinity poles,
since the scalar integrals yield simpler residues there.

\subsection{Octuply Pinched Genus-3 Curve}
Let us now relax the condition $m_1m_2\neq 0$ and analyze the analytic structure
of the loop integrand under those circumstances. The eight hepta-cut solutions
for this class of kinematics were determined in section~\ref{MASSLESSLIMITS}.
The topological picture is that of an octuply pinched genus-3 surface, where
tubes have been contracted along one vertical and two horizontal lines through
the center as shown in fig.~\ref{GENUS3}.

We can reproduce this situation locally from coincidence of residues.
Localizing the scalar master integral onto each of the eight Riemann spheres
parametrized by the hepta-cut solutions $\tS_1$ through $\tS_8$ yields four
pairs of one-dimensional contour integrals,
\begin{gather}
\X[1]_{\tS_{1,2}} = 
+\frac{\gamma_\star(1-\tau)}{
\tau\bxi_1(1+\mu\tau)}
\oint\frac{dz}{z(z+1)(z+1/\tau)} \;, 
\label{DEGENERATECUTRESIDUES1} \\[2mm]
\X[1]_{\tS_{3,4}} = 
-\frac{\gamma_\star(1-\tau)}{
\tau\bxi_1(1+\mu\tau)}
\oint\frac{dz}{z(z+1)(z+1/\tau)} \;, \\[2mm]
\X[1]_{\tS_{5,6}} = 
-\frac{\gamma_\star}{\bxi_1}
\oint\frac{dz}{(z+1)(z-\mu\tau)} \;, \\[2mm]
\X[1]_{\tS_{7,8}} = 
+\frac{\gamma_\star}{\bxi_1}
\oint\frac{dz}{(z+\mu)(z-1/\tau)} \;.
\label{DEGENERATECUTRESIDUES4}
\end{gather}
The residues at the poles in the displayed loop integrands again satisfy a rich
set of linear relations across the irreducible components of the genus-3 curve
and thus reflect the global structure of the unitarity cut. In fact,

\noindent
\begin{minipage}{0.50\textwidth}
\begin{align}
\Res_{z=0}\widetilde{J}_1(z)\Phi(z)|_{\tS_1} = {} &
-\Res_{z=0}\widetilde{J}_3(z)\Phi(z)|_{\tS_3}\,, \nn \\ 
\Res_{z=-1}\widetilde{J}_1(z)\Phi(z)|_{\tS_1} = {} &
-\Res_{z=-1}\widetilde{J}_6(z)\Phi(z)|_{\tS_6}\,, \nn \\ 
\Res_{z=-1/\tau}\widetilde{J}_1(z)\Phi(z)|_{\tS_1} = {} & 
-\Res_{z=-\mu}\widetilde{J}_8(z)\Phi(z)|_{\tS_8}\,, \nn \\ 
\Res_{z=-1}\widetilde{J}_3(z)\Phi(z)|_{\tS_3} = {} &
-\Res_{z=\mu\tau}\widetilde{J}_5(z)\Phi(z)|_{\tS_5}\,, \nn \\ 
\Res_{z=-1/\tau}\widetilde{J}_3(z)\Phi(z)|_{\tS_3} = {} &
-\Res_{z=1/\tau}\widetilde{J}_7(z)\Phi(z)|_{\tS_7}\,, \nn \\ \nn
\end{align}
\end{minipage}
\begin{minipage}{0.50\textwidth}
\begin{align}
\Res_{z=0}\widetilde{J}_2(z)\Phi(z)|_{\tS_2} = {} & 
-\Res_{z=0}\widetilde{J}_4(z)\Phi(z)|_{\tS_4}\,, \nn \\ 
\Res_{z=-1}\widetilde{J}_2(z)\Phi(z)|_{\tS_2} = {} &
-\Res_{z=-1}\widetilde{J}_5(z)\Phi(z)|_{\tS_5}\,, \nn \\ 
\Res_{z=-1/\tau}\widetilde{J}_2(z)\Phi(z)|_{\tS_2} = {} & 
-\Res_{z=-\mu}\widetilde{J}_7(z)\Phi(z)|_{\tS_7}\,, \nn \\ 
\Res_{z=-1}\widetilde{J}_4(z)\Phi(z)|_{\tS_4} = {} &
-\Res_{z=\mu\tau}\widetilde{J}_6(z)\Phi(z)|_{\tS_6}\,, \nn \\ 
\Res_{z=-1/\tau}\widetilde{J}_4(z)\Phi(z)|_{\tS_4} = {} & 
-\Res_{z=1/\tau}\widetilde{J}_8(z)\Phi(z)|_{\tS_8}\,. \nn \\
\end{align}
\end{minipage} 

In the purely massless limit, $m_1,\dots,m_4\to 0$, the kinematic quantities
defined above reduce as follows,
\begin{align}
\bxi_1\to1\;, \quad
\bxi_2\to0\;, \quad
\mu\to0\;, \quad
\tau\to\frac{\chi}{1+\chi}\;, \quad
\gamma_\star = \frac{1}{16\chi s_{12}^3}\;,
\end{align}
where $\chi\equiv s_{14}/s_{12}$. We point out that the hepta-cut contributions
in eqs.~\eqref{DEGENERATECUTRESIDUES1}-\eqref{DEGENERATECUTRESIDUES4} in
ref.~\cite{Sogaard:2013yga} after appropriate reparametrization,
\begin{gather}
\X[1]_{\tS_{5,6}} = 
-\frac{1}{16s_{12}^3}\oint\frac{dz}{z(z+\chi)}\;, \\[2mm]
\X[1]_{\tS_{7,8}} = 
-\frac{1}{16s_{12}^3}\oint\frac{dz}{z(z-\chi-1)}\;, \\[2mm]
\X[1]_{\tS_{1,2,3,4}} = 
-\frac{1}{16s_{12}^3}\oint\frac{dz}{z(z-\chi)(z-\chi-1)}\;.
\end{gather}

\section{Master Integral Projectors}
The hepta-cut localizes the two-loop crossed-box integrand onto a variety of
linked Riemann spheres associated with the joint solutions of the on-shell
equations. What remains is a one-dimensional complex contour integral whose
integrand has poles. We can now choose contours that extract residues of the
integrand and effectively obtain an octa-cut such that the integral is
completely localized to a point in $\C^8$. On top of that, consistency of the
unitarity method imposes nontrivial constraints on these contours, however.

The reason is that converting a real slice integral into a multidimensional
contour integral in general does not respect various relations among integrals.
Each identity leads to a constraint which at two loops always can be rearranged
and phrased as vanishing of a certain function upon integration over
$\R^D\times\R^D$. Schematically,
\begin{align}
\X[\Phi(\ell_1,\ell_2)] = 0 \;\LRa\; 
\X[\Phi(\ell_1,\ell_2)]\big|_{8-\cut} = 0\;.
\end{align}
It is easy to understand the nature of these relations if we imagine that we
compute an amplitude diagram by diagram. All contractions between loop momenta
and external vectors are expressible in terms of the eight fundamental scalar
products $\ell_i\cdot e_j$ for $i = 1,2$ and $j = 1,2,3,4$ where 
$e = (k_1,k_2,k_4,\omega)$. Here, $\omega$ is a spurious direction that is
perpendicular to the subspace spanned by the four external momenta. 

Odd powers of $\ell_1\cdot\omega$ and $\ell_2\cdot\omega$ vanish upon
integration, whereas even powers are reducible in four dimensions. It readily
follows that $\ell_1\cdot k_1$, $\ell_1\cdot k_2$ and $\ell_2\cdot k_4$ can be
written in terms of inverse propagators and external invariants. Moreover,
$\ell_2\cdot k_2$ depends linearly on $\ell_1\cdot k_4$ and $\ell_2\cdot k_1$.
The latter two may be selected as irreducible scalar products. Accordingly, the
general numerator polynomial for the problem at hand can be parametrized as
follows,
\begin{align}
N = {} & 
\sum_{\{\alpha_1,\dots,\alpha_4\}}c_{\alpha_1\cdots\,\alpha_4}
(\ell_1\cdot k_4)^{\alpha_1}
(\ell_2\cdot k_1)^{\alpha_2}
(\ell_1\cdot \omega)^{\alpha_3}
(\ell_2\cdot \omega)^{\alpha_4}\;.
\end{align}
The integrand reduction can be obtained by imposing renormalizability conditions
that constrain the exponents of the ISPs and performing the multivariate
polynomial division of $N$ modulo a Gr{\"o}bner basis constructed from the seven
inverse propagators. The latter part including identification of ISPs is carried
out automatically by the Mathematica package {\tt BasisDet} \cite{Zhang:2012ce}.
The integrand contains 19 parity-odd and 19 parity-even elements as previously
reported \cite{Badger:2012dp}.

The analysis above suggests that the amplitude contribution in question contains
19 genuine integrals of the form 
\begin{align}
\X[n,m]\equiv\X\bigg[
\bigg(
\frac{(\ell_1+k_4)^2}{2}\bigg)^{\!n}
\bigg(\frac{(\ell_2+k_1)^2}{2}\bigg)^{\!m}\bigg]\;.
\end{align}
Many integrals are expressible as linear combinations of integrals with
lower-rank tensors and fewer than seven propagators. This reduction is achieved
due to IBP relations that follow from inserting a total derivative into the loop
integrand and discarding the boundary term in $D$ dimensions. The relations take
the form
\begin{align}
\X[n,m] = \sum_{(p,q)\in{\text{Basis}}}c_{pq}\X[p,q]+\cdots\;,
\end{align}
where $\cdots$ means fewer-propagator topologies that have vanishing hepta-cuts,
and the consistency constraint thus reads
\begin{align}
\X[n,m]\big|_{8-\cut} = 
\sum_{(p,q)\in{\text{Basis}}}c_{pq}\X[p,q]\big|_{8-\cut}\;.
\end{align}

The IBP relations can be generated by various public computer codes. For this
project we used the Mathematica package {\tt FIRE} \cite{Smirnov:2013dia}. We
list below a few examples. There are two relations which hold for arbitrary
values of the external masses,
\begin{align}
\X[0,-1] = {} & A\cdot\X[0,0]+B\cdot\X[-1,0]\;, \nn \\ 
\X[-1,-1] = {} & A\cdot\X[-1,0]+B\cdot\X[-2,0]\;,
\end{align}
for $A$ and $B$ given by
\begin{align}
A = \frac{1}{4}\left(
s_{14}+\frac{m_1^2m_3^2-m_2^2m_4^2}{s_{12}}\right)\;, \quad
B = -\frac{1}{2}\left(
1+\frac{m_1^2-m_2^2}{s_{12}}\right)\;.
\end{align}
If for instance $m_2 = m_3 = m_4 = 0$ and $m_1$ is nonzero, then there is one
additional relation among these integrals,
\begin{align}
\X[-2,0] = 
-\frac{s_{12}^2s_{14}}{16(m_1^2-s_{12})}\X[0,0]
-\frac{1}{8}\left(3+\frac{2s_{14}}{m_1^2-s_{12}}\right)\X[-1,0]\;.
\end{align}

The parity-odd terms in the integrand basis of course vanish identically after
the loop momentum integration has been performed, but they are nonetheless very
important for integrand-level reduction and unitarity purposes. It is sufficient
to require that the full variety of integrals with Levi-Civita insertions, after
invoking momentum conservation,
\begin{gather}
\varepsilon(\ell_1,k_2,k_3,k_4)\;, \;\;\;
\varepsilon(\ell_2,k_2,k_3,k_4)\;, \;\;\;
\varepsilon(\ell_1,\ell_2,k_1,k_2)\;, \;\;\;
\varepsilon(\ell_1,\ell_2,k_1,k_3)\;, \;\;\;
\varepsilon(\ell_1,\ell_2,k_2,k_3)\;,
\end{gather}
continue to integrate to zero on general contours in $\C^4\times\C^4$.

Our goal of the rest of this paper is to massage the amplitude master equation
\eqref{MIEQ} into a form that allows us to project the master integral
coefficients.

\subsection{One-Mass Projectors}
The simplest configuration is the one-mass diagram with, say, $m_1\neq 0$. As
in the purely massless case \cite{Sogaard:2013yga,Badger:2012dp}, there are two
master integrals so the amplitude contribution can be expressed as
\begin{align}
\mc A^{(2)}_4 = c_1\X[1]+c_2\X[(\ell_1+k_4)^2/2]+\cdots\;,
\end{align}
where integrals with less than seven propagators are truncated. Since momentum
$k_4$ is massless, we have $\ell_1\cdot k_4 = (\ell_1+k_4)^2/2$. We decide to
encircle the following set of global poles,
\begin{align}
\{\tG_i\} = (
\tG_{1\cap6},\,
\tG_{1\cap8},\,
\tG_{3\cap5},\,
\tG_{3\cap7},\,
\tG_{1\cap3},\,
\tG_{2\cap5},\,
\tG_{2\cap7},\,
\tG_{4\cap6},\,
\tG_{4\cap8},\,
\tG_{2\cap4}
) \equiv (\tG_1,\dots,\tG_{10})\;,
\end{align}
and let $\tW$ denote the corresponding weights,
\begin{align}
\label{ONEMASSRESIDUELOCUS}
\W' = (
\tomega_{1\cap6},\,
\tomega_{1\cap8},\,
\tomega_{3\cap5},\,
\tomega_{3\cap7},\,
\tomega_{1\cap3},\,
\tomega_{2\cap5},\,
\tomega_{2\cap7},\,
\tomega_{4\cap6},\,
\tomega_{4\cap8},\,
\tomega_{2\cap4})
\equiv (\tomega_1,\dots,\tomega_{10})\;.
\end{align}
The hepta-cut two-loop crossed box integrals reduce to
\begin{gather}
\label{DEGENERATECUTM11}
\X[1]_{\tS_{1,2}} \longrightarrow
-\frac{m_1^2-s_{12}}{\chi s_{12}}
\oint\frac{dz}{z(z+1)(z+(s_{12}(1+\chi)-m_1^2)/(\chi s_{12}))} \;, \\[2mm]
\label{DEGENERATECUTM12}
\X[1]_{\tS_{3,4}} \longrightarrow
+\frac{m_1^2-s_{12}}{\chi s_{12}}
\oint\frac{dz}{z(z+1)(z+(s_{12}(1+\chi)-m_1^2)/(\chi s_{12}))} \;,
\end{gather}
and likewise for the remaining four solutions, which we do not explicitly need
here. Notice that we stripped off the overall factor. As expected we immediately
derive five linearly independent contour constraints arising from parity-odd
numerator insertions,
\begin{align}
\label{PARITYCONSTRAINTSDEG}
\left(
\begin{array}{lllllrrrrr}
1\;\; & 0\;\; & 0\;\; & 0\;\; & 0 & -1 & 0 & 0 & 0 & 0 \\
0 & 1 & 0 & 0 & 0 & 0 & -1 & 0 & 0 & 0 \\
0 & 0 & 1 & 0 & 0 & 0 & 0 & -1 & 0 & 0 \\
0 & 0 & 0 & 1 & 0 & 0 & 0 & 0 & -1 & 0 \\
0 & 0 & 0 & 0 & 1 & 0 & 0 & 0 & 0 & -1 \\
\end{array}
\right)
\left(
\begin{array}{c}
\tomega_{1\cap6} \\
\tomega_{1\cap8} \\
\tomega_{3\cap5} \\
\tomega_{3\cap7} \\
\tomega_{1\cap3} \\
\tomega_{2\cap5} \\
\tomega_{2\cap7} \\
\tomega_{4\cap6} \\
\tomega_{4\cap8} \\
\tomega_{2\cap4}
\end{array}
\right) = 0\;.
\end{align}
Moreover, there are three linearly independent IBP constraints,
\begin{align}
\tW\cdot
(0,\,1,\,0,\,1,\,0,\,0,\,1,\,0,\,1,\,0) = 0 \;, \\
\tW\cdot
(1,\,-1,\,1,\,-1,\,0,\,1,\,-1,\,1,\,-1,\,0) = 0\;, \\
\tW\cdot
(1,\,1,\,-1,\,-1,\,2,\,1,\,1,\,-1,\,-1,\,2) = 0 \;.
\end{align}
The residues computed by the master integrals around the global poles are
\begin{align}
{\textstyle\Res}_{\{\tG_i\}}\X[1] = {} & 
\bigg(
-1,\,-\frac{\chi s_{12}}{m_1^2-(1+\chi)s_{12}},\,1,\,
\frac{\chi s_{12}}{m_1^2-(1+\chi)s_{12}},\,
\frac{\chi s_{12}}{m_1^2-(1+\chi)s_{12}},\,
\nn \\[1mm] & \quad\!
-1,\,-\frac{\chi s_{12}}{m_1^2-(1+\chi)s_{12}},\,1,\,
\frac{\chi s_{12}}{m_1^2-(1+\chi)s_{12}},\,
\frac{\chi s_{12}}{m_1^2-(1+\chi)s_{12}}
\bigg)\;,\; \nn \\[2mm]
{\textstyle\Res}_{\{\tG_i\}}\X[\ell_1\cdot k_4] = {} &
\frac{\chi s_{12}^2}{2(m_1^2-(1+\chi)s_{12})}
(0,\,1,\,0,\,-1,\,-1,\,0,\,1,\,0,\,-1,\,-1)\;.
\end{align}
We exploit the freedom to choose contours after imposing the reduction
conditions and define two master integral projectors (also called master
contours) which extract of either of the master integral coefficients,
\begin{align}
\mc M_1\cdot\big(\Res_{\{\tG_i\}}\X[1],\,
\Res_{\{\tG_i\}}\X[\ell_1\cdot k_4]\big) = (1,0)\;, \\
\mc M_2\cdot\big(\Res_{\{\tG_i\}}\X[1],\,
\Res_{\{\tG_i\}}\X[\ell_1\cdot k_4]\big) = (0,1)\;.
\end{align}
Here, $\mc M_1$ and $\mc M_2$ are just particular lists of the winding numbers
of the corresponding global poles with the property that they only receive
contribution from one master integral, which is also normalized to unity.

The eight contour constraints together with either of the projectors are in
practice arranged as $10\times 10$ matrices. The rank is 10 and the solutions
for the weights are uniquely determined. We find that the projectors are
characterized by the 10-tuples
\begin{align}
\mc M_1 = {} & 
\frac{1}{16}
(-3,\,1,\,3,\,-1,\,2,\,-3,\,1,\,3,\,-1,\,2)\;, \\
\mc M_2 = {} & 
\frac{1}{8\chi s_{12}^2}
(-m_1^2+(1-2\chi)s_{12},\,
3m_1^2-(3+2\chi)s_{12},\,
m_1^2-(1-2\chi)s_{12},\, \nn \\[-2mm] & \qquad\quad\!
-3m_1^2+(3+2\chi)s_{12},\,
2(-m_1^2+(1+2\chi)s_{12}),\,
-m_1^2+(1-2\chi)s_{12},\, \nn \\[1mm] & \qquad\quad
3m_1^2-(3+2\chi)s_{12},\,
m_1^2-(1-2\chi)s_{12},\, 
-3m_1^2+(3+\chi)s_{12},\, \nn \\[1mm] & \qquad\;\;\;
2(-m_1^2+(1+2\chi)s_{12}))\;.
\end{align}
The master integral coefficients can be written compactly in terms of tree-level
data as
\begin{align}
c_i = \oint_{\mc M_i}dz\,\widetilde{J}(z)
\sum_{\substack{\text{helicities}\\\text{particles}}}
\prod_{k=1}^6 A_{(k)}^\tree(z)\;,
\end{align}
where the rescaled Jacobian for this configuration is defined by
\begin{align}
\widetilde{J}(z)\equiv
\pm\frac{m_1^2-s_{12}}{\chi s_{12}}
\frac{1}{z(z+1)(z+(s_{12}(1+\chi)-m_1^2)/(\chi s_{12})}\;.
\end{align}

The computation of the remaining one-mass configurations is essentially
equivalent to the one described here and the projectors are similar. The lack of
symmetry in the two-loop crossed box suggests that we also derive projectors for
the one-mass diagram with $m_4\neq 0$ (or $m_3\neq 0$). The hepta-cuts in the
limit $m_1,m_2,m_3\to 0$ follow from
eqs.~\eqref{DEGENERATECUTRESIDUES1}-\eqref{DEGENERATECUTRESIDUES4},
\begin{gather}
\label{DEGENERATECUTM41}
\X[1]_{\tS_{1,2}} \longrightarrow
+\frac{1}{\chi}
\oint\frac{dz}{z(z+1)(z+(s_{12}(1+\chi)-m_4^2)/(\chi s_{12}))} \;, \\[2mm]
\label{DEGENERATECUTM42}
\X[1]_{\tS_{3,4}} \longrightarrow
-\frac{1}{\chi}
\oint\frac{dz}{z(z+1)(z+(s_{12}(1+\chi)-m_4^2)/(\chi s_{12}))} \;,
\end{gather}
whereas the singular point locus and parity-odd vanishing constraints carry
over directly from the calculation above. The residues in the masters are
\begin{align}
&\textstyle\Res_{\{\tG_i\}}\X[1] = \nn \\[1mm] &\qquad
\frac{s_{12}}{(m_4^2-s_{12})(m_4^2-(1+\chi)s_{12})}
\big(m_4^2-(1+\chi)s_{12},\,\chi s_{12},\,
(1+\chi)s_{12}-m_4^2,\,
\nn \\[1mm] &\qquad\qquad\;
-\chi s_{12},\, 
s_{12}-m_4^2,\,
m_4^2-(1+\chi)s_{12},\,
\chi s_{12},\,
(1+\chi)s_{12}-m_4^2,\,
-\chi s_{12},\,s_{12}-m_4^2\big)\;, \\[-1mm]
&\textstyle\Res_{\{\tG_i\}}\X[(\ell_1+k_4)^2/2] = 
\dfrac{\chi s_{12}^2}{2(m_4^2-(1+\chi)s_{12})}
(0,\,1,\,0,\,-1,\,-1,\,0,\,1,\,0,\,-1,\,-1)\;,
\end{align}
and the unique projectors read
\begin{align}
\label{M4PROJECTORS}
\mc M_1 = {} &
\frac{m_4^2-s_{12}}{16s_{12}}
(3,\,-1,\,-3,\,1,\,-2,\,3,\,-1,\,-3,\,1,\,-2)\;, \\[1mm]
\mc M_2 = {} & 
\frac{1}{8\chi s_{12}^2}
((1-2\chi)s_{12}-m_4^2,\,
3m_4^2-(3+2\chi)s_{12},\,
m_4^2-(1-2\chi)s_{12},\,
\nn \\[-1mm] & \qquad\quad\,
(3+2\chi)s_{12}-3m_4^2,\,
2((1+2\chi)s_{12}-m_4^2),\, 
(1-2\chi)s_{12}-m_4^2,\, \nn \\[1mm] & \qquad\quad\,
3m_4^2-(3+2\chi)s_{12},\,
m_4^2-(1-2\chi)s_{12},\,
(3+2\chi)s_{12}-3m_4^2,\, \nn \\[1mm] & \qquad\quad\,
2((1+2\chi)s_{12}-m_4^2))\;.
\end{align}
These master contours respect the three linearly independent IBP constraints
\begin{align}
\tW\cdot
(0,\,1,\,0,\,1,\,0,\,0,\,1,\,0,\,1,\,0) = 0 \;, \\
\tW\cdot
(1,\,-1,\,1,\,-1,\,0,\,1,\,-1,\,1,\,-1,\,0) = 0\;, \\
\tW\cdot
(1,\,1,\,-1,\,-1,\,2,\,1,\,1,\,-1,\,-1,\,2) = 0 \;.
\end{align}

The expressions for the one-mass projectors derived here are consistent with the
purely massless calculation reported in ref.~\cite{Sogaard:2013yga}.

\subsection{Two-Mass Projectors}
As previously explained, there are four kinematically inequivalent distributions
of massless and massive external legs in the two-mass four-point crossed box.
Indeed, we distinguish between the two-mass short-side diagrams with either both
massive legs situated in the planar or nonplanar end and the long-side and
diagonal diagrams. From the point of view of the global structure of the
hepta-cut, the latter three are similar and can be treated within the regime of
the octuply pinched genus-3 curve whereas the first diagram is a variant of the
three- and four-mass case.

The long-side two-mass diagram can be studied by taking over the singular point
locus \eqref{ONEMASSRESIDUELOCUS}, basis integral decomposition as well as the
parity-odd contour constraints \eqref{PARITYCONSTRAINTSDEG} it turns out. We
assume that $m_1m_4\neq 0$ and $m_2 = m_3 = 0$. Under these circumstances, the
relevant hepta-cuts are 
\begin{gather}
\X[1]_{\tS_{1,2}} \longrightarrow
+\frac{s_{12}-m_1^2}{\chi s_{12}}
\oint\frac{dz}{z(z+1)(z+\lambda)} \;, \\[2mm]
\X[1]_{\tS_{3,4}} \longrightarrow
-\frac{s_{12}-m_1^2}{\chi s_{12}}
\oint\frac{dz}{z(z+1)(z+\lambda)} \;,
\end{gather}
where the pole location $\lambda$ is defined by
\begin{align}
\lambda \equiv
\frac{m_1^2(m_4^2-s_{12})+s_{12}(s_{12}(1+\chi)-m_4^2)}{\chi s_{12}^2}\;.
\end{align}
These hepta-cuts clearly coincide with
eqs.~\eqref{DEGENERATECUTM11}-\eqref{DEGENERATECUTM12} and
\eqref{DEGENERATECUTM41}-\eqref{DEGENERATECUTM42} in the respective limits,
$m_4\to0$ and $m_1\to0$.

The residues associated with the two master integrals are as follows,
\begin{align}
{\textstyle\Res}_{\{\tG_i\}}\X[1] = {} &
N_1(r_1,\,r_2,\,-r_1,\,-r_2,\,r_3,\,r_1,\,r_2,\,-r_1,\,-r_2,\,r_3)\;, \\[2mm]
{\textstyle\Res}_{\{\tG_i\}}\X[(\ell_1+k_4)^2/2] = {} &
N_2(0,\,-1,\,0,\,1,\,1,\,0,\,-1,\,0,\,1,\,1)\;,
\end{align}
where $N_1$ and $N_2$ are given by 
\begin{align}
N_1\equiv {} &
\frac{s_{12}}{(m_4^2-s_{12})((m_1^2+m_4^2)
s_{12}-(1+\chi)s_{12}^2-m_1^2m_4^2)}\;, \\[1mm]
N_2\equiv {} &
\frac{\chi s_{12}^3}{2(m_1^2-s_{12})(m_4^2-s_{12})+
2\chi s_{12}^2}\;,
\end{align}
and the $r_i$s are defined as
\begin{align}
r_1\equiv s_{12}(m_4^2-(1+\chi)s_{12})-m_1^2(m_4^2-s_{12})\;, \quad
r_2\equiv \chi s_{12}^2\;, \quad
r_3\equiv (m_1^2-s_{12})(m_4^2-s_{12})\;.
\end{align}
The projectors for the coefficients $c_1$ and $c_2$ become
\begin{align}
\mc M_1 = {} & 
\frac{m_4^2-s_{12}}{16s_{12}}
(3,\,-1,\,-3,\,1,\,-2,\,3,\,-1,\,-3,\,1,\,-2)\;, \\
\mc M_2 = {} &
\frac{1}{8\chi s_{12}^3}
(q_1,\,q_2,\,-q_1,\,-q_2,\,q_3,\,q_1,\,q_2,\,-q_1,\,-q_2,\,q_3)\;,
\end{align}
for constants $q_1$, $q_2$ and $q_3$ where
\begin{align}
q_1\equiv {} & +m_1^2(m_4^2-s_{12})-s_{12}(m_4^2-(1-2\chi)s_{12})\;, \\
q_2\equiv {} & -2\chi s_{12}^2-3(m_1^2-s_{12})(m_4^2-s_{12})\;, \\
q_3\equiv {} & +2(2\chi s_{12}^2+(m_1^2-s_{12})(m_4^2-s_{12}))\;.
\end{align}
For this external kinematics, the three independent consistency relations from
IBP identities are
\begin{align}
\tW\cdot
(0,\,1,\,0,\,1,\,0,\,0,\,1,\,0,\,1,\,0) = 0 \;, \\
\tW\cdot
(1,\,-1,\,1,\,-1,\,0,\,1,\,-1,\,1,\,-1,\,0) = 0\;, \\
\tW\cdot
(1,\,1,\,-1,\,-1,\,2,\,1,\,1,\,-1,\,-1,\,2) = 0 \;.
\end{align}

We have also derived projectors for the two-mass diagonal configuration. The
computation essentially resembles that of the two-mass long-side diagram,
meaning that the same singular point locus, integral basis and contour
constraints can be used. For $m_2 = m_4 = 0$ and $m_1m_3\neq 0$, the hepta-cuts
evaluated at the branches $\mc S_1,\dots,\mc S_4$ are
\begin{gather}
\X[1]_{\tS_{1,2}} \longrightarrow
+\frac{(m_1^2-s_{12})(m_3^2-s_{12})}{\chi s_{12}^2-m_1^2m_3^2}
\oint\frac{dz}{z(z+1)(z+\lambda)} \;, \\[2mm]
\X[1]_{\tS_{3,4}} \longrightarrow
-\frac{(m_1^2-s_{12})(m_3^2-s_{12})}{\chi s_{12}^2-m_1^2m_3^2}
\oint\frac{dz}{z(z+1)(z+\lambda)} \;,
\end{gather}
where the third pole $\lambda$ is now defined by
\begin{align}
\lambda \equiv
\frac{s_{12}(m_1^2+m_3^2-(1+\chi)s_{12})}{m_1^2m_3^2-\chi s_{12}^2}\;.
\end{align}
It can be shown that the master contours for this kind of integrals are
characterized by the following numbers,
\begin{align}
\mc M_1 = {} &
\frac{1}{16}
(3,\,-1,\,-3,\,1,\,-2,\,3,\,-1,\,-3,\,1,\,-2)\;, \\
\mc M_2 = {} & 
\frac{1}{8(m_3^2-s_{12})(m_1^2m_3^2-\chi s_{12}^2)}
(q_1,\,q_2,\,-q_1,\,-q_2,\,q_3,\,q_1,\,q_2,\,-q_1,\,-q_2,\,q_3)\;,
\end{align}
and the three independent weights $q_1$, $q_2$ and $q_3$ in $\mc M_2$ are
\begin{align}
q_1\equiv {} &
+m_1^2(3m_3^2-s_{12})-s_{12}(m_3^2-(1-2\chi)s_{12})\;, \nn \\
q_2\equiv {} &
-m_1^2(m_3^2-3s_{12})+s_{12}(3m_3^2-(3+2\chi)s_{12})\;, \nn \\
q_3\equiv {} &
-2(m_1^2(m_3^2+s_{12})+s_{12}(m_3^2-(1+2\chi)s_{12}))\;.
\end{align}

Our next example is the short-side two-mass diagram with $m_1m_2\neq 0$ and
$m_3 = m_4 = 0$ which is a smooth limit of the three- and four-mass case.
Accordingly, we now have the master equation
\begin{align}
\label{2MSHORTBASIS}
\mc A^{(2)}_4 = 
c_1\X[1]+c_2\X[(\ell_1+k_4)^2/2]+c_3\X[((\ell_1+k_4)^2/2)^2]+\cdots\;,
\end{align}
and we must choose a new singular point locus for this computation. It is
natural to encircle the following ten global poles,
\begin{align}
\label{2MSHORTLOCUS}
\{\mc G_i\} = (
\mc G_{1\cap4},\,
\mc G_{1\cap6},\,
\mc G_{1\cap3},\,
\mc G_{1\cap5},\,
\mc G_{1,\infty_{LR}},\,
\mc G_{2\cap3},\,
\mc G_{2\cap5},\,
\mc G_{2\cap4},\,
\mc G_{2\cap6},\,
\mc G_{2,\infty_{LR}}
) \equiv (\mc G_1,\dots,\mc G_{10})\;.
\end{align}
We remind that $(\mc G_1,\dots,\mc G_5)$ and $(\mc G_5,\dots,\mc G_{10})$, are
located at the following values of $z$ on the Riemann sphere parametrized by
$\mc S_1$ and $\mc S_2$ respectively,
\begin{align}
\left\{-1,-\frac{1}{\tau},
-\frac{\bxi_2}{\bxi_1},-\frac{\tau\bxi_2}{\bxi_1},0\right\}\;.
\end{align}
Evidently, on-shell branches three through six are eliminated and we retain only
the following two hepta-cut integrals,
\begin{align}
\X[1]_{\mc S_{1,2}} \longrightarrow
+N\oint dz\frac{z}{(z+1)(z+1/\tau)(z+\bxi_2/\bxi_1)(z+\tau\bxi_2/\bxi_1)} \;,
\end{align}
where the overall constant is given by
\begin{align}
N\equiv {} & \frac{(\gamma_{12}^2-m_1^2m_2^2)^3}
{\gamma_{12}^3(\gamma_{12}+m_2^2)(m_1^2m_2^2+\chi\gamma_{12}s_{12})}\;,
\end{align}
and the poles are located at
\begin{align}
1/\tau = 
\frac{\gamma_{12}(\gamma_{12}+\chi s_{12})}{m_1^2m_2^2+\chi\gamma_{12}s_{12}}
\;, \quad
\frac{\bxi_2}{\bxi_1} = 
\frac{m_1^2m_2^2(\gamma_{12}+\chi s_{12})}
{\gamma_{12}(m_1^2m_2^2+\chi\gamma_{12}\chi_{12})}\;, \quad
\frac{\tau\bxi_2}{\bxi_1} = {} & \frac{m_1^2m_2^2}{\gamma_{12}^2}\;.
\end{align}

Requiring that all parity-odd numerator insertions have vanishing hepta-cuts
translates into the statement
\begin{align}
\label{2MSHORTSPURIOUS}
\left(
\begin{array}{lllllrrrrr}
1\; & 0\; & 0\; & 0\; & 0 & -1 & 0 & 0 & 0 & 0 \\
0 & 1 & 0 & 0 & 0 & 0 & -1 & 0 & 0 & 0 \\
0 & 0 & 1 & 0 & 0 & 0 & 0 & -1 & 0 & 0 \\
0 & 0 & 0 & 1 & 0 & 0 & 0 & 0 & -1 & 0 \\
0 & 0 & 0 & 0 & 1 & 0 & 0 & 0 & 0 & -1 \\
\end{array}
\right)
\left(
\begin{array}{c}
\omega_{1\cap4} \\
\omega_{1\cap6} \\
\omega_{1\cap3} \\
\omega_{1\cap5} \\
\omega_{1,\infty_{LR}} \\
\omega_{2\cap3} \\
\omega_{2\cap5} \\
\omega_{2\cap4} \\
\omega_{2\cap6} \\
\omega_{2,\infty_{LR}}
\end{array}
\right) = 0\;.
\end{align}
There are two linearly independent constraints from IBP relations,
\begin{align}
\label{2MSHORTIBP1}
\W\cdot
(0,\,1,\,0,\,1,\,-1,\,0,\,1,\,0,\,1,\,-1) = 0 \;, \\
\label{2MSHORTIBP2}
\W\cdot
(1,\,-1,\,1,\,-1,\,0,\,1,\,-1,\,1,\,-1,\,0) = 0 \;.
\end{align}
Resolving the contour constraints leaves three contour weights undetermined,
exactly balancing the number of master integrals for this problem. The master
contours which pick up contribution from one basis integral and annihilate the
other two are,
\begin{align}
\mc M_1\cdot\big(\Res_{\{\mc G_i\}}\X[1],\,
\Res_{\{\mc G_i\}}\X[(\ell_1+k_4)^2/2],\,
\Res_{\{\mc G_i\}}\X[((\ell_1+k_4)^2/2)^2]
\big) = (1,0,0)\;, \nn \\
\mc M_2\cdot\big(\Res_{\{\mc G_i\}}\X[1],\,
\Res_{\{\mc G_i\}}\X[(\ell_1+k_4)^2/2],\,
\Res_{\{\mc G_i\}}\X[((\ell_1+k_4)^2/2)^2]
\big) = (0,1,0)\;, \nn \\
\label{2MSHORT}
\mc M_3\cdot\big(\Res_{\{\mc G_i\}}\X[1],\,
\Res_{\{\mc G_i\}}\X[(\ell_1+k_4)^2/2],\,
\Res_{\{\mc G_i\}}\X[((\ell_1+k_4)^2/2)^2]
\big) = (0,0,1)\;.
\end{align}

After changing the remaining one-dimensional contour into a linear combination
of small circles around the global poles, the three master integrals reduce to
the following residues,
\begin{align}
\textstyle\Res_{\{\mc G_i\}}\X[1] = {} &
N_1(r_1,-r_2,-r_1,r_2,0,r_1,-r_2,-r_1,r_2,0)\;, \\[1mm]
\textstyle\Res_{\{\mc G_i\}}\X[(\ell_1+k_4)^2/2] = {} &
N_2(0,\,-1,\,0,\,1,\,0,\,0,\,-1,\,0,\,1,\,0)\;, \\[1mm]
\textstyle\Res_{\{\mc G_i\}}\X[((\ell_1+k_4)^2/2)^2] = {} &
N_3(0,\,r_3,\,0,\,-r_3,\,r_1,\,0,\,r_3,\,0,\,-r_3,\,r_1)\;,
\end{align}
where
\begin{align}
N_1\equiv {} &
-\frac{m_1^2m_2^2+\chi\gamma_{12}s_{12}}{
\chi s_{12}(\gamma_{12}+m_2^2)(
\gamma_{12}^2+m_1^2m_2^2+\chi\gamma_{12}s_{12})}\;, \\[1mm]
N_2\equiv {} &
+\frac{(\gamma_{12}+m_1^2)(m_1^2m_2^2+\chi\gamma_{12}s_{12})}
{2(\gamma_{12}^2+m_1^2m_2^2+\chi \gamma_{12}s_{12})}\;, \\[1mm]
N_3\equiv {} &
+\frac{(\gamma_{12}+m_1^2)^2(\gamma_{12}+m_2^2)
(m_1^2m_2^2+\chi\gamma_{12}s_{12})}{
4\gamma_{12}(\gamma_{12}^2-m_1^2m_2^2)(
\gamma_{12}^2+m_1^2m_2^2+\chi\gamma_{12}s_{12})}\;,
\end{align}
and
\begin{align}
r_1\equiv 
\gamma_{12}^2+m_1^2m_2^2+\chi\gamma_{12}s_{12}\;, \quad
r_2\equiv 
\chi\gamma_{12}s_{12}\;, \quad
r_3\equiv 
\gamma_{12}^2-m_1^2m_2^2\;.
\end{align}
The projectors for the integral coefficients take the form
\begin{align}
\mc M_1 = {} & 
\frac{(\gamma_{12}+m_2^2)\chi s_{12}}
{4(m_1^2m_2^2+\gamma_{12}\chi s_{12})}
(-1,\,0,\,1,\,0,\,0,\,-1,\,0,\,1,\,0,\,0) \;, \\[1mm]
\mc M_2 = {} & 
\frac{1}{2(\gamma_{12}+m_1^2)(m_1^2m_2^2+\gamma_{12}\chi s_{12})}
(q_1,\,q_2,\,q_3,\,q_4,\,q_5,\,q_1,\,q_2,\,q_3,\,q_4,\,q_5)\;, \\[2mm]
\mc M_3 = {} & 
\frac{\gamma_{12}(\gamma_{12}^2-m_1^2m_2^2)}
{(\gamma_{12}+m_1^2)^2(\gamma_{12}+m_2^2)(m_1^2m_2^2+\gamma_{12}\chi s_{12})}
(1,\,1,\,1,\,1,\,2,\,1,\,1,\,1,\,1,\,2)\;.
\end{align}
where $q_1,\dots,q_5$ are defined by
\begin{align}
q_1\equiv {} & \gamma_{12}^2-m_1^2m_2^2-\gamma_{12}\chi s_{12}\;, &
q_2\equiv {} & -2m_1^2m_2^2-\gamma_{12}\chi s_{12}\;, & & \nn \\[1mm]
q_3\equiv {} & \gamma_{12}^2-m_1^2m_2^2+\gamma_{12}\chi s_{12}\;, &
q_4\equiv {} & q_1-q_2+q_3\;, &
q_5\equiv {} & q_1+q_3\;.
\end{align}

We point out that although the projectors are functions of irrational quantities
such as $\gamma_{12}$ which has a square root, the final integral coefficients
obtained in this way are rational in external invariants.

\subsection{Three-Mass Projectors}
The three-mass case with $m_1 = 0$ or $m_2 = 0$ and the remaining three external
masses nonzero is similar to the two-mass long-side calculation previously
presented. Here we will instead focus on the three-mass diagram with $m_4 = 0$
which has the two-mass short-side diagram with $m_1m_2\neq 0$ as a smooth limit.
This means that we can continue to use the integral basis \eqref{2MSHORTBASIS},
the singular point locus \eqref{2MSHORTLOCUS} and the parity-odd constraints
\eqref{2MSHORTSPURIOUS}. Moreover, the contour constraints from IBP relations
are identical to those in eqs.~\eqref{2MSHORTIBP1}-\eqref{2MSHORTIBP2}.

For this problem, we will use the hepta-cuts for $\mc S_1$ and $\mc S_2$ from
eq.~\eqref{4MJACOBIANS} with $\mu = 0$ and without the overall factor of
$\gamma_\star$,
\begin{align}
\X[1]_{\mc S_{1,2}} \longrightarrow 
\frac{(1-\tau)(\bxi_1-\tau\bxi_2)}{
\tau\bxi_1^2}
\oint dz\frac{z}{
(z+1)(z+1/\tau)(z+\bxi_2/\bxi_1)(z+\tau\bxi_2/\bxi_1)
}\;.
\end{align}
It is possible to obtain rather clean forms of the residues computed by the
hepta-cut master integrals at the singular point locus, if we prefer quantities
constructed from flat momenta, $\bxi_1$, $\bxi_2$ and $\tau$, instead of the
usual Mandelstam variables and external masses. Indeed, the residues can be
expressed as
\begin{align}
\textstyle\Res_{\{\mc G_i\}}\X[1] = {} & 
\bigg(
-\frac{1}{\bxi_1-\bxi_2},\,
\frac{\tau}{\bxi_1-\tau^2\bxi_2},\,
\frac{1}{\bxi_1-\bxi_2},\,
-\frac{\tau}{\bxi_1-\tau^2\bxi_2},\,
0,\, \nn \\ & \;\;\;
-\frac{1}{\bxi_1-\bxi_2},\,
\frac{\tau}{\bxi_1-\tau^2\bxi_2},\,
\frac{1}{\bxi_1-\bxi_2},\,
-\frac{\tau}{\bxi_1-\tau^2\bxi_2},\,
0\bigg)\;,
\end{align}
\begin{align}
\textstyle\Res_{\{\mc G_i\}}\X[(\ell_1+k_4)^2/2] = {} & 
\bigg(
0,\,
-\frac{(1-\tau)(\bxi-\tau\bxi_2)k_1^\flat\cdot k_4^\flat}
{\bxi_1-\tau^2\bxi_2},\,
0,\,
\frac{(1-\tau)(\bxi-\tau\bxi_2)k_1^\flat\cdot k_4^\flat}
{\bxi_1-\tau^2\bxi_2},\,
0,\, \nn \\ & \;\;\;
0,\,
-\frac{(1-\tau)(\bxi-\tau\bxi_2)k_1^\flat\cdot k_4^\flat}
{\bxi_1-\tau^2\bxi_2},\,
0,\,
\frac{(1-\tau)(\bxi-\tau\bxi_2)k_1^\flat\cdot k_4^\flat}
{\bxi_1-\tau^2\bxi_2},\,
0
\bigg)\;,
\end{align}
\vspace*{-1cm}
\begin{align}
\textstyle\Res_{\{\mc G_i\}}\X[((\ell_1+k_4)^2/2)^2] = {} 
\hspace*{8.5cm} \nn \\[1mm]
\bigg(
0,\,
\frac{(1-\tau)^2(\bxi-\tau\bxi_2)^2(k_1^\flat\cdot k_4^\flat)^2}
{\tau(\bxi_1-\tau^2\bxi_2)},\,
0,\,
-\frac{(1-\tau)^2(\bxi-\tau\bxi_2)^2(k_1^\flat\cdot k_4^\flat)^2}
{\tau(\bxi_1-\tau^2\bxi_2)},\, \; \nn \\
\frac{(1-\tau)(\bxi-\tau\bxi_2)(k_1^\flat\cdot k_4^\flat)^2}{\tau}\,,
0,\,
\frac{(1-\tau)^2(\bxi-\tau\bxi_2)^2(k_1^\flat\cdot k_4^\flat)^2}
{\tau(\bxi_1-\tau^2\bxi_2)},\,
0,\, \;\; \nn \\
-\frac{(1-\tau)^2(\bxi-\tau\bxi_2)^2(k_1^\flat\cdot k_4^\flat)^2}
{\tau(\bxi_1-\tau^2\bxi_2)},\,
\frac{(1-\tau)(\bxi-\tau\bxi_2)(k_1^\flat\cdot k_4^\flat)^2}{\tau}
\bigg)\;.
\end{align}
The master contours that respect all integral reduction identities and extract
either of the three master integrals are
\begin{align}
\mc M_1 = {} & 
N_1(1,\,0,\,-1,\,0,\,0,\,1,\,0,\,-1,\,0,\,0)\;, \\
\mc M_2 = {} & 
N_2(q_1,\,q_2,\,q_3,\,q_4,\,q_5,\,q_1,\,q_2,\,q_3,\,q_4,\,q_5)\;, \\
\mc M_3 = {} &
N_3(1,\,1,\,1,\,1,\,2,\,1,\,1,\,1,\,1,\,2)\;.
\end{align}
In these equations, the overall constants $N_1$, $N_2$ and $N_3$ are
\begin{align}
N_1\equiv {} &
\frac{\gamma_{12}m_1^2m_3^2-\chi(\gamma_{12}+m_1^2)(\gamma_{12}+m_2^2)s_{12}}
{4m_1^2(m_2^2(\gamma_{12}+m_1^2)-\gamma_{12}m_3^2)+
4\chi\gamma_{12}s_{12}(\gamma_{12}+m_1^2)}\;, \\[2mm]
N_2\equiv {} & 
\frac{1}{2(\gamma_{12}^2+m_1^2m_2^2+\gamma_{12}(m_1^2+m_2^2-m_3^2))}
\nn \\ & \qquad\times\frac{1}{
m_1^2(m_2^2(\gamma_{12}+m_1^2)-\gamma_{12}m_3^2)
+\chi\gamma_{12}(\gamma_{12}+m_1^2)s_{12}}\;, \\[2mm]
N_3\equiv {} & 
\frac{\gamma_{12}(\gamma_{12}^2-m_1^2m_2^2)}
{\gamma_{12}^2+m_1^2m_2^2+\gamma_{12}(m_1^2+m_2^2-m_3^2)} 
\nn \\ & \qquad\times\frac{1}{
m_1^4m_2^2+\chi\gamma_{12}^2 s_{12}
+\gamma_{12}m_1^2(m_2^2-m_3^2+\chi s_{12})}\;,
\end{align}
along with the residue weights $q_1,\dots,q_5$,
\begin{align}
q_1\equiv {} & 
\gamma_{12}^4-m_1^4m_2^4+\gamma_{12}^3(m_1^2+m_2^2-m_3^2-\chi s_{12})
\nn \\ &
-\gamma_{12}m_1^2m_2^2(m_1^2+m_2^2-m_3^2+\chi s_{12})+
\gamma_{12}^2(m_1^2m_3^2-\chi(m_1^2+m_2^2)s_{12})
\;, \\[2mm]
q_2\equiv {} & 
m_1^2(-2(\gamma_{12}+m_1^2)m_2^2(\gamma_{12}+m_2^2)
\nn \\ &
+\gamma_{12}(\gamma_{12}+2m_2^2)m_3^2)-
\chi\gamma_{12}(\gamma_{12}+m_1^2)(\gamma_{12}+m_2^2)s_{12}\;, \\[2mm]
q_3\equiv {} &
\gamma_{12}^4-m_1^4m_2^4+\gamma_{12}^3(m_1^2+m_2^2-m_3^2+\chi s_{12})
\nn \\ &
-\gamma_{12}m_1^2m_2^2(m_1^2+m_2^2-m_3^2-\chi s_{12})-
\gamma_{12}^2(m_1^2m_3^2-\chi(m_1^2+m_2^2)s_{12})\;, \\[2mm]
q_4\equiv {} & q_1-q_2+q_3\;, \\[1mm]
q_5\equiv {} & q_1+q_3\;.
\end{align}

We note that the three-mass projectors written here reduce to the two-mass
short-side formula \eqref{2MSHORT} in the limit $m_3\to 0$.

\subsection{Four-Mass Projectors}
We finally examine the principal kinematic configuration with four distinct
external masses $m_i^2 = k_i^2\neq 0$. The intermediate calculations are more
complicated because neither $\gamma_{12}$ nor $\gamma_{14}$ can be simplified to
rational expressions. In order to simplify the computation to the maximum extent
possible prior to solving for the projectors, we will encircle a slightly
different set of global poles compared to the previous examples. More
specifically, we exploit that integrand of the scalar master integral, evaluated
on branches three through six, has vanishing residues at infinity. 

It is convenient to arrange the poles so that $\mc G_i$ and $\mc G_{i+5}$ are
still parity conjugates of each other. The set of global poles is
\begin{align}
\label{4MNEWLOCUS}
\{\mc G_i\} = (
\mc G_{1\cap4},\,
\mc G_{1\cap 5},\,
\mc G_{3,\infty_R},\,
\mc G_{5,\infty_R},\,
\mc G_{1,\infty_{LR}},\,
\mc G_{2\cap3},\,
\mc G_{2\cap6},\,
\mc G_{4,\infty_R},\,
\mc G_{6,\infty_R},\,
\mc G_{2,\infty_{LR}}
)\;.
\end{align}
The residues at these poles can be streamlined by rescaling all hepta-cut
Jacobians $J_1,\dots,J_6$ by a common factor,
\begin{align}
J_i \to (1+\mu\tau)(\bxi_1-\bxi_2)J_i\;,
\end{align}
and this is what we will do implicitly below. This constant will eventually drop
out when we compute integral coefficients.

Without repeating the exercise, we know that the contour constraints from
parity-odd numerator insertions are
\begin{align}
\left(
\begin{array}{lllllrrrrr}
1\;\; & 0\;\; & 0\;\; & 0\;\; & 0 & -1 & 0 & 0 & 0 & 0 \\
0 & 1 & 0 & 0 & 0 & 0 & -1 & 0 & 0 & 0 \\
0 & 0 & 1 & 0 & 0 & 0 & 0 & -1 & 0 & 0 \\
0 & 0 & 0 & 1 & 0 & 0 & 0 & 0 & -1 & 0 \\
0 & 0 & 0 & 0 & 1 & 0 & 0 & 0 & 0 & -1 \\
\end{array}
\right)
\left(
\begin{array}{c}
\omega_{1\cap4} \\
\omega_{1\cap 5} \\
\omega_{3,\infty_R} \\
\omega_{5,\infty_R} \\
\omega_{1,\infty_{LR}} \\
\omega_{2\cap3} \\
\omega_{2\cap6} \\
\omega_{4,\infty_R} \\
\omega_{6,\infty_R} \\
\omega_{2,\infty_{LR}}
\end{array}
\right) = 0\;.
\end{align}

Prior to presenting the residues computed by the three master integrals, it
proves advantageous to define the following four constants constructed out of
various previously defined quantities,
\begin{align}
r_1\equiv {} &
-\frac{\tau(\bxi_1-\bxi_2)}{\bxi_1-\tau^2\bxi_2}\;, \\
r_2\equiv {} &
-\frac{m_4^2}{2}-\mu(1-\tau)(\bxi_1-\tau\bxi_2)k_1^\flat\cdot k_4^\flat\;, \\
r_3\equiv {} &
+\frac{m_4^2}{2}+(1-1/\tau)(\bxi_1-\tau\bxi_2)k_1^\flat\cdot k_4^\flat\;,
\\[1mm]
r_4\equiv {} &
-(1-1/\tau)(1+\mu\tau)^2(\bxi_1-\bxi_2)(\bxi_1-\tau\bxi_2)
(k_1^\flat\cdot k_4^\flat)^2\;.
\end{align}
In terms of the $r_i$s, the residues of the master integrals can be brought to a
particularly simple form,
\begin{align}
\textstyle\Res_{\{\mc G_i\}}\X[1] = {} &
(-1,\,r_1,\,0,\,0,\,0,\,-1,\,r_1,\,0,\,0,\,0)\;, \\[1mm]
\textstyle\Res_{\{\mc G_i\}}\X[(\ell_1+k_4)^2/2] = {} & 
(r_2,\,r_1r_3,\,0,\,0,\,0,\,r_2,\,r_1r_3,\,0,\,0,\,0)\;,\, \\[1mm]
\textstyle\Res_{\{\mc G_i\}}\X[((\ell_1+k_4)^2/2)^2] = {} & 
(-r_2^2,\,r_1r_3^2,\,0,\,0,\,r_4,\,-r_2^2,\,r_1r_3^2,\,0,\,0,\,r_4)\;.
\label{4MSIMPLERESIDUES}
\end{align}

There are again two linearly independent consistency equations arising from the
IBP identities,
\begin{align}
\W\cdot(1,0,2,0,-1,1,0,2,0,-1) = 0\;,
\label{4MCONSTRAINTSINF1} \\
\W\cdot(0,1,0,2,-1,0,1,0,2,-1) = 0\;.
\label{4MCONSTRAINTSINF2}
\end{align}
At first sight these constraints differ from those found in the two- and
three-mass calculations, see e.g.  eqs.~\eqref{2MSHORTIBP1}-\eqref{2MSHORTIBP2}.
However, the two pairs of equations enforce the same constraints, as can be
argued easily. We may express the constraints without imposing the global
residue theorem. For the four-mass case we then have
\begin{align}
&\omega_{1\cap3}+\omega_{1\cap4}+\omega_{2\cap3}+\omega_{2\cap4} \nn \\
&\qquad\quad-\omega_{1,\infty_{LR}}-\omega_{2,\infty_{LR}}
-\omega_{1,\infty}-\omega_{2,\infty}
+2\omega_{3,\infty}+2\omega_{4,\infty} = 0\;,
\label{OVERCOMPLETE1} \\[1mm]
&\omega_{1\cap5}+\omega_{1\cap6}+\omega_{2\cap5}+\omega_{2\cap6} \nn \\
&\qquad\quad-\omega_{1,\infty_{LR}}-\omega_{2,\infty_{LR}}
-\omega_{1,\infty}-\omega_{2,\infty}
+2\omega_{5,\infty}+2\omega_{6,\infty} = 0\;.
\label{OVERCOMPLETE2}
\end{align}
These equations encompass either form of the IBP constraints, i.e.
eqs.~\eqref{2MSHORTIBP1}-\eqref{2MSHORTIBP2} and
\eqref{4MCONSTRAINTSINF1}-\eqref{4MCONSTRAINTSINF2}. This can be shown by
writing out the latter explicitly, i.e.
\begin{align}
\omega_{1\cap3}+\omega_{1\cap4}+\omega_{2\cap3}+\omega_{2\cap4}
-\omega_{1,\infty_{LR}}-\omega_{2,\infty_{LR}} = 0\;, \\[1mm]
\omega_{1\cap5}+\omega_{1\cap6}+\omega_{2\cap5}+\omega_{2\cap6}
-\omega_{1,\infty_{LR}}-\omega_{2,\infty_{LR}} = 0\;,
\end{align}
and
\begin{align}
\omega_{1\cap4}+\omega_{2\cap3}
-\omega_{1,\infty_{LR}}-\omega_{2,\infty_{LR}}
-\omega_{1,\infty}-\omega_{2,\infty}
+2\omega_{3,\infty}+2\omega_{4,\infty} = 0\;, \\
\omega_{1\cap5}+\omega_{2\cap6}
-\omega_{1,\infty_{LR}}-\omega_{2,\infty_{LR}}
-\omega_{1,\infty}-\omega_{2,\infty}
+2\omega_{5,\infty}+2\omega_{6,\infty} = 0\;.
\end{align}
Then it is immediately clear that we obtain the same answer from
eqs.~\eqref{OVERCOMPLETE1}-\eqref{OVERCOMPLETE2} when we truncate to linearly
independent sets of residues by the global residue theorem.

Returning to actual calculation with the representation \eqref{4MSIMPLERESIDUES}
of the residues at hand, it is quite easy to solve the set of linear equations
to derive compact expressions for the three master integral projectors,
\begin{align}
\mc M_1 = {} & -\frac{1}{4r_1(r_2+r_3)r_4}\big(
2r_1r_3r_4,\,-2r_2r_4,\,r_1r_3(r_2^2+r_2r_3-r_4),\, \nn \\
& \qquad\qquad\qquad\qquad
r_2(r_1r_3(r_2+r_3)+r_4),\,2r_1r_2r_3(r_2+r_3),\, \nn \\[2mm]
& \qquad\qquad\qquad\qquad
2r_1r_3r_4,\,-2r_2r_4,\,r_1r_3(r_2^2+r_2r_3-r_4),\, \nn \\[2mm]
& \qquad\qquad\qquad\qquad
r_2(r_1r_3(r_2+r_3)+r_4),\,2r_1r_2r_3(r_2+r_3)\big)\;, \\[3mm]
\mc M_2 = {} & +\frac{1}{4r_1(r_2+r_3)r_4}\big(
2r_1r_4,\,2r_4,\,r_1(r_2^2-r_3^2-r_4),\,
r_1(r_2^2-r_3^2)-r_4,\,2r_1(r_2^2-r_3^2),\, \nn \\
& \qquad\qquad\qquad\qquad\;
2r_1r_4,\,2r_4,\,r_1(r_2^2-r_3^2-r_4),\,
r_1(r_2^2-r_3^2)-r_4,\,2r_1(r_2^2-r_3^2)\big)\;, \\
\mc M_3 = {} & \frac{1}{4r_4}
(0,\,0,\,1,\,1,\,2,\,0,\,0,\,1,\,1,\,2)\;.
\end{align}
This result completes our derivation of master integral projectors for the
two-loop crossed box with up to four massive external legs.

\section{Reduction of Integrals with Doubled Propagators}
\label{DOUBLEDSEC}
Feynman integrals with doubled and in general higher powers of propagators
frequently appear in loop amplitude computations, for instance in IBP
identities, Schwinger parametrizations or bubble insertions. It was recently
explained that generalized unitarity cuts of such integrals are naturally
treated as degenerate multivariate residues using computational algebraic
geometry \cite{Sogaard:2014ila}. In that connection, several examples were
given for one- and two-loop integrals with massless kinematics. This method
extends seamlessly to multiloop integrals with external masses, as we will
demonstrate shortly. However, the calculation can be accelerated by using the
\emph{Bezoutian matrix algorithm}.

\subsection{Unitarity Cut Algorithm: Bezoutian Matrix Method}
We very briefly review the unitarity cut algorithm for integrals with higher
powers of propagators. For more details and examples, please refer to
ref.~\cite{Sogaard:2014ila,Sogaard:2013fpa}. The main ingredient needed is
computational algebraic geometry.

Recall that a residue is {\it nondegenerate}, if the Jacobian at the pole $\xi$
is nonzero, i.e.,
\begin{equation} 
J(\xi)=\det_{i,j}\bigg(\pd{f_i}{z_j}\bigg)\bigg|_{z=\xi}\neq 0\;.
\end{equation}
In this case, the value of the residue is simply calculated by Cauchy's theorem
in higher dimensions, i.e. eq.~\eqref{LOCALIZATIONPROPERTY}. However, the
Jacobian clearly vanishes if there is one or more doubled propagators being cut
and the residue is {\it degenerate}, so this approach does not apply. To solve
the problem, we need techniques from algebraic geometry. There are two ways of
evaluating such residues:

\begin{enumerate}
\item The transformation law (see for instance ref.~\cite{MR0463157}). This
theorem can be used to convert a degenerate residue at the simultaneous zero of
the inverse propagators to a factorizable residue. The explicit transformation
matrix is found by the Gr\"obner bases method. The algorithm is described in
refs.~\cite{Sogaard:2013fpa,Sogaard:2014ila}. 
\item The Bezoutian matrix method. Here one determines the {\it duality
structure} \cite{MR507725} of the multivariate residues, which in turn can be
calculated easily. In general, the Bezoutian matrix method is considerably
faster than the transformation law for complicated cuts with many independent
external invariants.
\end{enumerate}
Our Mathematica package \texttt{MathematicaM2}\footnote{The package can be
downloaded from \tt https://bitbucket.org/yzhphy/mathematicam2.} is capable of
computing multivariate residues using either of these techniques. In what
follows, we outline the Bezoutian matrix approach and provide some basic
examples.
 
Let $I=\langle f_1,\dots,f_n \rangle$ be an ideal in the ring
$R=\C[z_1,\dots,z_n]$. Assume that $I$ is a zero-dimensional ideal, i.e. the
zero locus $\mc Z(I)=\{\xi_1,\dots,\xi_k\}$ consists of finite number of
discrete points. For a zero-dimensional ideal $I$, the quotient ring $R/I$ is a
finite dimensional $\mathbb C$-linear space. 

Before we calculate individual residues, we first examine the structure of the
sum of residues by Bezoutian Matrix. Then we eventually get individual residues
from {\it partition functions}. For a polynomial $h$ in $R$, we define the {\it
global residue} as
\begin{equation}
  \label{Global Residue}
  \Res(h)=\sum_{i}\Res{}_{\xi_i}\left(
  \frac{h dz_1 \wedge\cdots\wedge dz_n}{f_1\cdots f_n}
  \right)
\end{equation}
which is just the sum of all residues. By Stokes' theorem, the values of the
residues only depend on $h$'s equivalence class $[h]$ in $R/I$. Furthermore, we
can define an inner product $\langle\;,\,\rangle$ in $R/I$,
\begin{equation}
  \label{eq:2}
  \langle g,h \rangle \equiv \Res(g\cdot h)\;.
\end{equation}

\begin{thm}
  $\langle\;,\,\rangle$ is a nondegenerate inner product in $R/I$. 
\end{thm}
The proof of the theorem is given in ref.~\cite{MR0463157}. This theorem implies
that, given a linear basis $\{p_i\}$ for $R/I$, we can find its dual basis
$\{\Delta_i\}$ in $R/I$, such that
\begin{equation}
  \label{eq:3}
  \langle p_i,\Delta_j\rangle = \delta_{ij}\;.
\end{equation}
In practice, the basis and dual basis can be found by the Gr\"obner basis method
and the Bezoutian matrix \cite{MR2161985}. The procedure involves the following
steps:

\begin{enumerate}
\item Calculate $G$, the Gr\"obner basis of $I$ in the {\it DegreeLexicographic}
order. Denote the {\it leading terms} for all polynomials in $G$ as $LT(G)$.
Then all monomials in $R$ which are lower than $LT(G)$ constitute $\{p_i\}$,
which is the canonical linear basis for $R/I$.  \item Introduce a set of
auxiliary variables $\{y_1,\dots,y_n\}$ and define the Bezoutian matrix $B$ for
$I$ as,
\begin{equation}
  \label{eq:4}
  B_{ij}\equiv
  \frac{f_i(y_1,\dots,y_{j-1},z_j,\dots,z_n)-
  f_i(y_1,\dots,y_j,z_{j+1},\dots,z_n)}{z_j-y_j}\;.
  \end{equation}
Calculate its determinant, $\det B$. 
\item Define $\tilde G$ as the set $G$ after the replacement $z_i \to y_i$.
Perform the polynomial division of the Bezoutian determinant, $\det B$, over
$G\otimes \tilde G$. Then the remainder can be written,
\begin{equation}
  \label{eq:5}
  \sum_i a_i(y) p_i(z)\;,
\end{equation}
where the $p_i$s form the canonical linear basis for $R/I$, and the $a_i(y)$s
are polynomials in the $y$-variables only.

\item The dual basis $\{\Delta_i\}$, with respect to the inner product
$\langle\;,\,\rangle$, is defined as $\Delta_i=a_i(z)$.
\end{enumerate}

The dual basis explicitly characterizes the structure of global residues. Let
the decomposition of the unit $1$ over the dual basis be given as
\begin{equation}
  \label{eq:6}
  1=\sum_i \mu_i \Delta_i\;.
\end{equation}
Then for an arbitrary numerator $h$, expand $[h]$ over the canonical linear
basis,
\begin{equation}
  \label{eq:7}
  [h]=\sum_i \lambda_i p_i\;,
\end{equation}
and the global residue is given as \cite{MR2161985},
\begin{equation}
  \label{Global_Residue_Evaluation}
  \Res(h)=\sum_i \lambda_i \mu_i\;.
\end{equation}
This formula is the result of the definition of the dual basis, and provides a
very efficient way of calculating the residues.

To get individual residues, we can use the formula
\eqref{Global_Residue_Evaluation} and the new ingredient {\it partition
functions} of $\mathcal Z(I)$.

\begin{thm}
  Let $I$ be a zero-dimensional ideal and $\mc Z(I)=\{\xi_1,\ldots,\xi_k\}$ be
  its zero locus. Denote $\mc O_i$ as the local ring of $\xi$, 
  $\mc O_i=\{f/g\;|\;g(\xi_i)\not=0, f\in R,g\in R\}$, and $I\mc O_i$ as
  the ideal generated by $I$ in $O_i$. Then there is a set of {\it partition
  functions}, $e_1,\ldots,e_k$, each of which is an element in $R/I$, such that,
  \begin{enumerate}
    \item In $R/I$, $\sum_i e_i=1$, $e_i^2=e_i$ and $e_i e_j=0$ if $i\not=j$.
    \item $e_i \in I \mc O_j$ if $i\not =j$, and $e_i-1 \in I\mc O_i$. 
  \end{enumerate}
\end{thm}

This theorem can be proved by construction \cite{MR2122859}. Then for each
individual residue at $\xi_i$ \cite{MR2161985}, we have the result
\begin{equation}
  \label{Residue_Evaluation}
  \Res{}_{\xi_i}
  \left(\frac{h dz_1 \wedge\cdots\wedge dz_n}{f_1 \cdots f_n}\right) =
  \Res(h\cdot e_i)\;.
\end{equation}

Explicitly, the partition functions $e_1,\dots,e_k$ can be constructed by the
method of Lagrange interpolation. The computation via Bezoutian matrix method is
realized in our package, \texttt{MathematicaM2}. We demonstrate this computation
by a simple example before we return to generalized unitarity cuts of integrals
with doubled propagators. To compare with the transform law method described in
\cite{Sogaard:2013yga,Sogaard:2013fpa}, we present some one-loop and two-loop
residues computations in Example 2, 3.

\begin{example}
Let $I=\langle (z_1+z_2)^2, z_2^2+z_1^2-2\rangle$. There are two residues
located at $\xi_1=(-1,1)$ and $\xi_2=(1,-1)$. Both residues are degenerate.
From the Gr\"obner basis computation, the linear basis $\{p_i\}$ for $R/I$ is,
\begin{equation}
  \label{eq:8}
  \{z_2^2,z_1,z_2,1\}\;.
\end{equation}

The Bezoutian matrix is,
\begin{equation}
  B = \label{eq:1}
  \left(
  \begin{array}{cc}
    y_1+z_1+2 z_2 & 2 y_1+y_2+z_2 \\
    y_1+z_1 & y_2+z_2 \\
   \end{array}
   \right)\;.
\end{equation}
Its determinant is $\det B = -2 y_1 z_1+2 y_2 z_2-2 y_1^2+2 z_2^2$. So after the
polynomial division over the Gr\"obner basis of $I$, we have the dual basis, 
\begin{equation}
  \{\Delta_i\} = \{2,-2 z_1,2 z_2,2 \left(z_2^2-2\right)\}\;.
\end{equation}

Consider the numerator $h=z_2^2$. From the dual basis structure, we immediately
get decomposition,
\begin{equation}
  \label{eq:12}
  1=\frac{1}{2} \Delta_1\;.
\end{equation}
Hence $\{\mu_i\}=\{\frac{1}{2},0,0,0\}$. Now it is clear that,
\begin{equation}
  \label{eq:9}
  \Res(h)=\frac{1}{2}\;.
\end{equation}
Furthermore, we construct partition functions for this ideal according to
ref.~\cite{MR2122859}. By Lagrangian interpolation, we obtain two polynomials
\begin{equation}
 l_1=-\frac{1}{2}(z_1-1)\;, \quad l_2=\frac{1}{2}(z_1+1)\;,
\end{equation}
such that $l_i(\xi_j)=\delta_{ij}$, $i,j=1,2$. Since the two poles both have
multiplicity $2$, the partition functions are \cite{MR2122859},
\begin{eqnarray}
  \label{eq:10}
  e_1&=&1-(1-l_1^2)^2=\frac{1}{4}(2-2 z_1+z_2)  \mod I\;, \\ 
  e_2&=&1-(1-l_2^2)^2=\frac{1}{4} (2+2 z_1-z_2) \mod I\;.
\end{eqnarray}
Here, to simplify the expression, we performed a polynomial division over $I$.
Then the local residues are,
\begin{eqnarray}
  \label{eq:11}
  \Res{}_{\xi_1}\left(\frac{h dz_1 
  \wedge\cdots\wedge dz_n}{f_1 \cdots f_n}\right)&=&
  \Res(h\cdot e_1)=\frac{1}{4}\;, \\
  \Res{}_{\xi_2}\left(\frac{h dz_1 
  \wedge\cdots\wedge dz_n}{f_1 \cdots f_n}\right)&=&
  \Res(h\cdot e_2)=\frac{1}{4}\;.
\end{eqnarray}
\end{example}

\begin{example}
We calculate the two multivariate residues from the maximal cut of the one-loop
massless box diagram and then compare the efficiency of Bezoutian method and the
transformation law method. The Feynman integral of the box diagram is,
\begin{equation}
  \label{eq:13}
  I_\square(\sigma_1,\dots,\sigma_4)\equiv
\int_{\R^D}\frac{d^D\ell}{(2\pi)^D}
\prod_{k=1}^4\frac{1}{f_k^{\sigma_k}(\ell)}\;,
\end{equation}
where the denominators are,
\begin{equation}
  \label{eq:14}
  f_1=\ell^2\;,\quad
  f_2=(\ell-k_1)^2\;,\quad
  f_3=(\ell-k_1-k_2)^2\;,\quad
  f_4=(\ell+p_4)^2\;.
\end{equation}
We parametrize the loop momentum $\ell$ as,
\begin{equation}
  \label{eq:15}
  \ell^\mu=\alpha_1 k_1^\mu + \alpha_2 k_2^\mu +
  \frac{\alpha_3 s_{12}}{2\spaa{14}\spbb{42}}\spvec{1}{\gamma^\mu}{2}+
  \frac{\alpha_4 s_{12}}{2\spaa{24}\spbb{42}}\spvec{2}{\gamma^\mu}{1}\;.
\end{equation}
The box integrand has two quadruple-cut poles,
\begin{equation}
  \label{eq:16}
  (\alpha_1,\alpha_2,\alpha_3,\alpha_4)=(1,0,0,-\chi)\equiv\xi_1\;,\quad
  (\alpha_1,\alpha_2,\alpha_3,\alpha_4)=(1,0,-\chi,0)\equiv\xi_2\;.
\end{equation}
Consider the residue of the triple propagator integral $I_\square(3,1,1,1)$. The
dual basis $\{\Delta_i\}$ from the Bezoutian matrix computation is
\begin{equation}
  \{\Delta_i\}=\frac{s_{12}^6}{\chi^3 (1+\chi)}
  \left\{-1,\alpha_3,-\alpha_4-3\chi,\alpha_3(\alpha_3+3\chi),
  -\alpha_4^2-3\alpha_4\chi-3\chi^2,-(\alpha_4+\chi)^3\right\}\;,
\end{equation}
and the partition functions are,
\begin{equation}
  \label{eq:18}
  \{e_1, e_2\}=\frac{1}{\chi^3}\left\{-
  \alpha_4\left(\alpha_4^2+3\alpha_4\chi+3\chi^2\right),
  (\alpha_4+\chi)^3\right\}\;.
\end{equation}
Then we have
\begin{eqnarray}
  \Res{}_{\xi_1}\left(\frac{d\alpha_1
  \wedge\cdots\wedge d\alpha_4}{f_1^3 f_2 f_3 f_4}\right)&=&
  \Res(e_1)=+\frac{1+\chi}{s_{12}^6}\;, \\
  \Res{}_{\xi_2}\left(\frac{d\alpha_1 
  \wedge\cdots\wedge d\alpha_4}{f_1^3 f_2 f_3 f_4}\right)&=&
  \Res( e_2)=-\frac{1+\chi}{s_{12}^6}\;.
\end{eqnarray}
This is a very simple example. Using the package \texttt{MathematicaM2}, the
whole computation takes $0.54$ seconds, via Bezoutian method. The computation
based on transformation law \cite{Sogaard:2014ila}, gives the same result, but
takes $1.07$ seconds. There is no significant efficiency difference between the
two methods, for this one-loop example.
\end{example}

\begin{example}
We calculate one of the multivariate residues from the maximal cut of the
two-loop massless double-box diagram and again compare the efficiency of
Bezoutian method and the transformation law method. The Feynman integral of the
double-box diagram is,
\begin{equation}
  P_{2,2}^{**}(\sigma_1,\dots,\sigma_7)\equiv
  \int_{\R^D}\frac{d^D\ell_1}{(2\pi)^D} \int_{\R^D}\frac{d^D\ell_2}{(2\pi)^D}
  \prod_{k=1}^7\frac{1}{f_k^{\sigma_k}(\ell)}\;,
\end{equation}
where the denominators are,
\begin{align}
f_1 = {} & \ell_1^2\;, &\;\;
f_2 = {} & (\ell_1-k_1)^2\;, &\;\;
f_3 = {} & (\ell_1-k_1-k_2)^2\;, \nn \\
f_4 = {} & \ell_2^2\;, &\;\;
f_5 = {} & (\ell_2-k_4)^2\;, &\;\; 
f_6 = {} & (\ell_2-k_3-k_4)^2\;, &\;\;
f_7 = {} & (\ell_1+\ell_2)^2\;. &
\end{align}
The loop momenta are parametrized as,
\begin{align}
  \ell_1^\mu={}&\alpha_1k_1^\mu+\alpha_2k_2^\mu+
    \frac{\alpha_3s_{12}}{\spaa{14}\spbb{42}}\spvec{1}{\gamma^\mu}{2}+
    \frac{\alpha_4s_{12}}{\spaa{24}\spbb{41}}\spvec{2}{\gamma^\mu}{1}\;, \nn \\
 \ell_2^\mu={}&\beta_1k_3^\mu+\beta_2k_4^\mu+\beta_3+
    \frac{\beta_3s_{12}}{\spaa{31}\spbb{14}}\spvec{3}{\gamma^\mu}{4}+
    \frac{\beta_4s_{12}}{\spaa{41}\spbb{13}}\spvec{4}{\gamma^\mu}{3}\;.
\end{align}
Consider the residue at 
\begin{equation}
  (a_1,a_2,a_3,a_4,b_1,b_2,b_3,b_4)=(1,0,-\chi,0,0,1,z,0)\equiv \xi.
\end{equation}
By the Bezoutian matrix method, we find that
\begin{equation}
  \Res{}_{\xi}\left(\frac{d\alpha_1 
  \wedge\cdots\wedge d\alpha_4\wedge d\beta_1 
  \wedge\cdots\wedge d\beta_4}
  {f_1f_2f_3f_4f_5f_6f_7^3}\right)=\frac{\chi^2(1+\chi)^2}{z(z+\chi)^3}\;.
\end{equation}
The computation takes $1.28$ seconds with a numerical value of $\chi$. The
transformation law methods takes $10.7$ seconds with a numerical value of
$\chi$. So for this two-loop example, the Bezoutian method is about $8$ times
faster.
\end{example}

\subsection{Example: One-Mass Two-Loop Crossed Box}
The previous examples show that the Bezoutian matrix algorithm is significantly
faster than the transformation law method for massless two-loop problems. The
difference is even more profound for nonplanar diagrams with external masses, as
we shall see shortly.

We will be slightly more general than in eq.~\eqref{INVPROP} and define the
two-loop crossed box integral with arbitrary integer powers
$(\sigma_1,\dots,\sigma_9)$ of propagators and irreducible numerators as
\begin{align}
\X(\sigma_1,\dots,\sigma_9)\equiv
\int_{\R^D}\frac{d^D\ell_1}{(2\pi)^D}
\int_{\R^D}\frac{d^D\ell_2}{(2\pi)^D}
\prod_{k=1}^9\frac{1}{f_k^{\sigma_k}(\ell_1,\ell_2)}\;,
\end{align}
where the seven propagators $f_1,\dots,f_7$ can be found in eq.~\eqref{XBOXINT}
with $k_5 = k_6 = 0$ and 
\begin{align}
f_8\equiv \frac{1}{2}(\ell_1+k_4)^2\;, \quad 
f_9\equiv \frac{1}{2}(\ell_2+k_1)^2\;.
\end{align}
In order to unambiguously define the degenerate multivariate residue associated
with the maximal cut, the inverse propagators are grouped into seven factors
$g_i\equiv f_i^{\sigma_i}$. As in \cite{Sogaard:2014ila} we will for technical
simplicity only consider cuts in strictly four dimensions, postponing the
analysis in $D = 4-2\epsilon$ dimensions to future work.
\begin{figure}[!h]
\bc
\includegraphics[scale=0.75]{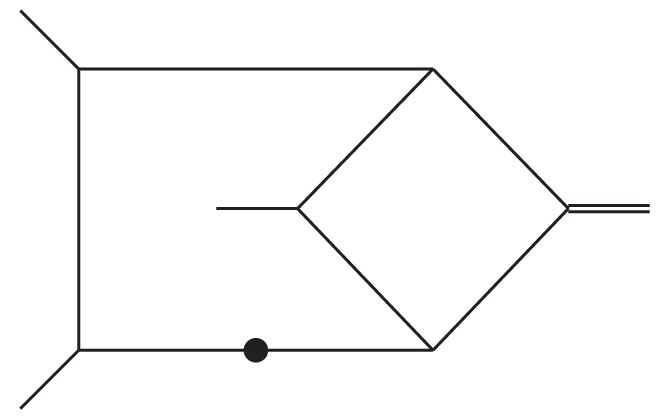}
\put(-250,-2){$k_1$}
\put(-250,149){$k_2$}
\put(-180,74){$k_3$}
\put(0,74){$k_4$}
\caption{The four-point two-loop crossed-box diagram with one external massive
leg and a doubled propagator, which is depicted by a black dot. The massive
momentum $k_4$ is marked by a doubled line.}
\label{1MDOUBLED}
\ec
\end{figure}
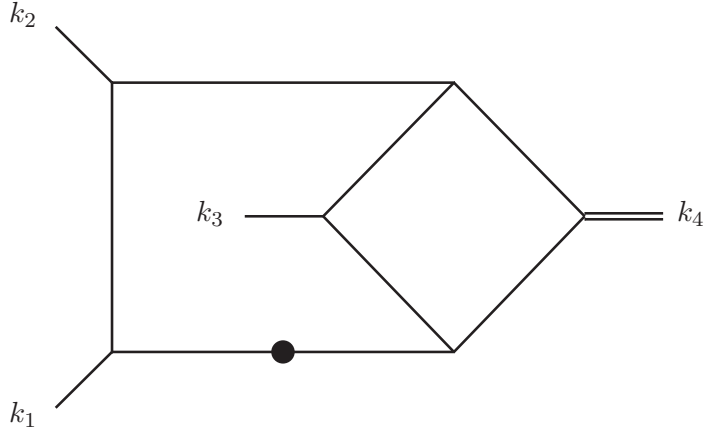

For simplicity we assume for now that $m_1m_2 = 0$ so that a general integral
can be expanded onto a basis of two masters,
\begin{align}
\X(\sigma_1,\dots,\sigma_9) = 
c_1\X(1,\dots,1,0,0)+c_2\X(1,\dots,1,-1,0)+\cdots\;.
\end{align}
As an example, we will examine the one-mass doubled propagator integral with
indices $(2,1,\dots,1,0,0)$ (see fig.~\ref{1MDOUBLED}) and reconstruct the
coefficients $c_1$ and $c_2$ to leading order in dimensional regularization
using the projectors derived in the previous section. Let $m_1 = m_2 = m_3 = 0$
and $m_4\neq 0$. The degenerate multivariate residues at the simultaneous zeros
of the inverse propagators specified by $\mc S_1',\dots,\mc S_8'$ are computed
using our package, {\tt MathematicaM2},
\begin{gather}
\X(2,1,\dots,1,0,0)_{\tS_{1,2}} = 
+\frac{\gamma_\star}{\chi^2 s_{12}^2}
\oint dz\frac{s_{12}(2(1+\chi)+(1+2\chi)z)-m_4^2}
{z(z+1)^2(z+(s_{12}(1+\chi)-m_4^2)/(\chi s_{12}))^2} \;, \nn \\[2mm]
\X(2,1,\dots,1,0,0)_{\tS_{3,4}} = 
-\frac{\gamma_\star}{\chi^2 s_{12}^2}
\oint dz\frac{s_{12}(2(1+\chi)+(1+2\chi)z)-m_4^2}
{z(z+1)^2(z+(s_{12}(1+\chi)-m_4^2)/(\chi s_{12}))^2} \;, \nn \\[2mm]
\X(2,1,\dots,1,0,0)_{\tS_{5,6}} = 
-\frac{\gamma_\star}{s_{12}}
\oint dz\frac{2+z}{(z+1)^2(z+m_4^2/s_{12})}\;, \nn \\[2mm]
\X(2,1,\dots,1,0,0)_{\tS_{7,8}} = 
-\frac{\gamma_\star}{m_4^2}
\oint dz\frac{h(z)}{(z+\mu)^3(z-1/\tau)^2}\;.
\label{XBOX1MDEGENERATERESIDUES}
\end{gather}
In these equations, $h$ is a cubic polynomial,
\begin{align}
h(z) = 
\mu\tau z^3-\mu(2-(3+\tau)\mu\tau)z^2-\mu^2(4+\tau(1-\mu\tau))z-\mu^3\tau\;.
\end{align}
The variables $\mu$ and $\tau$ as functions of $\chi$, $s_{12}$ and $m_4$ are
\begin{align}
\mu = \frac{m_4^2(m_4^2-(1+\chi)s_{12})}{\chi s_{12}^2}\;, \quad
\tau = \frac{\chi s_{12}}{(1+\chi)s_{12}-m_4^2}\;.
\end{align}

Actually, we do not explicitly need the doubled propagator hepta-cuts for 
$\tS_5,\dots,\tS_8$ as the Jacobian poles are located on the intersections
between the on-shell branches and no further poles are generated
\cite{Sogaard:2014ila}. Given these hepta-cut integrals, it is easy to
reproduce the residue relations.

It is convenient to strip off the overall factor 
$\gamma_\star = (16\chi s_{12}^3)^{-1}$ from all hepta-cuts as we will do now.
The residues computed by the doubled propagator scalar integral at the singular
point locus \eqref{ONEMASSRESIDUELOCUS} thus read 
\begin{align}
{\textstyle\Res_{\{\mc G_i\}}}\X(2,1,\dots,1,0,0)
\hspace*{10cm} \nn \\[1mm] = 
\frac{1}{(m_4^2-s_{12})^2(m_4^2-(1+\chi)s_{12})^2}
(r_1,\,r_2,\,-r_1,\,-r_2,\,r_3,\,r_1,\,r_2,\,-r_1,\,-r_2,\,r_3)\;,\,
\end{align}
for $r_1$, $r_2$ and $r_3$ given by
\begin{align}
r_1\equiv {} &
(m_4^2-2s_{12})(m_4^2-(1+\chi)s_{12})^2\;, \nn \\[1mm]
r_2\equiv {} &
\chi s_{12}^2(2(1+\chi)s_{12}-(2+\chi)m_4^2)\;, \nn \\[1mm]
r_3\equiv {} &
(m_4^2-s_{12})^2(2(1+\chi)s_{12}-m_4^2)\;.
\end{align}
This information allows us to derive the desired coefficients by applying the
relevant projectors \eqref{M4PROJECTORS}. The result is as follows,
\begin{align}
\X(2,1,\dots,1,0,0) = {} &
+\frac{m_4^2-2s_{12}}{(m_4^2-s_{12})s_{12}}
\X(1,\dots,1,0,0) \nn \\ &
-\frac{2m_4^2}{(m_4^2-s_{12})(m_4^2-(1+\chi)s_{12})s_{12}}
\X(1,\dots,1,-1,0)+\cdots
\end{align}
which is consistent with the $D = 4$ limit of the following IBP relation in 
$D = 4-2\epsilon$ dimensions generated by {\tt FIRE} \cite{Smirnov:2013dia},
\begin{align}
&\X(2,1,\dots,1,0,0) = \nn \\[1mm]
&\quad +\frac{
(1+2\epsilon)(2(1+\chi)(1+\epsilon)s_{12}^2+
m_4^2(m_4^2(1+\epsilon)-(\chi+3(1+\epsilon))s_{12})}
{(1+\epsilon)s_{12}(m_4^2-s_{12})(m_4^2-(1+\chi)s_{12}))}
\X(1,\dots,1,0,0) \nn \\[1mm]
&\qquad\quad-\frac{2(1+2\epsilon)(1+4\epsilon)m_4^2}
{(1+\epsilon)(m_4^2-s_{12})(m_4^2-(1+\chi)s_{12})s_{12}}
\X(1,\dots,1,-1,0)+\cdots\;.
\end{align}

It only takes a few seconds to obtain each of the degenerate multivariate
residues \eqref{XBOX1MDEGENERATERESIDUES} in {\tt Mathematica} using the
Bezoutian matrix algorithm and numerical values for the external invariants.
Remarkably, the Bezoutian matrix algorithm is at least $20-25$ times faster
compared to the transformation law method for this problem and in purely
analytic mode, the difference is even more significant.

\section{Discussion and Conclusion}
In the present paper we have extended the four-dimensional maximal unitarity
method \cite{Kosower:2011ty,CaronHuot:2012ab} to two-loop integrals with crossed
box topology with one through four massive external legs. In practice, these
integrals cover most of the nonplanar basis at four points \cite{Caola:2014lpa}.
In generalized unitarity, amplitudes are expanded onto a basis of master
integrals with rational coefficients which are then extracted systematically by
taking cuts that promote multiple internal lines simultaneously to on-shell
regions of momentum space. As this in general involves complex kinematics, cuts
are realized by replacing real slice integrations by multidimensional contours
that encircle the global poles of the loop integrand. These contours are subject
to the consistency requirement that the unitarity procedure respects the
reduction onto master integrals which relies on vanishing of parity-odd
integrands and total derivatives upon integration \cite{Kosower:2011ty}.

Our principal result is unique analytic contours for all basis integral
coefficients in all inequivalent configurations of massive and massless external
momenta in the four-point two-loop crossed box, valid to $\mc O(\epsilon^0)$ in
the dimensional regulator. The content of this paper is also relevant for
higher-multiplicity scattering of massless particles. The maximal cut defines a
nodal algebraic curve associated with a hextuply or octuply pinched genus-3
Riemann surface whose components are Riemann spheres.  The first category
includes contributions where both legs in the planar end of the diagram are
massive, whereas the second covers the rest. The number of linearly independent
residues is always ten as expected. We find that for the sixfold degenerate
curve, the projectors for all three master integrals are unique once we impose
five linearly independent Levi-Civita constraints and two linearly independent
IBP conditions. The Levi-Civita constraints are resolved for weights that
respect parity. In the four-mass case, unlike the situation for the four-mass
double box \cite{Johansson:2013sda}, the IBP constraints are not satisfied
automatically for the two-loop crossed box.
\begin{table}[!h]
\bc
\begin{tabular}{|c|c|c|c|c|c|c|c|c|}
\hline
$m_1$ & $m_2$ & $m_3$ & $m_4$ & $|\mc S|$ & Res & Odd & Even & MIs \\
\hline
$\neq0$ & $0$ & $0$ & $0$ & $8$ & 10 & 5 & 3 & $(0,0),\;(1,0)$ \\
\hline
$0$ & $0$ & $0$ & $\neq 0$ & $8$ & 10 & 5 & 3 & $(0,0),\;(1,0)$ \\
\hline
$\neq 0$ & $0$ & $\neq 0$ & $0$ & $8$ & 10 & 5 & 3 & $(0,0),\;(1,0)$ \\
\hline
$\neq 0$ & $0$ & $0$ & $\neq 0$ & $8$ & 10 & 5 & 3 & $(0,0),\;(1,0)$ \\
\hline
$\neq 0$ & $0$ & $\neq 0$ & $\neq 0$ & $8$ & 10 & 5 & 3 & $(0,0),\;(1,0)$ \\
\hline
$\neq 0$ & $\neq 0$ & $0$ & $0$ & $6$ & 10 & 5 & 2 & $(0,0),\;(1,0),\;(2,0)$ \\
\hline
$\neq 0$ & $\neq 0$ & $\neq 0$ & $0$ & $6$ & 10 & 5 & 2 & $(0,0),\;(1,0),\;(2,0)$ \\
\hline
$\neq 0$ & $\neq 0$ & $\neq 0$ & $\neq 0$ & $6$ & 10 & 5 & 2 & $(0,0),\;(1,0),\;(2,0)$ \\
\hline
\end{tabular}
\ec
\caption{Classification of all kinematically distinct diagrams from the
viewpoint of the maximal cut. The columns list whether the external masses
$m_1$, $m_2$, $m_3$ and $m_4$ are zero or not, the number $|\mc S|$ of hepta-cut
solutions, the number of independent residues, the number of parity-odd and
parity-even contour constraints and finally the set of master integrals. The
notation for the master integrals refers to the powers of the two irreducible
numerator insertions.}
\label{TABLEINT}
\end{table}

Overall, the results exhibit a very interesting and naively unexpected
simplicity, which clearly deserves more attention. Indeed, the systematics of
the contour constraints (e.g. eqs.~\eqref{2MSHORTIBP1}-\eqref{2MSHORTIBP2}) are
remarkable. Instead of being a set of disconnected calculations, the
inequivalent kinematic configurations related through a rich underlying
structure that seems to be governed by the global picture of the hepta-cut.  As
summarized in table~\ref{TABLEINT}, we find that the contour constraints are
identical for all configurations within a particular class of hepta-cut
solutions, for example all the way from the three-mass diagram with $m_1m_2 = 0$
to the purely massless case. The IBP contours seem to be even more systematic.
In all cases with six hepta-cut solutions we find two linearly independent IBP
constraints. The chiral branching from $6\to 8$ hepta-cut solutions triggers the
emergence of an additional IBP constraint. Interestingly, as for the planar
double box \cite{Johansson:2013sda}, the IBP constraints are inherited through
chiral branchings. To see this, let us instead consider hepta-cuts from the
eightfold degenerate genus-3 curve, with the set of global poles,
\begin{align}
(
\tG_{1\cap6},\,
\tG_{1\cap8},\,
\tG_{3\cap5},\,
\tG_{3\cap7},\,
\tG_{3,\infty_R},\,
\tG_{2\cap5},\,
\tG_{2\cap7},\,
\tG_{4\cap6},\,
\tG_{4\cap8},\,
\tG_{4,\infty_R}
)\;.
\end{align}
As pointed out in appendix~\ref{POLESTRUCTURE}, these poles exactly correspond
to poles on the hextuply pinched genus-3 curve. It is now an easy task to check
that two of the three IBP constraints are inherited. The same observation
applies to the four-mass computation. 

In view of the complexity of the hepta-cut expressions and the typical amount of
effort required to generate IBP relations for high-rank integrals with many
external relations, it is also striking that the constraints coefficients are
simply integers. This also applies to the planar double box with up to four
external masses \cite{Kosower:2011ty,Johansson:2012zv,Johansson:2013sda} and the
planar triple box \cite{Sogaard:2013yga}. This is a clear hint of a general
principle that may be explained by algebraic geometry.

The last part of this paper described a new algorithm based on the Bezoutian
matrix and Gr\"obner bases to compute degenerate multivariate residues which
typically appear in more complicated calculations. This algorithm was applied to
a few simple examples and to the reduction of a massive two-loop crossed box
integral with a doubled propagator onto master integrals with only single
propagators. Our tests have shown that the Bezoutian matrix algorithm is
considerably faster than the transformation law method \cite{Sogaard:2013fpa}
for two-loop problems. The multivariate residues from the massive two-loop
crossed box integral are computed at least $20-25$ times faster. Accordingly, we
expect the Bezoutian method to become increasingly valuable for multiscale
problems involving two-loop topologies with fewer propagators and at three loops
and beyond (see e.g. ref.~\cite{Sogaard:2013fpa}).

We end this paper by suggesting interesting projects for future research. It is
desirable to understand the nature of the contour constraints in complete
detail. In particular, is it possible to fully determine constraints arising
from integration-by-parts identities directly from the underlying algebraic
geometry? Recent progress for the planar double box shows that discrete
symmetries to some extent determine these constraints \cite{Johansson:2013sda}.
Such symmetries seem to be less constraining at higher genera. We also find it
urgent to extend maximal unitarity to $D$ dimensions to recover terms missed in
strictly $4D$. Another very important next step is to extend the method to basis
integrals with five external legs, for example the pentabox and turtle-box and
the related nonplanar diagrams.  First of all from a phenomenological point of
view, but we also hope that a generalization beyond four external particles will
offer insight in uniqueness of projectors \cite{CaronHuot:2012ab}. We expect the
one-mass hepta-cuts presented in this paper to be valuable in that direction,
because octa-cuts may be evaluated as hepta-cuts followed by a particular choice
of contour that puts the last propagator on-shell. Theoretically speaking, the
master integral coefficients for all-massive six-point planar and nonplanar
double boxes are exciting to compute because in those cases the on-shell
parametrization is irrational and hence the maximally cut integrals suffer from
genuine branch cuts \cite{CaronHuot:2012ab}. Ultimately, it would be intriguing
to implement the formalism numerically. We are looking forward to address some
of these questions soon.

\acknowledgments
We have benefited from many discussions with Emil Bjerrum-Bohr, Simon
Caron-Huot, Poul Henrik Damgaard, Rijun Huang, Henrik Johansson and David
Kosower. A special thanks goes to Hjalte Frellesvig and Kasper Larsen for
careful reading of the manuscript in draft stage and helpful comments. Both
authors are grateful to Institut de Physique Th\'eorique, CEA Saclay, and
especially David Kosower for hospitality during stages of this project. The work
of YZ is supported by Danish Council for Independent Research (FNU) grant
11-107241.

\clearpage
\appendix
\section{Explicit Parametrization of Hepta-Cut Solutions}
\label{EXPLICITSOLUTIONSAPP}
In this appendix we provide explicit forms of the loop momenta in all branches
for the two distinct classes of hepta-cut solutions considered in the present
paper. We also shed light on the pole structure.

Before we begin, recall that $\ell_1$ and $\ell_2$ are written in a basis of
mutually flattened momenta and parametrized by the complex variables $\zeta_i$,
$\xi_i$ and $\xi_i'$,
\begin{align}
\ell_1^\mu(\zeta_i,\xi_i,\xi_i') \,=\, {} & \zeta_1\eta_1^\mu+
\xi_1\xi_1'k_1^{\flat,\mu}+\xi_2\xi_2'
\frac{k_1^\flat\cdot k_4^\flat}{k_2^\flat\cdot k_4^\flat}k_2^{\flat,\mu}
\nn \\ {} &
+\frac{\xi_1\xi_2'}{2}\frac{\spbb{1^\flat 4^\flat}}{\spbb{2^\flat 4^\flat}}
\spvec{1^{\flat,-}}{\gamma^\mu}{2^{\flat,-}}+
\frac{\xi_2\xi_1'}{2}\frac{\spaa{1^\flat 4^\flat}}{\spaa{2^\flat 4^\flat}}
\spvec{2^{\flat,-}}{\gamma^\mu}{1^{\flat,-}}\;, \\[2mm]
\ell_2^\mu(\zeta_i,\xi_i,\xi_i') \,=\, {} & \zeta_2\eta_2^\mu+\xi_3\xi_3'
\frac{k_1^\flat\cdot k_4^\flat}{k_1^\flat\cdot k_3^\flat}k_3^{\flat,\mu}+
\xi_4\xi_4'k_4^{\flat,\mu}
\nn \\ {} &
+\frac{\xi_3\xi_4'}{2}\frac{\spaa{1^\flat 4^\flat}}{\spaa{1^\flat 3^\flat}}
\spvec{3^{\flat,-}}{\gamma^\mu}{4^{\flat,-}}+
\frac{\xi_4\xi_3'}{2}
\frac{\spbb{1^\flat 4^\flat}}{\spbb{1^\flat 3^\flat}}
\spvec{4^{\flat,-}}{\gamma^\mu}{3^{\flat,-}}\;.
\end{align}
After imposing the constraints of the maximal cut we find two distinct classes
of multiply degenerate kinematic solutions. In all cases, 
$\zeta_1 = \zeta_2 = 0$. For the sake of completeness we list the expansions
of all inverse propagators in parameter space with $\eta_1 = k_2^\flat$ and
$\eta_2 = k_4^\flat$ utilized in the computation of maximally cut integrals.
\begin{align}
\ell_1^2 = {} & \gamma_{12}\zeta_1\xi_1\xi_1'\;, \\
\ell_2^2 = {} & \gamma_{34}\frac{k_1^\flat\cdot k_4^\flat}
{k_1^\flat\cdot k_3^\flat}\zeta_2\xi_3\xi_3'\;, \\
(\ell_1-k_1)^2 = {} & m_1^2+\gamma_{12}\zeta_1\xi_1\xi_1'-
\gamma_{12}\zeta_1-m_1^2\xi_1\xi_1'-
\gamma_{12}\frac{k_1^\flat\cdot k_4^\flat}{k_2^\flat\cdot k_4^\flat}
\xi_2\xi_2'\;, \\
(\ell_1-k_{12})^2 = {} &
m_1^2+\gamma_{12}\zeta_1\xi_1\xi_1'-
\gamma_{12}\zeta_1-m_1^2\xi_1\xi_1'-
\gamma_{12}\frac{k_1^\flat\cdot k_4^\flat}{k_2^\flat\cdot k_4^\flat}
\xi_2\xi_2'\;, \\
(\ell_2-k_4)^2 = {} &
m_4^2+\gamma_{34}\frac{k_1^\flat\cdot k_4^\flat}
{k_1^\flat\cdot k_3^\flat}\zeta_2\xi_3\xi_3'-
m_4^2(\zeta_2+\xi_4\xi_4')-\gamma_{34}\frac{k_1^\flat\cdot k_4^\flat}
{k_1^\flat\cdot k_3^\flat}\xi_3\xi_3'\;, \\
(\ell_1+\ell_2)^2 = {} &
\gamma_{12}\zeta_1\xi_1\xi_1'+
\gamma_{34}\frac{k_1^\flat\cdot k_4^\flat}
{k_1^\flat\cdot k_3^\flat}\zeta_2\xi_3\xi_3' \\ & 
+2k_1^\flat\cdot k_4^\flat(\xi_1(\xi_3+\xi_4)+\xi_2(\tau\xi_3+\xi_4))\times
(\xi_1'(\xi_3'+\xi_4')+\xi_2'(\tau\xi_3'+\xi_4'))\;, \nn \\
(\ell_1+\ell_2+k_3)^2 = {} &
m_3^2+\gamma_{12}\zeta_1\xi_1\xi_1'+
\gamma_{34}\frac{k_1^\flat\cdot k_4^\flat}
{k_1^\flat\cdot k_3^\flat}\zeta_2\xi_3\xi_3' \nn \\ & 
+2k_1^\flat\cdot k_4^\flat(\xi_1(\xi_3+\xi_4)+\xi_2(\tau\xi_3+\xi_4))\times
(\xi_1'(\xi_3'+\xi_4')+\xi_2'(\tau\xi_3'+\xi_4')) \nn \\ & 
+2\bigg\{k_1^\flat\cdot k_3 \xi_1\xi_1'+
\frac{m_3^2k_1^\flat\cdot k_4^\flat}{2k_1^\flat\cdot k_3^\flat}\xi_3\xi_3'
+\frac{\gamma_{34}}{2}\xi_4\xi_4' \nn \\ & \qquad\qquad
+\bigg[\tau k_1^\flat\cdot k_3^\flat+
\frac{m_3^2 k_1^\flat\cdot k_4^\flat}{\gamma_{34}}
\bigg](\xi_1\xi_2'+\xi_1'\xi_2)
+\frac{k_1^\flat\cdot k_4^\flat k_2^\flat\cdot k_3}{k_2^\flat\cdot k_4^\flat}
\xi_2\xi_2'\bigg\}\;.
\end{align}

\subsection{The Four-Mass Case and Smooth Massless Limits}
The first class of solutions to the hepta-cut equations covers the four-mass
case but also applies in the limit where one or two external legs in the crossed
end of the diagram become massless. We simplify expressions by virtue of
introducing a new variable $w(z)$ defined by
\begin{align}
w(z) = \frac{1+z}{1+\tau z} \;\; \LLra \;\;
z(w) = -\frac{1-w}{1-\tau w}\;.
\end{align}
Then we have the following six inequivalent solutions,
\begin{align}
\ell_1^\mu|_{\mc S_1} \,=\, {} &
\bxi_1 k_1^{\flat,\mu}+\bxi_2
\frac{k_1^\flat\cdot k_4^\flat}{k_2^\flat\cdot k_4^\flat}k_2^{\flat,\mu}
+\frac{\bxi_2}{2z}\frac{\spbb{1^\flat 4^\flat}}{\spbb{2^\flat 4^\flat}}
\spvec{1^{\flat,-}}{\gamma^\mu}{2^{\flat,-}}+
\frac{\bxi_1z}{2}\frac{\spaa{1^\flat 4^\flat}}{\spaa{2^\flat 4^\flat}}
\spvec{2^{\flat,-}}{\gamma^\mu}{1^{\flat,-}}\;, \nn \\
\ell_2^\mu|_{\mc S_1} \,=\, {} &
-\frac{1+\tau z}{(1-\tau)(\bxi_1-\tau\bxi_2)z}\Big\{
\mu\tau w(z)(\bxi_1z+\bxi_2)
\frac{k_1^\flat\cdot k_4^\flat}{k_1^\flat\cdot k_3^\flat}k_3^{\flat,\mu}
-(\bxi_1z+\tau\bxi_2)k_4^{\flat,\mu}
\nn \\[1mm] {} &
-\frac{1}{2}\mu\tau w(z)(\bxi_1z+\tau\bxi_2)
\frac{\spaa{1^\flat 4^\flat}}{\spaa{1^\flat 3^\flat}}
\spvec{3^{\flat,-}}{\gamma^\mu}{4^{\flat,-}}+
\frac{1}{2}(\bxi_1z+\bxi_2)
\frac{\spbb{1^\flat 4^\flat}}{\spbb{1^\flat 3^\flat}}
\spvec{4^{\flat,-}}{\gamma^\mu}{3^{\flat,-}}\Big\}\;, \\[2mm]
\ell_1^\mu|_{\mc S_2} \,=\, {} &
\bxi_1 k_1^{\flat,\mu}+\bxi_2
\frac{k_1^\flat\cdot k_4^\flat}{k_2^\flat\cdot k_4^\flat}k_2^{\flat,\mu}
+\frac{\bxi_1z}{2}\frac{\spbb{1^\flat 4^\flat}}{\spbb{2^\flat 4^\flat}}
\spvec{1^{\flat,-}}{\gamma^\mu}{2^{\flat,-}}+
\frac{\bxi_2}{2z}\frac{\spaa{1^\flat 4^\flat}}{\spaa{2^\flat 4^\flat}}
\spvec{2^{\flat,-}}{\gamma^\mu}{1^{\flat,-}}\;, \nn \\
\ell_2^\mu|_{\mc S_2} \,=\, {} &
-\frac{1+\tau z}{(1-\tau)(\bxi_1-\tau\bxi_2)z}\Big\{
\mu\tau w(z)(\bxi_1z+\bxi_2)
\frac{k_1^\flat\cdot k_4^\flat}{k_1^\flat\cdot k_3^\flat}k_3^{\flat,\mu}
-(\bxi_1z+\tau\bxi_2)k_4^{\flat,\mu}
\nn \\[1mm] {} &
-\frac{1}{2}\mu\tau w(z)(\bxi_1z+\tau\bxi_2)
\frac{\spaa{1^\flat 4^\flat}}{\spaa{1^\flat 3^\flat}}
\spvec{3^{\flat,-}}{\gamma^\mu}{4^{\flat,-}}+
\frac{1}{2}(\bxi_1z+\bxi_2)
\frac{\spbb{1^\flat 4^\flat}}{\spbb{1^\flat 3^\flat}}
\spvec{4^{\flat,-}}{\gamma^\mu}{3^{\flat,-}}\Big\}\;, \\
\ell_1^\mu|_{\mc S_3} \,=\, {} &
\bxi_1 k_1^{\flat,\mu}+\bxi_2
\frac{k_1^\flat\cdot k_4^\flat}{k_2^\flat\cdot k_4^\flat}k_2^{\flat,\mu}
-\frac{\bxi_1}{2}\frac{\spbb{1^\flat 4^\flat}}{\spbb{2^\flat 4^\flat}}
\spvec{1^{\flat,-}}{\gamma^\mu}{2^{\flat,-}}
-\frac{\bxi_2}{2}\frac{\spaa{1^\flat 4^\flat}}{\spaa{2^\flat 4^\flat}}
\spvec{2^{\flat,-}}{\gamma^\mu}{1^{\flat,-}}\;, \nn \\ 
\ell_2^\mu|_{\mc S_3} \,=\, {} & 
k_4^{\flat,\mu}
+\frac{z}{2}\frac{\spaa{1^\flat 4^\flat}}{\spaa{1^\flat 3^\flat}}
\spvec{3^{\flat,-}}{\gamma^\mu}{4^{\flat,-}}\;, \\[2mm]
\ell_1^\mu|_{\mc S_4} \,=\, {} &
\bxi_1 k_1^{\flat,\mu}+\bxi_2
\frac{k_1^\flat\cdot k_4^\flat}{k_2^\flat\cdot k_4^\flat}k_2^{\flat,\mu}
-\frac{\bxi_2}{2}\frac{\spbb{1^\flat 4^\flat}}{\spbb{2^\flat 4^\flat}}
\spvec{1^{\flat,-}}{\gamma^\mu}{2^{\flat,-}}
-\frac{\bxi_1}{2}\frac{\spaa{1^\flat 4^\flat}}{\spaa{2^\flat 4^\flat}}
\spvec{2^{\flat,-}}{\gamma^\mu}{1^{\flat,-}}\;, \nn \\ 
\ell_2^\mu|_{\mc S_4} \,=\, {} & 
k_4^{\flat,\mu}
+\frac{z}{2}\frac{\spbb{1^\flat 4^\flat}}{\spbb{1^\flat 3^\flat}}
\spvec{4^{\flat,-}}{\gamma^\mu}{3^{\flat,-}}\;, \\[2mm]
\ell_1^\mu|_{\mc S_5} \,=\, {} &
\bxi_1 k_1^{\flat,\mu}+\bxi_2
\frac{k_1^\flat\cdot k_4^\flat}{k_2^\flat\cdot k_4^\flat}k_2^{\flat,\mu}
-\frac{\bxi_1}{2\tau}\frac{\spbb{1^\flat 4^\flat}}{\spbb{2^\flat 4^\flat}}
\spvec{1^{\flat,-}}{\gamma^\mu}{2^{\flat,-}}
-\frac{\bxi_2\tau}{2}\frac{\spaa{1^\flat 4^\flat}}{\spaa{2^\flat 4^\flat}}
\spvec{2^{\flat,-}}{\gamma^\mu}{1^{\flat,-}}\;, \nn \\ 
\ell_2^\mu|_{\mc S_5} \,=\, {} & 
\mu\frac{k_1^\flat\cdot k_4^\flat}{k_1^\flat\cdot k_3^\flat}k_3^{\flat,\mu}
+\frac{z}{2}\frac{\spbb{1^\flat 4^\flat}}{\spbb{1^\flat 3^\flat}}
\spvec{4^{\flat,-}}{\gamma^\mu}{3^{\flat,-}}\;, \\[2mm]
\ell_1^\mu|_{\mc S_6} \,=\, {} &
\bxi_1 k_1^{\flat,\mu}+\bxi_2
\frac{k_1^\flat\cdot k_4^\flat}{k_2^\flat\cdot k_4^\flat}k_2^{\flat,\mu}
-\frac{\tau\bxi_2}{2}\frac{\spbb{1^\flat 4^\flat}}{\spbb{2^\flat 4^\flat}}
\spvec{1^{\flat,-}}{\gamma^\mu}{2^{\flat,-}}
-\frac{\bxi_1}{2\tau}\frac{\spaa{1^\flat 4^\flat}}{\spaa{2^\flat 4^\flat}}
\spvec{2^{\flat,-}}{\gamma^\mu}{1^{\flat,-}}\;, \nn \\ 
\ell_2^\mu|_{\mc S_6} \,=\, {} & 
\mu\frac{k_1^\flat\cdot k_4^\flat}{k_1^\flat\cdot k_3^\flat}k_3^{\flat,\mu}
+\frac{z}{2}\frac{\spaa{1^\flat 4^\flat}}{\spaa{1^\flat 3^\flat}}
\spvec{3^{\flat,-}}{\gamma^\mu}{4^{\flat,-}}\;.
\end{align}
The displayed kinematic constants were defined in
eqs.~\eqref{TAUEXP}-\eqref{XI3XI3P} in the main text. It follows that solutions
$\mc S_1$ and $\mc S_2$ give rise to a singularity in both the left and right
loop momentum when $z\to 0$, whereas $\mc S_3$ through $\mc S_6$ have no poles
for finite values of $z$ and thus produce holomorphic integrands insertions.

\subsection{Degenerate Massless Limits}
As explained in the body of this paper, $\bxi_2 = 0$ for $m_1m_2 = 0$ and for
this class of kinematics the zero locus of the inverse propagators decomposes
into a union of eight irreducible components. Moreover, we have
$\bxi_1 = 1-m_1^2/s_{12}$ along with various other simplifications which we do
not show here in detail. Explicitly,
\begin{align}
\ell_1^\mu|_{\tS_1} \,=\, {} &
(1-m_1^2/s_{12})\Big\{k_1^{\flat,\mu}+
\frac{z}{2}\frac{\spaa{1^\flat 4^\flat}}{\spaa{2^\flat 4^\flat}}
\spvec{2^{\flat,-}}{\gamma^\mu}{1^{\flat,-}}\Big\}\;, \nn \\
\ell_2^\mu|_{\tS_1} \,=\, {} &
-\frac{1+\tau z}{1-\tau}\Big\{
\mu\tau w(z)
\frac{k_1^\flat\cdot k_4^\flat}{k_1^\flat\cdot k_3^\flat}k_3^{\flat,\mu}
-k_4^{\flat,\mu} \nn \\[1mm] {} & \hspace*{2cm}
+\frac{\mu\tau w(z)}{2}\frac{\spaa{1^\flat 4^\flat}}{\spaa{1^\flat 3^\flat}}
\spvec{3^{\flat,-}}{\gamma^\mu}{4^{\flat,-}}-
\frac{\spbb{1^\flat 4^\flat}}{\spbb{1^\flat 3^\flat}}
\spvec{4^{\flat,-}}{\gamma^\mu}{3^{\flat,-}}\Big\}\; \\[2mm]
\ell_1^\mu|_{\mc S_2} \,=\, {} &
(1-m_1^2/s_{12})\Big\{k_1^{\flat,\mu}
+\frac{z}{2}\frac{\spbb{1^\flat 4^\flat}}{\spbb{2^\flat 4^\flat}}
\spvec{1^{\flat,-}}{\gamma^\mu}{2^{\flat,-}}\Big\} \nn \\
\ell_2^\mu|_{\mc S_2} \,=\, {} &
-\frac{1+\tau z}{1-\tau}\Big\{
\mu\tau w(z)
\frac{k_1^\flat\cdot k_4^\flat}{k_1^\flat\cdot k_3^\flat}k_3^{\flat,\mu}
-k_4^{\flat,\mu}
\nn \\[1mm] {} & \hspace*{2cm}
+\frac{\spaa{1^\flat 4^\flat}}{\spaa{1^\flat 3^\flat}}
\spvec{3^{\flat,-}}{\gamma^\mu}{4^{\flat,-}}+
\frac{\mu\tau w(z)}{2}
\frac{\spbb{1^\flat 4^\flat}}{\spbb{1^\flat 3^\flat}}
\spvec{4^{\flat,-}}{\gamma^\mu}{3^{\flat,-}}\Big\}\;, \\
\ell_1^\mu|_{\tS_3} \,=\, {} &
(1-m_1^2/s_{12})\Big\{k_1^{\flat,\mu}+
\frac{z}{2}\frac{\spaa{1^\flat 4^\flat}}{\spaa{2^\flat 4^\flat}}
\spvec{2^{\flat,-}}{\gamma^\mu}{1^{\flat,-}}\Big\}\;, \nn \\
\ell_2^\mu|_{\tS_3} \,=\, {} &
-\frac{1+\tau z}{1-\tau}\Big\{
\mu\tau w(z)
\frac{k_1^\flat\cdot k_4^\flat}{k_1^\flat\cdot k_3^\flat}k_3^{\flat,\mu}
-k_4^{\flat,\mu} \nn \\[1mm] {} & \hspace*{2cm}
-\frac{\mu\tau}{2}\frac{\spaa{1^\flat 4^\flat}}{\spaa{1^\flat 3^\flat}}
\spvec{3^{\flat,-}}{\gamma^\mu}{4^{\flat,-}}+
\frac{w(z)}{2}\frac{\spbb{1^\flat 4^\flat}}{\spbb{1^\flat 3^\flat}}
\spvec{4^{\flat,-}}{\gamma^\mu}{3^{\flat,-}}\Big\}\;, \\[2mm]
\ell_1^\mu|_{\tS_4} \,=\, {} &
(1-m_1^2/s_{12})\Big\{k_1^{\flat,\mu}
+\frac{z}{2}\frac{\spaa{1^\flat 4^\flat}}{\spaa{2^\flat 4^\flat}}
\spvec{2^{\flat,-}}{\gamma^\mu}{1^{\flat,-}}\Big\} \nn \\
\ell_2^\mu|_{\tS_4} \,=\, {} &
-\frac{1+\tau z}{1-\tau}\Big\{
\mu\tau w(z)
\frac{k_1^\flat\cdot k_4^\flat}{k_1^\flat\cdot k_3^\flat}k_3^{\flat,\mu}
-k_4^{\flat,\mu}
\nn \\[1mm] {} & \hspace*{2cm}
+\frac{w(z)}{2}
\frac{\spaa{1^\flat 4^\flat}}{\spaa{1^\flat 3^\flat}}
\spvec{3^{\flat,-}}{\gamma^\mu}{4^{\flat,-}}
-\frac{\mu\tau}{2}
\frac{\spbb{1^\flat 4^\flat}}{\spbb{1^\flat 3^\flat}}
\spvec{4^{\flat,-}}{\gamma^\mu}{3^{\flat,-}}\Big\}\;, \\[2mm]
\ell_1^\mu|_{\tS_5} \,=\, {} &
(1-m_1^2/_{s12})\Big\{
k_1^{\flat,\mu}
-\frac{1}{2}\frac{\spbb{1^\flat 4^\flat}}{\spbb{2^\flat 4^\flat}}
\spvec{1^{\flat,-}}{\gamma^\mu}{2^{\flat,-}}\Big\}\;, \nn \\
\ell_2^\mu|_{\tS_5} \,=\, {} & 
k_4^{\flat,\mu}
+\frac{z}{2}\frac{\spaa{1^\flat 4^\flat}}{\spaa{1^\flat 3^\flat}}
\spvec{3^{\flat,-}}{\gamma^\mu}{4^{\flat,-}}\;,
\end{align}
\begin{align}
\ell_1^\mu|_{\mc S_6} \,=\, {} &
(1-m_1^2/s_{12})\Big\{k_1^{\flat,\mu}
-\frac{1}{2}\frac{\spaa{1^\flat 4^\flat}}{\spaa{2^\flat 4^\flat}}
\spvec{2^{\flat,-}}{\gamma^\mu}{1^{\flat,-}}\Big\}\;, \nn \\
\ell_2^\mu|_{\tS_6} \,=\, {} & 
k_4^{\flat,\mu}
+\frac{z}{2}\frac{\spbb{1^\flat 4^\flat}}{\spbb{1^\flat 3^\flat}}
\spvec{4^{\flat,-}}{\gamma^\mu}{3^{\flat,-}}\;, \\[2mm]
\ell_1^\mu|_{\tS_7} \,=\, {} &
(1-m_1^2/s_{12}\Big\{k_1^{\flat,\mu}
-\frac{\bxi_1}{2\tau}\frac{\spbb{1^\flat 4^\flat}}{\spbb{2^\flat 4^\flat}}
\spvec{1^{\flat,-}}{\gamma^\mu}{2^{\flat,-}}\Big\}\;, \nn \\ 
\ell_2^\mu|_{\tS_7} \,=\, {} & 
\mu\frac{k_1^\flat\cdot k_4^\flat}{k_1^\flat\cdot k_3^\flat}k_3^{\flat,\mu}
+\frac{z}{2}\frac{\spbb{1^\flat 4^\flat}}{\spbb{1^\flat 3^\flat}}
\spvec{4^{\flat,-}}{\gamma^\mu}{3^{\flat,-}}\;, \\[2mm]
\ell_1^\mu|_{\tS_8} \,=\, {} &
(1-m_1^2/s_{12})\Big\{k_1^{\flat,\mu}
-\frac{\bxi_1}{2\tau}\frac{\spaa{1^\flat 4^\flat}}{\spaa{2^\flat 4^\flat}}
\spvec{2^{\flat,-}}{\gamma^\mu}{1^{\flat,-}}\Big\}\;, \nn \\ 
\ell_2^\mu|_{\tS_8} \,=\, {} & 
\mu\frac{k_1^\flat\cdot k_4^\flat}{k_1^\flat\cdot k_3^\flat}k_3^{\flat,\mu}
+\frac{z}{2}\frac{\spaa{1^\flat 4^\flat}}{\spaa{1^\flat 3^\flat}}
\spvec{3^{\flat,-}}{\gamma^\mu}{4^{\flat,-}}\;.
\end{align}

\section{Structure of Global Poles}
\label{POLESTRUCTURE}
With the discussion of the previous section in mind, we explain that the global
poles are inherited through chiral branchings. To that end, define the following
eight Jacobian global poles in class (a),
\begin{align}
\mc G_1\equiv {} & \mc S_1\cap S_4\;, &
\mc G_2\equiv {} & \mc S_1\cap S_6\;, &
\mc G_3\equiv {} & \mc S_1\cap S_3\;, &
\mc G_4\equiv {} & \mc S_1\cap S_5\;, \nn \\
\mc G_5\equiv {} & \mc S_2\cap S_3\;, &
\mc G_6\equiv {} & \mc S_2\cap S_5\;, &
\mc G_7\equiv {} & \mc S_2\cap S_4\;, &
\mc G_8\equiv {} & \mc S_2\cap S_6\;,
\end{align}
along with the additional poles in numerator insertions,
\begin{align}
\mc G_5\equiv {} & \mc S_1|_{z=0}\;, &
\mc G_{10}\equiv {} & \mc S_1|_{z=0}\;, &
\mc G_{11}\equiv {} & \mc S_1|_{z=\infty}\;, &
\mc G_{12}\equiv {} & \mc S_2|_{z=\infty}\;, \nn \\
\mc G_{13}\equiv {} & \mc S_3|_{z=\infty}\;, &
\mc G_{14}\equiv {} & \mc S_4|_{z=\infty}\;, &
\mc G_{15}\equiv {} & \mc S_5|_{z=\infty}\;, &
\mc G_{16}\equiv {} & \mc S_6|_{z=\infty}\;.
\end{align}
As the total sum of residues for a meromorphic differential form on $\CP^1$
vanishes, only ten residues evaluated at, say, $\{\mc G_1,\dots,\mc G_{10}\}$
are independent.

Moreover, in class (b) we define the global poles
\begin{align}
\mc G_1'\equiv {} & \mc S_1'\cap S_6'\;, &
\mc G_2'\equiv {} & \mc S_1'\cap S_8'\;, &
\mc G_3'\equiv {} & \mc S_3'\cap S_5'\;, &
\mc G_4'\equiv {} & \mc S_3'\cap S_7'\;, &
\mc G_5'\equiv {} & \mc S_1'\cap S_3'\;, \nn \\
\mc G_6'\equiv {} & \mc S_2'\cap S_5'\;, &
\mc G_7'\equiv {} & \mc S_2'\cap S_7'\;, &
\mc G_8'\equiv {} & \mc S_4'\cap S_6'\;, &
\mc G_9'\equiv {} & \mc S_4'\cap S_8'\;, &
\mc G_{10'}\equiv {} & \mc S_2'\cap S_4'\;.
\end{align}
All primed branches are obviously holomorphically parametrized and hence none of
them have singularities for additional finite values of $z$. However, there are
possibly eight residues at infinity,
\begin{align}
\tG_{11}\equiv {} & \mc S_1|_{z=\infty}\;, &
\tG_{12}\equiv {} & \mc S_2|_{z=\infty}\;, &
\tG_{13}\equiv {} & \mc S_3|_{z=\infty}\;, &
\tG_{14}\equiv {} & \mc S_4|_{z=\infty}\;, \nn \\
\tG_{15}\equiv {} & \mc S_5|_{z=\infty}\;, &
\tG_{16}\equiv {} & \mc S_6|_{z=\infty}\;, &
\tG_{17}\equiv {} & \mc S_7|_{z=\infty}\;, &
\tG_{18}\equiv {} & \mc S_8|_{z=\infty}\;.
\end{align}

At this stage, it is not hard to realize that the eight Jacobian poles in class
(a) are inherited by class (b). The map goes as follows,
\begin{align}
(\mc G_1,\mc G_2,\mc G_3,\mc G_4,
\mc G_6,\mc G_7,\mc G_8,\mc G_9)
\;\xrightarrow{\bxi_2\to 0}\;
(\mc G_1,\mc G_2,\mc G_3,\mc G_4,
\mc G_6,\mc G_7,\mc G_8,\mc G_9)
\end{align}
and similarly for the residues at $z = 0$ and $z = \infty$,
\begin{align}
(\mc G_{11},\mc G_{12},\mc G_5,\mc G_{10},
\mc G_{13},\mc G_{14},\mc G_{15},\mc G_{16})
\;\xrightarrow{\bxi_2\to 0}\;
(\tG_{11},\tG_{12},\tG_{13},\tG_{14},
\tG_{15},\tG_{16},\tG_{17},\tG_{18})\;.
\end{align}
The remaining two Jacobian global poles $\tG_5$ and $\tG_{10}$ are located at
the nodal points $\mc S_1\cap \mc S_3$ and $\mc S_2\cap \mc S_4$ respectively,
and are thus generated by chiral branching from $6\to 8$ hepta-cut solutions. By
the one-dimensional Global Residue Theorem, the number of independent residues
is clearly invariant.


\clearpage
\begin{thebibliography}{99}
\bibitem{Britto:2004ap}
  R.~Britto, F.~Cachazo and B.~Feng,
  ``New recursion relations for tree amplitudes of gluons,''
  Nucl.\ Phys.\ B {\bf 715}, 499 (2005)
  [hep-th/0412308].

\bibitem{Britto:2005fq}
  R.~Britto, F.~Cachazo, B.~Feng and E.~Witten,
  ``Direct proof of tree-level recursion relation in Yang-Mills theory,''
  Phys.\ Rev.\ Lett.\  {\bf 94}, 181602 (2005)
  [hep-th/0501052].

\bibitem{Bern:1994cg}
  Z.~Bern, L.~J.~Dixon, D.~C.~Dunbar and D.~A.~Kosower,
  ``Fusing gauge theory tree amplitudes into loop amplitudes,''
  Nucl.\ Phys.\ B {\bf 435}, 59 (1995)
  [hep-ph/9409265].

\bibitem{Bern:1994zx}
  Z.~Bern, L.~J.~Dixon, D.~C.~Dunbar and D.~A.~Kosower,
  ``One loop n point gauge theory amplitudes, unitarity and collinear limits,''
  Nucl.\ Phys.\ B {\bf 425}, 217 (1994)
  [hep-ph/9403226].

\bibitem{Bern:1995db}
  Z.~Bern and A.~G.~Morgan,
  ``Massive loop amplitudes from unitarity,''
  Nucl.\ Phys.\ B {\bf 467}, 479 (1996)
  [hep-ph/9511336].

\bibitem{Bern:1997sc}
  Z.~Bern, L.~J.~Dixon and D.~A.~Kosower,
  ``One loop amplitudes for e+ e- to four partons,''
  Nucl.\ Phys.\ B {\bf 513}, 3 (1998)
  [hep-ph/9708239].

\bibitem{Britto:2004nc}
  R.~Britto, F.~Cachazo and B.~Feng,
  ``Generalized unitarity and one-loop amplitudes in N=4 super-Yang-Mills,''
  Nucl.\ Phys.\ B {\bf 725}, 275 (2005)
  [hep-th/0412103].

\bibitem{Britto:2004nj}
  R.~Britto, F.~Cachazo and B.~Feng,
  ``Computing one-loop amplitudes from the holomorphic anomaly of unitarity
  cuts,''
  Phys.\ Rev.\ D {\bf 71}, 025012 (2005)
  [hep-th/0410179].

\bibitem{Bern:2005hh}
  Z.~Bern, N.~E.~J.~Bjerrum-Bohr, D.~C.~Dunbar and H.~Ita,
  ``Recursive calculation of one-loop QCD integral coefficients,''
  JHEP {\bf 0511}, 027 (2005)
  [hep-ph/0507019].

\bibitem{Bidder:2005ri}
  S.~J.~Bidder, N.~E.~J.~Bjerrum-Bohr, D.~C.~Dunbar and W.~B.~Perkins,
  ``One-loop gluon scattering amplitudes in theories with N < 4
  supersymmetries,''
  Phys.\ Lett.\ B {\bf 612}, 75 (2005)
  [hep-th/0502028].

\bibitem{Britto:2005ha}
  R.~Britto, E.~Buchbinder, F.~Cachazo and B.~Feng,
  ``One-loop amplitudes of gluons in SQCD,''
  Phys.\ Rev.\ D {\bf 72}, 065012 (2005)
  [hep-ph/0503132].

\bibitem{Britto:2006sj}
  R.~Britto, B.~Feng and P.~Mastrolia,
  ``The Cut-constructible part of QCD amplitudes,''
  Phys.\ Rev.\ D {\bf 73}, 105004 (2006)
  [hep-ph/0602178].

\bibitem{Mastrolia:2006ki}
  P.~Mastrolia,
  ``On Triple-cut of scattering amplitudes,''
  Phys.\ Lett.\ B {\bf 644}, 272 (2007)
  [hep-th/0611091].

\bibitem{Brandhuber:2005jw}
  A.~Brandhuber, S.~McNamara, B.~J.~Spence and G.~Travaglini,
  ``Loop amplitudes in pure Yang-Mills from generalised unitarity,''
  JHEP {\bf 0510}, 011 (2005)
  [hep-th/0506068].

\bibitem{Ossola:2006us}
  G.~Ossola, C.~G.~Papadopoulos and R.~Pittau,
  ``Reducing full one-loop amplitudes to scalar integrals at the integrand
  level,''
  Nucl.\ Phys.\ B {\bf 763}, 147 (2007)
  [hep-ph/0609007].

\bibitem{Anastasiou:2006gt}
  C.~Anastasiou, R.~Britto, B.~Feng, Z.~Kunszt and P.~Mastrolia,
  ``Unitarity cuts and Reduction to master integrals in d dimensions for
  one-loop amplitudes,''
  JHEP {\bf 0703}, 111 (2007)
  [hep-ph/0612277].

\bibitem{Bern:2007dw}
  Z.~Bern, L.~J.~Dixon and D.~A.~Kosower,
  ``On-Shell Methods in Perturbative QCD,''
  Annals Phys.\  {\bf 322}, 1587 (2007)
  [arXiv:0704.2798 [hep-ph]].

\bibitem{Forde:2007mi}
  D.~Forde,
  ``Direct extraction of one-loop integral coefficients,''
  Phys.\ Rev.\ D {\bf 75}, 125019 (2007)
  [arXiv:0704.1835 [hep-ph]].

\bibitem{Badger:2008cm}
  S.~D.~Badger,
  ``Direct Extraction Of One Loop Rational Terms,''
  JHEP {\bf 0901}, 049 (2009)
  [arXiv:0806.4600 [hep-ph]].

\bibitem{Giele:2008ve}
  W.~T.~Giele, Z.~Kunszt and K.~Melnikov,
  ``Full one-loop amplitudes from tree amplitudes,''
  JHEP {\bf 0804}, 049 (2008)
  [arXiv:0801.2237 [hep-ph]].

\bibitem{Britto:2006fc}
  R.~Britto and B.~Feng,
  ``Unitarity cuts with massive propagators and algebraic expressions for
  coefficients,''
  Phys.\ Rev.\ D {\bf 75}, 105006 (2007)
  [hep-ph/0612089].

\bibitem{Britto:2007tt}
  R.~Britto and B.~Feng,
  ``Integral coefficients for one-loop amplitudes,''
  JHEP {\bf 0802}, 095 (2008)
  [arXiv:0711.4284 [hep-ph]].

\bibitem{Bern:2010qa}
  Z.~Bern, J.~J.~Carrasco, T.~Dennen, Y.~-t.~Huang and H.~Ita,
  ``Generalized Unitarity and Six-Dimensional Helicity,''
  Phys.\ Rev.\ D {\bf 83}, 085022 (2011)
  [arXiv:1010.0494 [hep-th]].

\bibitem{Anastasiou:2006jv}
  C.~Anastasiou, R.~Britto, B.~Feng, Z.~Kunszt and P.~Mastrolia,
  ``D-dimensional unitarity cut method,''
  Phys.\ Lett.\ B {\bf 645}, 213 (2007)
  [hep-ph/0609191].

\bibitem{Mastrolia:2009dr} 
  P.~Mastrolia,
  ``Double-Cut of Scattering Amplitudes and Stokes' Theorem,''
  Phys.\ Lett.\ B {\bf 678}, 246 (2009)
  [arXiv:0905.2909 [hep-ph]].

\bibitem{Ellis:2007br}
  R.~K.~Ellis, W.~T.~Giele and Z.~Kunszt,
  ``A Numerical Unitarity Formalism for Evaluating One-Loop Amplitudes,''
  JHEP {\bf 0803}, 003 (2008)
  [arXiv:0708.2398 [hep-ph]].

\bibitem{Berger:2008sj}
  C.~F.~Berger, Z.~Bern, L.~J.~Dixon, F.~Febres Cordero, D.~Forde, H.~Ita,
  D.~A.~Kosower and D.~Maitre,
  ``An Automated Implementation of On-Shell Methods for One-Loop Amplitudes,''
  Phys.\ Rev.\ D {\bf 78}, 036003 (2008)
  [arXiv:0803.4180 [hep-ph]].

\bibitem{Ossola:2007ax}
  G.~Ossola, C.~G.~Papadopoulos and R.~Pittau,
  ``CutTools: A Program implementing the OPP reduction method to compute
  one-loop amplitudes,''
  JHEP {\bf 0803}, 042 (2008)
  [arXiv:0711.3596 [hep-ph]].

\bibitem{Mastrolia:2008jb}
  P.~Mastrolia, G.~Ossola, C.~G.~Papadopoulos and R.~Pittau,
  ``Optimizing the Reduction of One-Loop Amplitudes,''
  JHEP {\bf 0806}, 030 (2008)
  [arXiv:0803.3964 [hep-ph]].

\bibitem{Giele:2008bc}
  W.~T.~Giele and G.~Zanderighi,
  ``On the Numerical Evaluation of One-Loop Amplitudes: The Gluonic Case,''
  JHEP {\bf 0806}, 038 (2008)
  [arXiv:0805.2152 [hep-ph]].

\bibitem{Berger:2009zg}
  C.~F.~Berger, Z.~Bern, L.~J.~Dixon, F.~Febres Cordero, D.~Forde, T.~Gleisberg,
H.~Ita and D.~A.~Kosower {\it et al.},
  ``Precise Predictions for $W$ + 3 Jet Production at Hadron Colliders,''
  Phys.\ Rev.\ Lett.\  {\bf 102}, 222001 (2009)
  [arXiv:0902.2760 [hep-ph]].

\bibitem{Badger:2010nx}
  S.~Badger, B.~Biedermann and P.~Uwer,
  ``NGluon: A Package to Calculate One-loop Multi-gluon Amplitudes,''
  Comput.\ Phys.\ Commun.\  {\bf 182}, 1674 (2011)
  [arXiv:1011.2900 [hep-ph]].

\bibitem{Berger:2010zx}
  C.~F.~Berger, Z.~Bern, L.~J.~Dixon, F.~Febres Cordero, D.~Forde, T.~Gleisberg,
  H.~Ita and D.~A.~Kosower {\it et al.},
  ``Precise Predictions for W + 4 Jet Production at the Large Hadron
  Collider,''
  Phys.\ Rev.\ Lett.\  {\bf 106}, 092001 (2011)
  [arXiv:1009.2338 [hep-ph]].

\bibitem{Hirschi:2011pa}
  V.~Hirschi, R.~Frederix, S.~Frixione, M.~V.~Garzelli, F.~Maltoni and
  R.~Pittau,
  ``Automation of one-loop QCD corrections,''
  JHEP {\bf 1105}, 044 (2011)
  [arXiv:1103.0621 [hep-ph]].

\bibitem{Bern:1997nh}
  Z.~Bern, J.~S.~Rozowsky and B.~Yan,
  ``Two loop four gluon amplitudes in N=4 superYang-Mills,''
  Phys.\ Lett.\ B {\bf 401}, 273 (1997)
  [hep-ph/9702424].

\bibitem{Bern:2000dn}
  Z.~Bern, L.~J.~Dixon and D.~A.~Kosower,
  ``A Two loop four gluon helicity amplitude in QCD,''
  JHEP {\bf 0001}, 027 (2000)
  [hep-ph/0001001].

\bibitem{Glover:2001af}
  E.~W.~N.~Glover, C.~Oleari and M.~E.~Tejeda-Yeomans,
  ``Two loop QCD corrections to gluon-gluon scattering,''
  Nucl.\ Phys.\ B {\bf 605}, 467 (2001)
  [hep-ph/0102201].

\bibitem{Bern:2002tk}
  Z.~Bern, A.~De Freitas and L.~J.~Dixon,
  ``Two loop helicity amplitudes for gluon-gluon scattering in QCD and
  supersymmetric Yang-Mills theory,''
  JHEP {\bf 0203}, 018 (2002)
  [hep-ph/0201161].

\bibitem{Anastasiou:2000kg}
  C.~Anastasiou, E.~W.~N.~Glover, C.~Oleari and M.~E.~Tejeda-Yeomans,
  ``Two-loop QCD corrections to the scattering of massless distinct quarks,''
  Nucl.\ Phys.\ B {\bf 601}, 318 (2001)
  [hep-ph/0010212].

\bibitem{Anastasiou:2000ue}
  C.~Anastasiou, E.~W.~N.~Glover, C.~Oleari and M.~E.~Tejeda-Yeomans,
  ``Two loop QCD corrections to massless identical quark scattering,''
  Nucl.\ Phys.\ B {\bf 601}, 341 (2001)
  [hep-ph/0011094].

\bibitem{Anastasiou:2001sv}
  C.~Anastasiou, E.~W.~N.~Glover, C.~Oleari and M.~E.~Tejeda-Yeomans,
  ``Two loop QCD corrections to massless quark gluon scattering,''
  Nucl.\ Phys.\ B {\bf 605}, 486 (2001)
  [hep-ph/0101304].

\bibitem{Buchbinder:2005wp}
  E.~I.~Buchbinder and F.~Cachazo,
  ``Two-loop amplitudes of gluons and octa-cuts in N=4 super Yang-Mills,''
  JHEP {\bf 0511}, 036 (2005)
  [hep-th/0506126].

\bibitem{Cachazo:2008vp}
  F.~Cachazo,
  ``Sharpening The Leading Singularity,''
  arXiv:0803.1988 [hep-th].

\bibitem{Gluza:2010ws}
  J.~Gluza, K.~Kajda and D.~A.~Kosower,
  ``Towards a Basis for Planar Two-Loop Integrals,''
  Phys.\ Rev.\ D {\bf 83}, 045012 (2011)
  [arXiv:1009.0472 [hep-th]].

\bibitem{Schabinger:2011dz}
  R.~M.~Schabinger,
  ``A New Algorithm For The Generation Of Unitarity-Compatible Integration By
  Parts Relations,''
  JHEP {\bf 1201}, 077 (2012)
  [arXiv:1111.4220 [hep-ph]].

\bibitem{Kosower:2011ty}
  D.~A.~Kosower and K.~J.~Larsen,
  ``Maximal Unitarity at Two Loops,''
  Phys.\ Rev.\ D {\bf 85}, 045017 (2012)
  [arXiv:1108.1180 [hep-th]].

\bibitem{CaronHuot:2012ab}
  S.~Caron-Huot and K.~J.~Larsen,
  ``Uniqueness of two-loop master contours,''
  JHEP {\bf 1210}, 026 (2012)
  [arXiv:1205.0801 [hep-ph]].

\bibitem{Johansson:2012zv}
  H.~Johansson, D.~A.~Kosower and K.~J.~Larsen,
  ``Two-Loop Maximal Unitarity with External Masses,''
  Phys.\ Rev.\ D {\bf 87}, 025030 (2013)
  [arXiv:1208.1754 [hep-th]].

\bibitem{Johansson:2013sda} 
  H.~Johansson, D.~A.~Kosower and K.~J.~Larsen,
  ``Maximal Unitarity for the Four-Mass Double Box,''
  arXiv:1308.4632 [hep-th].

\bibitem{Sogaard:2013yga} 
  M.~Sogaard,
  ``Global Residues and Two-Loop Hepta-Cuts,''
  JHEP {\bf 1309}, 116 (2013)
  [arXiv:1306.1496 [hep-th]].

\bibitem{Sogaard:2013fpa} 
  M.~Sogaard and Y.~Zhang,
  ``Multivariate Residues and Maximal Unitarity,''
  JHEP {\bf 1312}, 008 (2013)
  [arXiv:1310.6006 [hep-th]].

\bibitem{Sogaard:2014ila} 
  M.~Sogaard and Y.~Zhang,
  ``Unitarity Cuts of Integrals with Doubled Propagators,''
  arXiv:1403.2463 [hep-th].

\bibitem{Larsen:2012sx}
  K.~J.~Larsen,
  ``Global Poles of the Two-Loop Six-Point N=4 SYM integrand,''
  Phys.\ Rev.\ D {\bf 86}, 085032 (2012)
  [arXiv:1205.0297 [hep-th]].

\bibitem{Badger:2012dp}
  S.~Badger, H.~Frellesvig and Y.~Zhang,
  ``Hepta-Cuts of Two-Loop Scattering Amplitudes,''
  JHEP {\bf 1204}, 055 (2012)
  [arXiv:1202.2019 [hep-ph]].

\bibitem{Mastrolia:2011pr}
  P.~Mastrolia and G.~Ossola,
  ``On the Integrand-Reduction Method for Two-Loop Scattering Amplitudes,''
  JHEP {\bf 1111}, 014 (2011)
  [arXiv:1107.6041 [hep-ph]].

\bibitem{Badger:2012dv}
  S.~Badger, H.~Frellesvig and Y.~Zhang,
  ``An Integrand Reconstruction Method for Three-Loop Amplitudes,''
  JHEP {\bf 1208}, 065 (2012)
  [arXiv:1207.2976 [hep-ph]].

\bibitem{Zhang:2012ce}
  Y.~Zhang,
  ``Integrand-Level Reduction of Loop Amplitudes by Computational Algebraic
  Geometry Methods,''
  JHEP {\bf 1209}, 042 (2012)
  [arXiv:1205.5707 [hep-ph]].

\bibitem{Badger:2013gxa}
  S.~Badger, H.~Frellesvig and Y.~Zhang,
  ``A Two-Loop Five-Gluon Helicity Amplitude in QCD,''
  arXiv:1310.1051 [hep-ph].

\bibitem{Feng:2012bm}
  B.~Feng and R.~Huang,
  ``The classification of two-loop integrand basis in pure four-dimension,''
  JHEP {\bf 1302}, 117 (2013)
  [arXiv:1209.3747 [hep-ph]].

\bibitem{Mastrolia:2012an}
  P.~Mastrolia, E.~Mirabella, G.~Ossola and T.~Peraro,
  ``Scattering Amplitudes from Multivariate Polynomial Division,''
  Phys.\ Lett.\ B {\bf 718}, 173 (2012)
  [arXiv:1205.7087 [hep-ph]].

\bibitem{Mastrolia:2012wf}
  P.~Mastrolia, E.~Mirabella, G.~Ossola and T.~Peraro,
  ``Integrand-Reduction for Two-Loop Scattering Amplitudes through Multivariate
  Polynomial Division,''
  arXiv:1209.4319 [hep-ph].

\bibitem{Mastrolia:2012du}
  P.~Mastrolia, E.~Mirabella, G.~Ossola, T.~Peraro and H.~van Deurzen,
  ``The Integrand Reduction of One- and Two-Loop Scattering Amplitudes,''
  PoS LL {\bf 2012} (2012) 028
  [arXiv:1209.5678 [hep-ph]].

\bibitem{Kleiss:2012yv}
  R.~H.~P.~Kleiss, I.~Malamos, C.~G.~Papadopoulos and R.~Verheyen,
  ``Counting to One: Reducibility of One- and Two-Loop Zmplitudes at the
  Integrand Level,''
  JHEP {\bf 1212}, 038 (2012)
  [arXiv:1206.4180 [hep-ph]].

\bibitem{Huang:2013kh}
  R.~Huang and Y.~Zhang,
  ``On Genera of Curves from High-loop Generalized Unitarity Cuts,''
  arXiv:1302.1023 [hep-ph].

\bibitem{Mastrolia:2013kca}
  P.~Mastrolia, E.~Mirabella, G.~Ossola and T.~Peraro,
  ``Multiloop Integrand Reduction for Dimensionally Regulated Amplitudes,''
  arXiv:1307.5832 [hep-ph].

\bibitem{Feng:2014nwa} 
  B.~Feng, J.~Zhen, R.~Huang and K.~Zhou,
  ``Integral Reduction by Unitarity Method for Two-loop Amplitudes: A Case
  Study,''
  arXiv:1401.6766 [hep-th].

\bibitem{Bern:2009xq}
  Z.~Bern, J.~J.~M.~Carrasco, H.~Ita, H.~Johansson and R.~Roiban,
  ``On the Structure of Supersymmetric Sums in Multi-Loop Unitarity Cuts,''
  Phys.\ Rev.\ D {\bf 80}, 065029 (2009)
  [arXiv:0903.5348 [hep-th]].

\bibitem{Sogaard:2011pr}
  M.~Sogaard,
  ``Supersums for all supersymmetric amplitudes,''
  Phys.\ Rev.\ D {\bf 84}, 065011 (2011)
  [arXiv:1106.3785 [hep-th]].

\bibitem{Caola:2014lpa} 
  F.~Caola, J.~M.~Henn, K.~Melnikov and V.~A.~Smirnov,
  ``Non-planar master integrals for the production of two off-shell vector
  bosons in collisions of massless partons,''
  arXiv:1404.5590 [hep-ph].

\bibitem{Smirnov:2013dia}
  A.~V.~Smirnov and V.~A.~Smirnov,
  ``FIRE4, LiteRed and accompanying tools to solve integration by parts
  relations,''
  arXiv:1302.5885 [hep-ph].

\bibitem{MR507725}
  P.~Griffiths, J.~Harris, ``Principles of Algebraic Geometry".
  Wiley-Interscience [John Wiley \& Sons], New York,
  1978.

\bibitem{MR0463157}
  R.~Hartshorne, ``Algebraic Geometry". Springer-Verlag, New York,
  1977. Graduate Texts in Mathematics, No. 52.

\bibitem{M2}
  D.~R. Grayson and M.~E. Stillman, ``Macaulay2, a software system for research
  in algebraic geometry''. Available at {\tt
  http://www.math.uiuc.edu/Macaulay2/}.

\bibitem{MR2161985}
  E.~Cattani and A.~Dickenstein. ``Introduction to residues and
  resultants: Solving polynomial equations''. Springer Berlin
  Heidelberg, 2005. 

\bibitem{MR2122859}
  D.~A.Cox. and J.~Little, and Donal .~O'Shea, ``Using algebraic
  geometry'', Springer, New York, 2005
\end{thebibliography}
\end{document}